\documentclass[a4paper,11pt]{article}
\pdfoutput=1 % 

\usepackage{jcappub}
\usepackage{amsmath,amssymb,graphicx,float,slashed,xcolor,multicol}
\usepackage{tabularx}
\usepackage{url}
\usepackage{footmisc}
\usepackage{amsfonts}
\usepackage{cancel}
\usepackage[toc,page]{appendix}
\usepackage{color}
\usepackage{pifont}
\usepackage{epstopdf}
\usepackage{comment}
\usepackage{booktabs} 
\usepackage{natbib}
\usepackage{array}
\usepackage{mathrsfs}
\usepackage{mathtools}
\usepackage[normalem]{ulem}
\usepackage{bbold}
\usepackage{enumitem}
\usepackage{cleveref}
\usepackage{multirow}
\usepackage{upgreek}
\usepackage{romannum}
\allowdisplaybreaks
\usepackage{bbm}    % for \mathbbm{1} (unit matrix)
\usepackage{xspace}				% For spacing after command

\usepackage{tikz}
\usetikzlibrary{arrows,shapes}
\usetikzlibrary{trees}
\usetikzlibrary{matrix} 
\usetikzlibrary{positioning}				% For "above of=" commands
\usetikzlibrary{calc,through}				% For coordinates
\usetikzlibrary{decorations.pathreplacing}  % For curly braces
\usepackage{pgffor}							% For repeating patterns

\newcommand*{\rom}[1]{\expandafter\@slowromancap\romannumeral #1@}

\def\beq{\begin{equation}}
\def\eeq{\end{equation}}
\def\beqa{\begin{eqnarray}}
\def\eeqa{\end{eqnarray}}
\title{Ultra-High-Energy Cosmic Ray Boosted Relic Neutrinos}
\author[a]{Jiajie Zhang,}
\author[a]{Jiajun Liao}
\affiliation[a]{School of Physics, Sun Yat-sen University, No. 135, Xingang Xi Road, Guangzhou, 510275, P. R. China }
\emailAdd{zhangjj253@mail2.sysu.edu.cn}
\emailAdd{liaojiajun@mail.sysu.edu.cn}

\abstract{Ultra-high-energy cosmic rays (UHECRs) can boost relic neutrinos to high energies through Standard Model (SM) neutral-current interactions, providing an indirect probe of the cosmic neutrino background (C$\nu$B). 
In this work, we perform a systematic study of the diffuse UHECR-boosted C$\nu$B flux including elastic neutrino--nucleon scattering (ES), coherent elastic neutrino--nucleus scattering (COH), incoherent neutrino--nucleus scattering (INCOH), baryon-resonance production (RES), and deep inelastic scattering (DIS). 
For the UHECR flux, we use mixed-composition spectra obtained from the UHECR propagation code PriNCe and from the H3a and H4a implementations of the Hillas model, together with SFR, QSO, and GRB source evolution models. 
We find a clear hierarchy of scattering channels as a function of boosted neutrino energy. 
The coherent scattering dominates at low-energy neutrino flux for heavy nuclear components, while ES and INCOH become important once individual nucleons are resolved.
The RES channel gives a non-negligible contribution in the high-energy region, and DIS appears only at the highest energies and is most visible for the H4a case.
Using current IceCube and Pierre Auger Observatory data, we obtain upper limits on the C$\nu$B overdensity.
Our results show that reliable predictions of the UHECR-boosted C$\nu$B signal require a combined treatment of the relevant SM scattering channels, UHECR composition, source evolution, and the neutrino mass spectrum.} 

\begin{document}

\maketitle

\flushbottom

%%%%%%%%%%%%%%%%%%%%%%%%%%%
\section{Introduction}
\label{sec:Intro}

The cosmic neutrino background (C$\nu$B), or relic neutrino background, is a robust prediction of the standard hot Big Bang cosmology~\cite{Lesgourgues:2006nd,TopicalConvenersKNAbazajianJECarlstromATLee:2013bxd}.
In the standard thermal history of the Universe, neutrinos were in thermal equilibrium at early times and decoupled from the thermal bath when the temperature dropped to $\mathcal{O}(1)~{\rm MeV}$~\cite{Mangano:2005cc,Lesgourgues:2013sjj}. 
After decoupling, they propagated freely with their momenta redshifting as the Universe expanded, forming a nearly thermal background that persists until today~\cite{Mangano:2005cc,Lesgourgues:2013sjj}. 
Despite its fundamental importance, the C$\nu$B has not yet been detected directly, mainly because its present-day energy scale is extremely low and they only interact with the SM particles via the weak interactions. 
The present-day C$\nu$B temperature and number density are approximately~\cite{Lesgourgues:2006nd,Giunti:2007ry,Lesgourgues:2013sjj}
\begin{equation}
T_{\nu,0}
=
\left(\frac{4}{11}\right)^{1/3}T_{\gamma,0}
\simeq 1.95~{\rm K}
\simeq 1.7\times10^{-4}~{\rm eV},
\qquad
n_{\nu,0}\simeq 56~{\rm cm}^{-3},
\end{equation}
where $T_{\nu,0}$ and $T_{\gamma,0}$ denote the present temperatures of the C$\nu$B and the cosmic microwave background (CMB), respectively, and $n_{\nu,0}$ is the neutrino number density per flavor and per helicity state.

A variety of strategies have been proposed to probe the C$\nu$B. 
Early ideas focused on the coherent scattering of relic neutrinos from macroscopic material targets. 
Although macroscopic coherence can enhance the collective response of the target~\cite{Opher:1974drq,Lewis:1979mu}, subsequent analyses showed that the energy or momentum transfer to realistic detectors is extremely small, making these approaches  experimentally challenging~\cite{Akhmedov:2018wlf,Langacker:1982ih}. 
Another class of proposals uses ultra-high-energy cosmogenic neutrinos as probes of the relic background: resonant annihilation on the C$\nu$B can generate absorption dips in the cosmic-neutrino spectrum, for example through the $Z$ resonance~\cite{Eberle:2004ua} or through lower-mass SM vector-meson resonances~\cite{Brdar:2022kpu}, and can lead to a cutoff in the cosmogenic neutrino flux and a possible flux of highly boosted dark-sector particles~\cite{Yin:2018yjn}.
More recently, it has also been proposed that relic neutrinos may induce parametric fluorescence in atomic or molecular systems, where coherent scattering with molecular energy levels can produce an infrared signal photon under suitable resonance and phase-matching conditions~\cite{Huang:2025yqu}. 
In the terrestrial laboratory, PTOLEMY aims to detect the C$\nu$B through neutrino capture on tritium, but this method faces stringent requirements on tritium localization and the endpoint energy resolution~\cite{PTOLEMY:2018jst,PTOLEMY:2022ldz,PTOLEMY:2024lzs}.
At present, the strongest direct laboratory constraint on the local relic-neutrino overdensity is provided by KATRIN, which sets an upper bound of $\eta<9.7\times10^{10}$
at the $90\%$ confidence level~\cite{KATRIN:2022kkv}.

The prospects for detecting the C$\nu$B depend on the relic-neutrino abundance.
The possible enhancement of the C$\nu$B density is also subject to cosmological constraints. For a diffuse cosmological neutrino population, requiring its energy density to remain below the present critical density gives $\langle E_\nu\rangle n_\nu\lesssim\rho_c$.
Together with the fermionic phase-space constraint, Ref.~\cite{Bondarenko:2023ukx} obtained $\eta\lesssim2\times10^4$ as the maximal permissible cosmological overdensity under the most accommodating assumptions.
Nevertheless, a variety of nonstandard cosmological histories and interactions beyond the Standard Model can substantially modify the relic-neutrino background relative to the standard $\Lambda$CDM prediction. 
For example, long-lived dark-sector particles or Majorons can decay into neutrinos after neutrino decoupling, generating additional low-energy or nonthermal relic-neutrino populations
~\cite{McKeen:2018xyz,Chacko:2018uke,Bondarenko:2020vta,Batell:2024hzo}.
More recently, late-time production of cold active neutrinos from the decay of a light scalar field has also been proposed as a mechanism for enhancing the diffuse cosmological C$\nu$B
~\cite{Cline:2026tkp}.
New neutrino interactions can also generate large local relic-neutrino overdensities.
In particular, an attractive interaction mediated by a new light scalar can lead to the formation of stable neutrino bound systems or $\nu$-clusters once the relic neutrinos become nonrelativistic~\cite{Smirnov:2022sfo,Kaplan:2024ydw}. Such bound systems can reach local overdensities of order $\eta\sim10^7$ for $m_\nu\sim0.1~{\rm eV}$~\cite{Smirnov:2022sfo}.

Cosmic ray (CR) scattering provides a complementary indirect probe of the C$\nu$B by boosting relic neutrinos to energies relevant for high-energy neutrino telescopes. This possibility was discussed early by Hara and Sato and has recently been revisited in several studies~\cite{Hara:1979he,Hara:1980mz,Ciscar-Monsalvatje:2024tvm,Herrera:2024upj,DeMarchi:2024zer,Zhang:2025rqh,Herrera:2026pzj}. 
The existing literature has clarified several parts of this phenomenology, but the scattering processes are often restricted to a subset of scattering channels and to specific assumptions about the UHECR composition.
For a proton-dominated CR composition, elastic neutrino-proton scattering and DIS have been used to calculate the diffuse boosted C$\nu$B flux and assess its detectability at future high-energy neutrino telescopes~\cite{Herrera:2026pzj}. 
That study also showed that the charged-current contribution is subdominant compared with the neutral-current contribution.
For nuclear cosmic rays, Ref.~\cite{Zhang:2025rqh} showed that coherent scattering dominates at low boosted-neutrino energies, while elastic and incoherent scattering processes become important at higher energies. 
A recent propagation-based Monte Carlo study further investigated the impact of detailed UHECR propagation and showed that, for $m_\nu=0.1~{\rm eV}$, C$\nu$B overdensities below $\eta\sim10^8$ have little effect on the UHECR energy spectrum~\cite{Azeredo:2026qnc}. 
% Cosmic-ray reservoirs have also been considered as possible sites for boosted C$\nu$B production, where the local cosmic-ray density and nuclear composition can enhance the signal~\cite{DeMarchi:2024zer}. 
Related CR boosting mechanisms have also been considered for diffuse supernova neutrinos~\cite{Herrera:2025pdn,Sandrock:2025nzb}, and possible electromagnetic signatures from secondary particles produced in charged-current interactions have been investigated~\cite{Herrera:2026bie}. 
These studies provide important ingredients for the study of CR-boosted C$\nu$B, but a mixed-composition UHECR calculation requires a consistent treatment that tracks how the relevant neutral-current description changes with momentum transfer and hadronic invariant mass.
For the direct neutral-current boosted relic-neutrino signal considered here, the relevant scattering channels should include elastic neutrino--nucleon scattering (ES), coherent elastic neutrino--nucleus scattering (COH), incoherent neutrino--nucleus scattering (INCOH), baryon-resonance production (RES), and deep inelastic scattering (DIS). 
These channels are closely related to the energy and composition of the incident UHECRs: low momentum transfer favors coherent nuclear scattering, while higher-energy UHECRs can access nucleon-level elastic scattering, baryon-resonance production, and eventually the DIS regime.

Since the relative strength of these scattering channels is dependent on the incident UHECR energy and composition, the prediction of boosted C$\nu$B signals strongly relies on the UHECR flux model.
Some previous calculations assumed a purely protonic composition, whereas Pierre Auger data favor a mixed composition that becomes heavier toward the highest energies, with the proton fraction dropping below $\sim 10\%$ above $\sim10~{\rm EeV}$~\cite{PierreAuger:2022atd,Ehlert:2023btz}. 
In this work, we therefore consider three descriptions of the UHECR flux in order to quantify the dependence of the boosted C$\nu$B signal on the UHECR spectrum, composition, and maximum rigidity. 
The first is based on propagation calculations with the UHECR propagation code PriNCe~\cite{Heinze:2019jou}, which provides redshift-dependent mixed-composition spectra. 
The other two are based on the Hillas model, namely the H3a and H4a implementations~\cite{Hillas:2005cs,Gaisser:2011klf,Gaisser:2013bla}.  
This allows us to assess how different assumptions about the UHECR composition and maximum rigidity will affect the boosted C$\nu$B flux and the relative strength of different scattering channels. 
For each UHECR flux description, we consider three representative source evolution models: the star formation rate (SFR), quasi-stellar objects (QSO), and gamma-ray bursts (GRB). 
This setup allows us to examine how the source evolution affects the predicted boosted C$\nu$B signal.

In this work, we present a systematic calculation of the diffuse UHECR-boosted C$\nu$B flux over the relevant SM neutral-current scattering regimes. 
We compute the ES, COH, INCOH, RES, and DIS contributions within a common framework, covering the coherent nuclear regime at low momentum transfer, the nucleon-resolved regime, the baryon-resonance region, and the deep-inelastic partonic regime at large momentum transfer.
This channel-by-channel treatment allows us to identify the dominant scattering mechanisms in different boosted-neutrino energy ranges and for different UHECR flux models. 
For $m_\nu=0.1~{\rm eV}$, we find that the COH contribution is important at low boosted-neutrino energies, while ES and INCOH become important at higher energies. 
The RES channel provides a visible contribution in the high-energy region for all UHECR flux models considered in this work. 
The DIS contribution is negligible for the PriNCe and H3a flux models, but becomes important in the H4a case.

The remainder of this paper is organized as follows.
In Sec.~\ref{sec:cross_sections}, we present the SM neutral-current cross sections for ES, COH, INCOH, RES, and DIS processes relevant to UHECR--C$\nu$B interactions.
In Sec.~\ref{sec:cr_flux}, we introduce the UHECR flux models and source-evolution benchmarks used in the calculation.
In Sec.~\ref{sec:boosted_flux}, we present the resulting boosted C$\nu$B spectra, decompose the contributions from different scattering channels and UHECR components. 
In Sec.~\ref{sec:eta_constraints}, we obtain constraints on the C$\nu$B overdensity using current data from IceCube and the Pierre Auger Observatory, and determine allowed regions from the observed KM3NeT event.
Finally, we summarize our conclusions in Sec.~\ref{sec:Conclusions}.

%%%%%%%%%%%%%%%%%%%%%%%%%%%%%
\section{Standard Model neutral-current scattering channels}
\label{sec:cross_sections}

UHECRs can boost the C$\nu$B through Standard Model neutral-current interactions with cosmic-ray protons or nuclei.
The appropriate description of the scattering depends on the momentum transfer and on which substructure of the incident CR is resolved, ranging from the nucleus as a whole to individual nucleons and, at sufficiently high energies, to quark and antiquark constituents.

For an outgoing boosted C$\nu$B neutrino with laboratory-frame energy $E_\nu$, we define the positive momentum transfer as $Q^2\equiv -q^2\simeq 2m_\nu E_\nu$, where $m_\nu$ is the relic-neutrino mass.
The value of $Q^2$ determines which scattering regime is relevant.
In the low-$Q^2$ region, the exchanged $Z$ boson cannot resolve the internal structure of the nucleus, so the neutrino scatters coherently off the entire nucleus; this gives the coherent elastic neutrino--nucleus scattering (CE$\nu$NS) contribution.
Once the momentum transfer becomes large enough to resolve individual nucleons, the coherent enhancement is suppressed and the incoherent nuclear contribution becomes relevant.
For larger hadronic excitation energies, neutrino--nucleon scattering can produce baryon resonances, which we denote as the RES contribution.
When both the momentum transfer and the hadronic invariant mass are sufficiently large, the scattering enters the deep inelastic scattering regime, where the neutrino probes quark and antiquark constituents of the nucleon.
Accordingly, for nuclear UHECR components we include COH, INCOH, RES, and DIS, whereas for proton UHECRs only ES, RES, and DIS are present.

For the relic-neutrino sector, we focus on the lightest neutrino mass eigenstate with $m_1 \gtrsim 10^{-3}~{\rm eV}$.
In this regime, the relic neutrinos are nonrelativistic today, since their masses are larger than the present C$\nu$B temperature, $T_{\nu,0}\simeq 1.7\times10^{-4}~{\rm eV}$, and can be approximated as being at rest in the laboratory frame~\cite{Zhang:2025rqh}. 
If the lightest state were still relativistic, $m_1\ll T_{\nu,0}$, the cross section would scale as $\sigma\propto T_{\nu,0}$. 
This would give a smaller cross section than in the nonrelativistic case considered here, where $\sigma\propto m_\nu$.
Therefore, even if $m_1$ were relativistic, the total boosted C$\nu$B signal would still be dominated by the heavier eigenstates $m_2$ and $m_3$, which remain nonrelativistic.
We assume Dirac neutrinos in the following calculation.
The corresponding Majorana case was studied in Ref.~\cite{Zhang:2025rqh}, where it was shown that the difference between Dirac and Majorana scattering cross sections is negligible for the UHECR kinematics considered here.
We consider a lepton-symmetric cosmology with equal number densities of relic neutrinos and antineutrinos~\cite{Zhang:2025rqh,Grohs:2020xxd,Long:2014zva}.
All cross sections below are defined as the sum of the neutrino and antineutrino contributions.

\subsection{Elastic neutrino--nucleon scattering}
\label{subsec:ES}

Elastic neutrino--nucleon scattering is the basic neutral-current process once the momentum transfer resolves an individual proton or neutron. 
It contributes directly to the proton UHECR and also provides the nucleon-level building block for the incoherent nuclear contribution when the momentum transfer is large enough to resolve nucleons inside a cosmic-ray nucleus. 
We therefore first introduce the elastic cross section for a nucleon target, $N=p,n$, before using it in Sec.~\ref{subsec:incoherent} to construct the INCOH contribution for nuclear UHECR components.

In the laboratory frame, where the incoming C$\nu$B neutrino is approximately at rest, the neutral-current elastic differential cross section summed over the neutrino and antineutrino contributions can be written as~\cite{Giunti:2007ry,DeMarchi:2024zer,Zhang:2025rqh}
\begin{equation}
\frac{d\sigma_{\nu N}^{\rm ES}}{dE_\nu}
=
\frac{G_F^2 m_\nu m_N^4}{\pi (s-m_N^2)^2}
\left[
A_N(Q^2)
+
C_N(Q^2)\frac{(s-u)^2}{m_N^4}
\right]\,,
\label{eq:ES_general}
\end{equation}
where $m_N$ is the nucleon mass, and $E_\nu$ is the energy of the outgoing boosted C$\nu$B neutrino.
The incoming nucleon energy is denoted by $E_N$: for a proton UHECR it is the proton energy, while for a nuclear UHECR, it corresponds to the energy carried by the struck nucleon in the incoherent approximation. 
The Mandelstam variables are approximated as
$
s\simeq m_N^2+2m_\nu E_N
$,
$
t\equiv -Q^2
$,
and
$
u\simeq 2m_\nu^2+2m_N^2-s+Q^2
$.
For the boosted-neutrino energies considered here, the momentum transfer satisfies
$
Q^2\simeq 2m_\nu(E_\nu-m_\nu)\simeq 2m_\nu E_\nu
$.
The coefficient functions $A_N(Q^2)$ and $C_N(Q^2)$ are expressed in terms of the neutral-current nucleon form factors and are summarized in Appendix~\ref{app:ES_coefficients}.
For ES, the kinematic upper limit for the outgoing boosted relic neutrino energy is
$ E_\nu^{\max}(E_{N_i}) = E_{N_i}^2/ \left[ E_{N_i}+m_N^2/(2m_\nu) \right] $.

\subsection{Coherent elastic neutrino--nucleus scattering}
\label{subsec:coherent}

Coherent elastic neutrino--nucleus scattering (also known as CE$\nu$NS) occurs when the momentum transfer in the scattering process is small enough that the interaction cannot resolve the individual nucleons inside the nucleus. 
In the low momentum transfer regime, the neutral-current amplitudes from different nucleons add coherently, and the cross section can be substantially enhanced~\cite{Freedman:1973yd}. 
In the standard CE$\nu$NS experiment, an incident neutrino scatters off a nuclear target that is approximately at rest in the laboratory frame.
This process was first observed by the COHERENT Collaboration in 2017 using accelerator-produced neutrinos~\cite{COHERENT:2017ipa}. 
It has since been further measured by COHERENT with several target materials~\cite{COHERENT:2020iec,COHERENT:2024axu,COHERENT:2021xmm,COHERENT:2026yje}, and complementary evidence has also been reported using solar neutrinos and reactor antineutrinos in terrestrial detectors~\cite{XENON:2024ijk,PandaX:2024muv,Ackermann:2025obx}. 
In the CR-boosted C$\nu$B process, by contrast, a highly energetic CR nucleus scatters off a relic neutrino that is approximately at rest. 
Although the laboratory-frame kinematics are reversed, the coherence condition is still determined by the momentum transfer $Q$: coherent scattering is maintained only when the nucleus is not resolved, approximately $Q \lesssim 1/R_A$, where $R_A$ is the nuclear radius.

For a cosmic-ray nucleus $\mathcal{N}_i$, the coherent differential cross section in the relic-neutrino rest frame is~\cite{Zhang:2025rqh}
\begin{equation}
\frac{d\sigma_{\nu \mathcal{N}_i}^{\rm COH}}{dE_\nu}
=
\frac{G_F^2 m_\nu}{\pi}
Q_{W,i}^2
\left(
1-\frac{E_\nu}{E_{\mathcal{N}_i}}
-\frac{m_{\mathcal{N}_i}^2 E_\nu}{2m_\nu E_{\mathcal{N}_i}^2}
\right)
F^2(Q^2)\,,
\label{eq:coherent}
\end{equation}
where $m_{\mathcal{N}_i}$ and $E_{\mathcal{N}_i}$ are the mass and energy of the incoming cosmic-ray nucleus, respectively.
The quantity $Q_{W,i}=Z_i g_V^p+N_i g_V^n$ is the nuclear weak charge, with $Z_i$ and $N_i$ denoting the proton and neutron numbers of $\mathcal{N}_i$. 
The vector couplings are $g_V^p=\frac{1}{2}-2\sin^2\theta_W$ and $g_V^n=-\frac{1}{2}$, and we take $\sin^2\theta_W=0.238$ in the calculation~\cite{ParticleDataGroup:2024cfk}. 
For COH, elastic kinematics gives
$
E_\nu^{\max}(E_{\mathcal{N}_i})
=
{E_{\mathcal{N}_i}^2}/[
{E_{\mathcal{N}_i}+m_{\mathcal{N}_i}^2/(2m_\nu)}]\,.
$
For the nuclear form factor, we use the Klein--Nystrand parametrization~\cite{Klein:1999gv}, which can be written as
$F(Q^2)=3j_1(QR_i)/(QR_i)\,[1+a_k^2Q^2]^{-1}$.
Here, $R_i=1.2A_i^{1/3}\,{\rm fm}$, $a_k=0.7\,{\rm fm}$, and $j_1$ denotes the spherical Bessel function of the first kind.
In the fully coherent limit, where the momentum transfer is much smaller than the inverse nuclear size, $F(Q^2)\simeq 1$; as the momentum transfer increases, $F(Q^2)$ suppresses the coherent contribution.

\subsection{Incoherent neutrino--nucleus scattering}
\label{subsec:incoherent}

As the momentum transfer increases beyond the coherent regime, the scattering becomes sensitive to the internal nucleon structure of the nucleus. The condition $q\lesssim 1/R_A$ is no longer satisfied, and the amplitudes from different nucleons do not add coherently.
Instead, the contributions from the protons and neutrons inside the nucleus add incoherently. 
In this regime, the incoherent neutrino--nucleus contribution becomes relevant.

For a CR nucleus $\mathcal{N}_i$ with proton number $Z_i$ and neutron number $N_i$, we write the incoherent differential cross section as~\cite{Zhang:2025rqh}
\begin{equation}
\frac{d\sigma_{\nu \mathcal{N}_i}^{\rm INCOH}}{dE_\nu}
=
\left[
Z_i \frac{d\sigma_{\nu p}^{\rm ES}}{dE_\nu}
+
N_i \frac{d\sigma_{\nu n}^{\rm ES}}{dE_\nu}
\right]
\left[1-F^2(Q^2)\right]\,,
\label{eq:incoherent}
\end{equation}
Here $d\sigma_{\nu p}^{\rm ES}/dE_\nu$ and $d\sigma_{\nu n}^{\rm ES}/dE_\nu$ are the elastic neutrino--proton and neutrino--neutron cross sections given in Eq.~\eqref{eq:ES_general}, while $F(Q^2)$ is the nuclear form factor used in the Klein--Nystrand parametrization~\cite{Klein:1999gv}. Since INCOH is constructed from nucleon-level elastic scattering, it has the same kinematic upper limit as ES.
The factor $1-F^2(Q^2)$ controls the transition from the coherent to the incoherent regime. 
For sufficiently small momentum transfer, $F^2(Q^2)\simeq 1$ and hence $1-F^2(Q^2)\simeq 0$, so the incoherent contribution is strongly suppressed and CE$\nu$NS dominates. 
At larger momentum transfer, $F^2(Q^2)$ decreases and $1-F^2(Q^2)\simeq 1$, in which case the scattering approaches a linear superposition of nucleon-level contributions and the incoherent channel becomes dominant.

\subsection{Resonance region}
\label{subsec:RES}

As the momentum transfer and hadronic invariant mass increase beyond the elastic and low-$Q^2$ nuclear-scattering regimes, neutrino--nucleon scattering can excite baryon resonances. 
These resonance-production channels constitute the RES contribution to the boosted C$\nu$B flux. 
In this work, the neutral-current RES cross sections are evaluated using the resonance-production model implemented in the GENIE neutrino event generator~\cite{GENIE:2021npt}. 

For a nucleon target $N=p,n$, the differential RES cross section with respect to the boosted-neutrino energy is obtained by integrating the double-differential cross sections in the hadronic invariant mass $W$ and the positive momentum transfer $Q^2$,
\begin{equation}
\frac{d\sigma_{\nu N}^{\rm RES}}{dE_\nu}
=
2m_\nu
\int dW\,
\frac{d^2\sigma_{\nu N}^{\rm RES}}{dW\,dQ^2}
\bigg|_{Q^2=2m_\nu E_\nu}\,,
\label{eq:RES_dEdnu}
\end{equation}
Here the total double-differential RES cross section entering Eq.~\eqref{eq:RES_dEdnu} is constructed as the sum over the resonance states included in the GENIE resonance model,
\begin{equation}
\frac{d^2\sigma_{\nu N}^{\rm RES}}{dW\,dQ^2}
=
\sum_{\mathcal{R}=1}^{17}
\frac{d^2\sigma_{\nu N\to \nu \mathcal{R}}}{dW\,dQ^2}\,,
\label{eq:RES_sum}
\end{equation}
where $\mathcal{R}$ is the individual baryon resonance, and we sum over the full set of 17 baryon resonance states as used in the default GENIE implementation~\cite{GENIE:2021npt}.

For an individual baryon resonance $\mathcal{R}$, we use the Rein--Sehgal helicity-amplitude formalism ~\cite{Rein:1980wg,Feynman:1971wr}, in which the weak transition $N\to\mathcal{R}$ is expressed in terms of helicity amplitudes associated with the left- and right-handed transverse polarizations, and the scalar polarization component of the exchanged weak current. 
We adopt the GENIE Berger--Sehgal RES implementation, a realization of the Rein--Sehgal resonance framework that incorporates the KLN finite-lepton-mass and polarization treatment, together with the Berger--Sehgal pion-pole update~\cite{Kuzmin:2003ji,Berger:2007rq,GENIE:2021npt}.
In this representation, the neutrino and antineutrino cross sections are related by interchanging the left- and right-handed transverse terms. 
Since all cross sections in this work are defined as the sum of neutrino and antineutrino contributions, we use
\begin{equation}
\frac{d^2\sigma_{\nu N\to \nu \mathcal{R}}}{dW\,dQ^2}
=
\frac{1}{2}\,
\sigma_0
\left[
(U^2+V^2)
\left(
\sigma_{\rm R}^{N}
+
\sigma_{\rm L}^{N}
\right)
+
4UV\sigma_{\rm S}^{N}
\right]
\mathcal{B}_{\mathcal{R}}(W)\,
\Theta(W_{\rm cut}-W)\,,
\label{eq:RES_d2sigma_single}
\end{equation}
Here $U$ and $V$ are kinematic factors, $\mathcal{B}_{\mathcal{R}}(W)$ is the normalized Breit--Wigner distribution for the resonance $\mathcal{R}$, and $\sigma_0$, $\sigma_{\rm L}^{N}$, $\sigma_{\rm R}^{N}$, and $\sigma_{\rm S}^{N}$ are defined in Appendix~\ref{app:RES_delta}. 
Following the default GENIE implementation, we take $W_{\rm cut}=1.7~{\rm GeV}$ to separate the RES region from the PDF-based DIS region~\cite{GENIE:2021zuu,GENIE:2021npt}. 
This $W$ separation is commonly used when combining RES and DIS contributions in neutrino--nucleon cross-section calculations~\cite{Kuzmin:2005bm,Jeong:2023hwe}. 
In the present calculation, the RES contribution is restricted by the factor $\Theta(W_{\rm cut}-W)$ in Eq.~\eqref{eq:RES_d2sigma_single}, while the DIS contribution is evaluated in the complementary region $W>W_{\rm cut}$.
The total RES cross section in Eq.~\eqref{eq:RES_sum} is obtained by summing Eq.~\eqref{eq:RES_d2sigma_single} over the 17 resonance states implemented in GENIE~\cite{GENIE:2021npt}.
Note that the factor of $1/2$ in Eq.~(\ref{eq:RES_d2sigma_single}) accounts for the fact that present-day C$\nu$B neutrinos are nonrelativistic helicity eigenstates, containing approximately equal left- and right-chiral components~\cite{Long:2014zva}, while only the left-chiral component participates in the weak interaction~\cite{Zhang:2025rqh}. 

\begin{figure*}[t]
    \centering
    \begin{minipage}{0.48\textwidth}
        \centering
        \includegraphics[width=\textwidth]{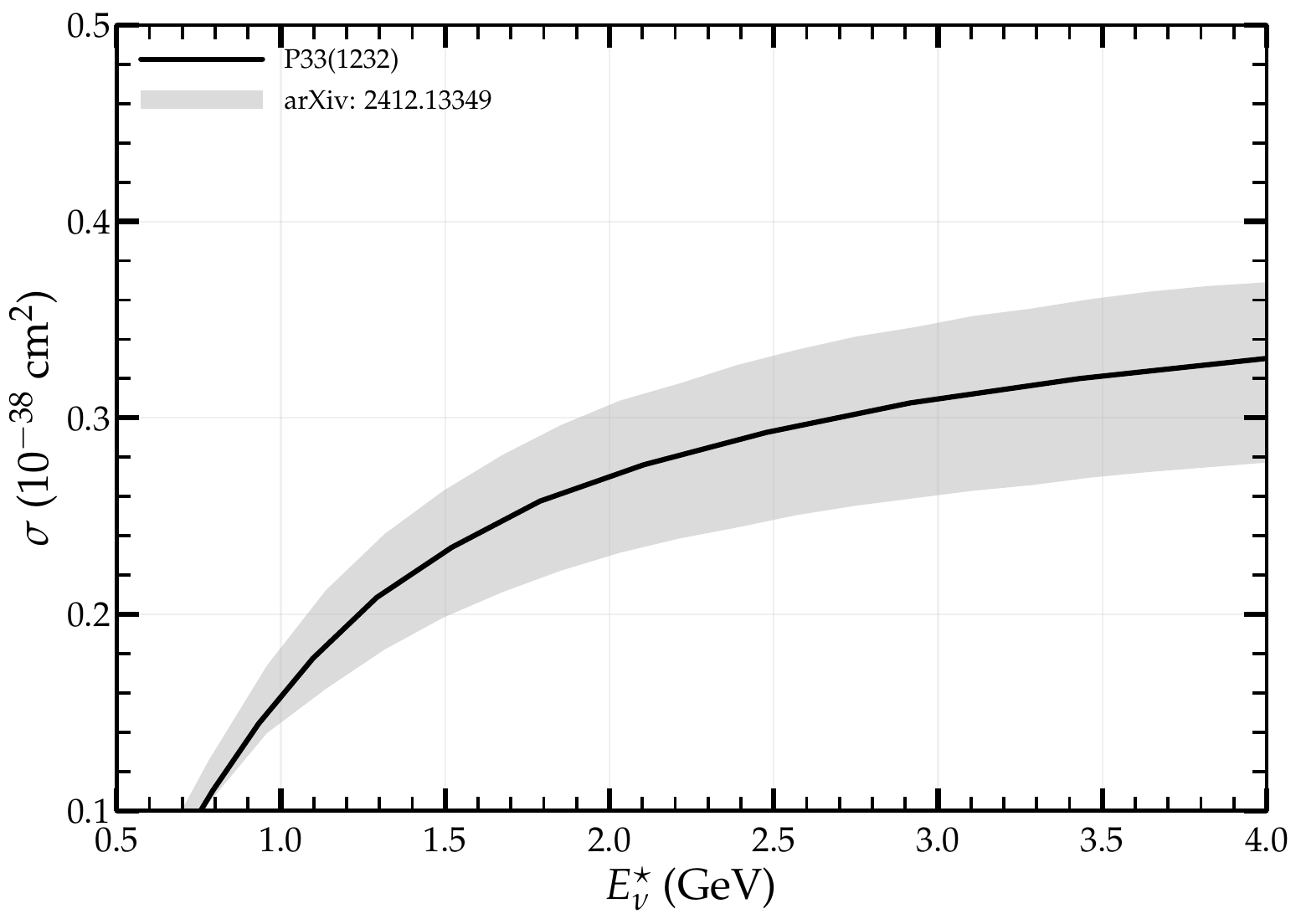}
    \end{minipage}
    \hfill
    \begin{minipage}{0.48\textwidth}
        \centering
        \includegraphics[width=\textwidth]{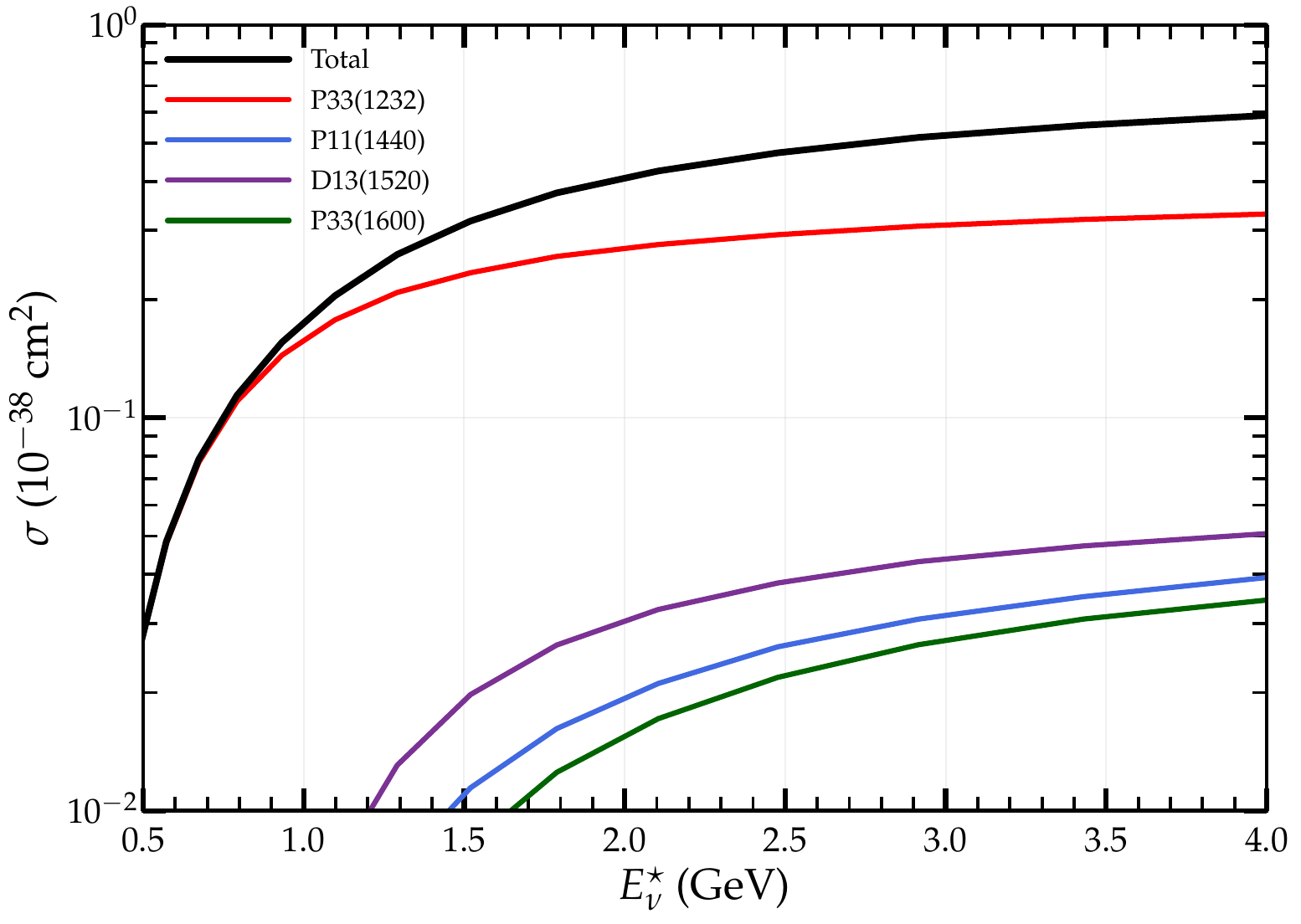}
    \end{minipage}
    \caption{
    Resonance cross sections in the nucleon rest frame as functions of the incident neutrino energy $E_\nu^\star$.
    The left panel compares our $P_{33}(1232)$ result with the range reported in Ref.~\cite{Abbaslu:2024jzo}.
    The right panel shows the total RES cross section along with representative resonance states, including $P_{33}(1232)$, $P_{11}(1440)$, $D_{13}(1520)$, and $P_{33}(1600)$.
    }
    \label{fig:res_validation}
\end{figure*}

We have performed both a consistency check of the resonance calculation and a comparison of different resonance states, as shown in Fig.~\ref{fig:res_validation}. 
The left panel of Fig.~\ref{fig:res_validation} shows that our implemented $P_{33}(1232)$ total cross section is consistent with the result shown in Fig.~7 of Ref.~\cite{Abbaslu:2024jzo}, and it follows the same dependence on the incident neutrino energy $E_\nu^\star$, where $E_\nu^\star$ denotes the neutrino energy in the nucleon rest frame. 
The right panel of Fig.~\ref{fig:res_validation} compares the total RES cross section with the representative states $P_{33}(1232)$, $P_{11}(1440)$, $D_{13}(1520)$, and $P_{33}(1600)$, showing that $P_{33}(1232)$ is the largest single contribution over most of the energy range, while sub-leading states still affect the total RES normalization once all 17 resonance states are summed.
Therefore, in Appendix~\ref{app:RES_delta} we present the neutral-current $P_{33}(1232)$ production channel as a representative example to illustrate the corresponding cross-section formulae. 
In our calculation, we include the full set of 17 default resonance states in the GENIE implementation.
We have explicitly verified that our numerical implementation reproduces the corresponding GENIE results for all 17 resonance states.

For a nuclear UHECR component $\mathcal{N}_i$, we approximate the RES contribution as an incoherent sum of resonance production on the protons and neutrons inside the nucleus,
\begin{equation}
\frac{d\sigma_{\nu \mathcal{N}_i}^{\rm RES}}{dE_\nu}
=
Z_i \frac{d\sigma_{\nu p}^{\rm RES}}{dE_\nu}
+
N_i \frac{d\sigma_{\nu n}^{\rm RES}}{dE_\nu}\,.
\label{eq:RES_nucleus}
\end{equation}
This approximation treats resonance production as a nucleon-level process inside the nucleus and neglects nuclear-medium corrections.

\subsection{Deep inelastic scattering}
\label{subsec:DIS}

When the momentum transfer and hadronic invariant mass become sufficiently large, the neutrino resolves the partonic structure of the nucleon. 
This regime is described by neutral-current DIS, in which the relic neutrino scatters from quarks and antiquarks inside the nucleon. 
For a nucleon target $N=p,n$, the DIS differential cross section with respect to the boosted-neutrino energy is written as~\cite{DeMarchi:2024zer}
\begin{equation}
\frac{d\sigma_{\nu N}^{\rm DIS}}{dE_\nu}
\simeq
\sum_{a=q,\bar q}
\frac{G_F^2 \left[(g_V^a)^2+(g_A^a)^2\right]}{4\pi E_N}
\int_{y_{\rm DIS}^{\rm min}}^{y_{\rm max}}
\frac{dy}{y^2}\,
\frac{Q^2 f_a^N(x,Q^2)}
{\left(1+Q^2/M_Z^2\right)^2}\,
g(y,Q^2)\,,
\label{eq:DIS}
\end{equation}
where $f_a^N(x,Q^2)$ is the parton distribution function (PDF) for a quark or antiquark in the nucleon $N$, and $g(y,Q^2)=y^2-2y+2-2m_N^2x^2y^2/Q^2$. 
The Bjorken variable $x=(E_\nu-m_\nu)/(yE_N)\simeq E_\nu/(yE_N)$, and the inelasticity is $y\simeq E_\nu/(xE_N)$. 
Note that the DIS differential cross section used here is one half of the expression used in Ref.~\cite{DeMarchi:2024zer}, because the latter does not include the chirality suppression associated with nonrelativistic C$\nu$B targets.

As discussed above, to avoid double counting the low-mass resonance region, we include the PDF-based DIS contribution only for hadronic invariant masses above the resonance--DIS separation scale, i.e., we require the hadronic invariant mass to satisfy $W>W_{\rm cut}$ and take $W_{\rm cut}=1.7~{\rm GeV}$~\cite{GENIE:2021zuu,GENIE:2021npt}, the same value used in the RES calculation.
In the transition region between RES and DIS, a nonresonant DIS component can also contribute for $W<W_{\rm cut}$~\cite{Andreopoulos:2015wxa}.

The hadronic invariant mass is related to the DIS variables by~\cite{Jeong:2023hwe}
\begin{equation}
W^2
=
m_N^2+Q^2\left(\frac{1}{x}-1\right)\,,
\label{eq:DIS_W_relation}
\end{equation}
Using Eq.~\eqref{eq:DIS_W_relation}, the condition $W>W_{\rm cut}$ is implemented as
\begin{equation}
x<
x_W(Q^2)
\equiv
\frac{Q^2}{Q^2+W_{\rm cut}^2-m_N^2},
\qquad
y_{\rm DIS}^{\rm min}
=
\frac{E_\nu}{E_Nx_W(Q^2)}\,.
\end{equation}

In our calculation, the PDFs are evaluated using the nCTEQ15 parton distribution function set~\cite{Kovarik:2015cma} with the ManeParse package~\cite{Clark:2016jgm}.
The nCTEQ15 PDFs are provided for
$Q^2\geq Q_{\rm DIS}^2=1.69~{\rm GeV}^2$;
at lower $Q^2$, nonperturbative effects such as higher-twist corrections become increasingly important, making a standard perturbative PDF description less reliable~\cite{Segarra:2020gtj}.
For a fixed $Q^2$, Eq.~\eqref{eq:DIS_W_relation} directly gives
$W<W_{\rm cut}$ for $x>x_W(Q^2)$.
Already at the lowest $Q^2$ covered by the nCTEQ15 PDFs, $Q^2=1.69~{\rm GeV}^2$, this condition requires $x\gtrsim0.46$, and the required value of $x$ becomes larger as $Q^2$ increases. Since the parton distributions decrease rapidly at large Bjorken $x$~\cite{McFarland:2008xd,Li:2014vka},
the PDF-based DIS contribution from the $W<W_{\rm cut}$ region is strongly suppressed.
We have checked this suppression explicitly by repeating the PDF-based DIS calculation without imposing the $W>W_{\rm cut}$ condition while retaining $Q^2\geq1.69~{\rm GeV}^2$.
The additional low-$W$ contribution is found to be negligible, and we therefore neglect the nonresonant contribution for $W<W_{\rm cut}=1.7~{\rm GeV}$ in our calculation.

For RES and DIS, the final-state hadronic system has invariant mass $W$, and the kinematic upper limit for the outgoing boosted relic neutrino energy is
\begin{equation}
E_\nu^{\max}(E_{N}) = \frac{E_{N}^2}
{E_{N}+m_{N}^2/(2m_\nu)}
\left(1-\frac{W^2-m_N^2}{2m_\nu E_{N}}
\right)\,.
\end{equation}
The PDF evaluation is further restricted to $x\geq x_{\rm min}$ with $x_{\rm min}=5.0\times10^{-6}$~\cite{Kovarik:2015cma}. Accordingly, the upper limit of the $y$ integration in Eq.~\eqref{eq:DIS} is
$y_{\rm max}=\min[
1-E_\nu m_N^2/(2m_\nu E_N^2),
E_\nu/(E_N x_{\rm min})
]$.

For a nuclear UHECR component $\mathcal{N}_i$, a large momentum transfer resolves the nucleus into its nucleonic constituents. 
We approximate the DIS contribution as an incoherent sum over proton and neutron DIS cross sections,
\begin{equation}
\frac{d\sigma_{\nu \mathcal{N}_i}^{\rm DIS}}{dE_\nu}
\simeq
Z_i \frac{d\sigma_{\nu p}^{\rm DIS}}{dE_\nu}
+
N_i \frac{d\sigma_{\nu n}^{\rm DIS}}{dE_\nu}\,.
\label{eq:DIS_nucleus}
\end{equation}
% where $Z_i$ and $N_i$ are the proton and neutron numbers of $\mathcal{N}_i$, respectively. 
This approximation treats nuclear DIS as a linear superposition of nucleon-level DIS contributions and neglects nuclear PDF corrections.

\subsection{Comparison of cross sections from different channels}
\label{subsec:cross_section_summary}

\begin{figure*}[t]
    \centering

    \begin{minipage}{0.48\textwidth}
        \centering
        \includegraphics[width=\textwidth]{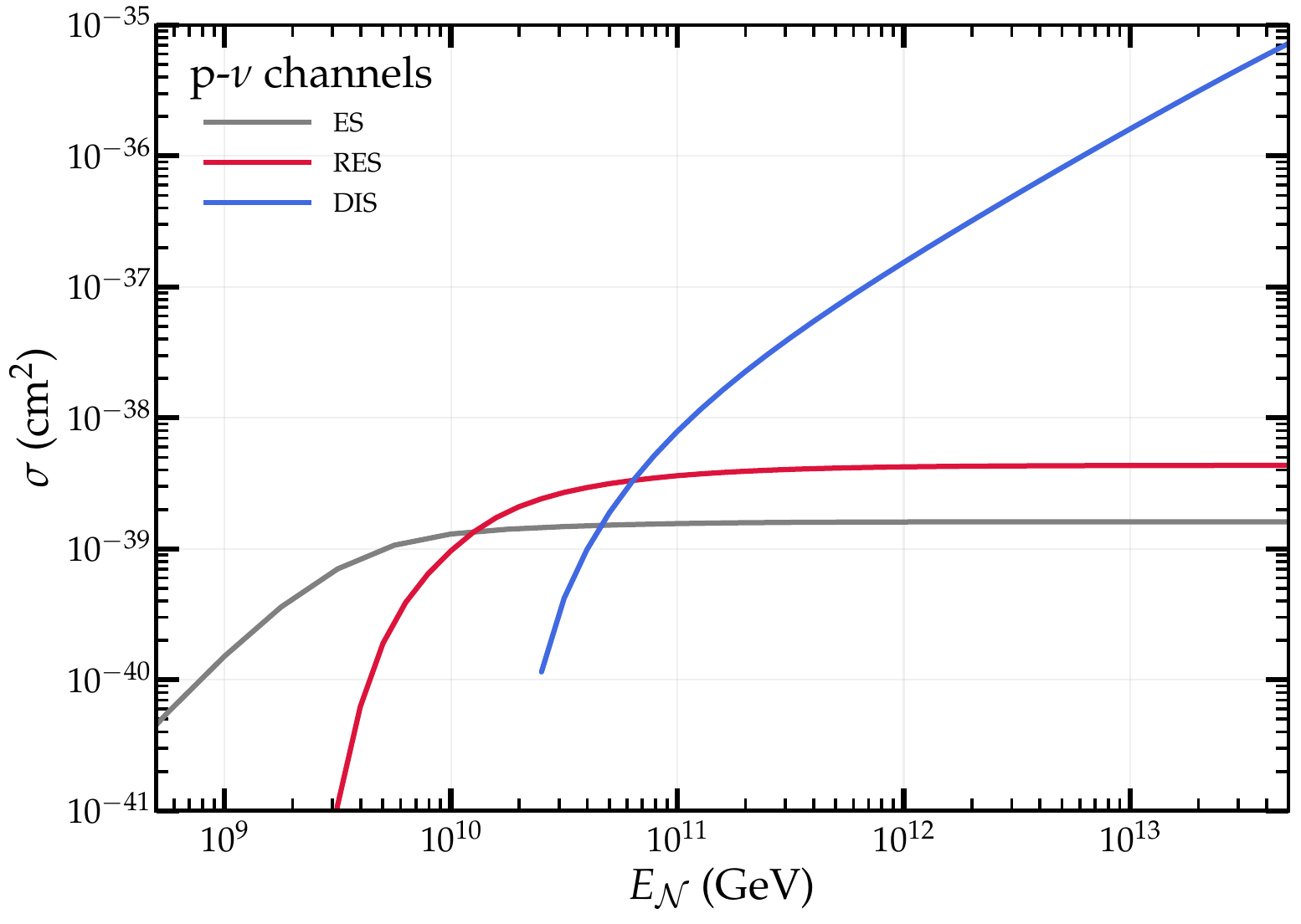}
    \end{minipage}
    \hfill
    \begin{minipage}{0.48\textwidth}
        \centering
        \includegraphics[width=\textwidth]{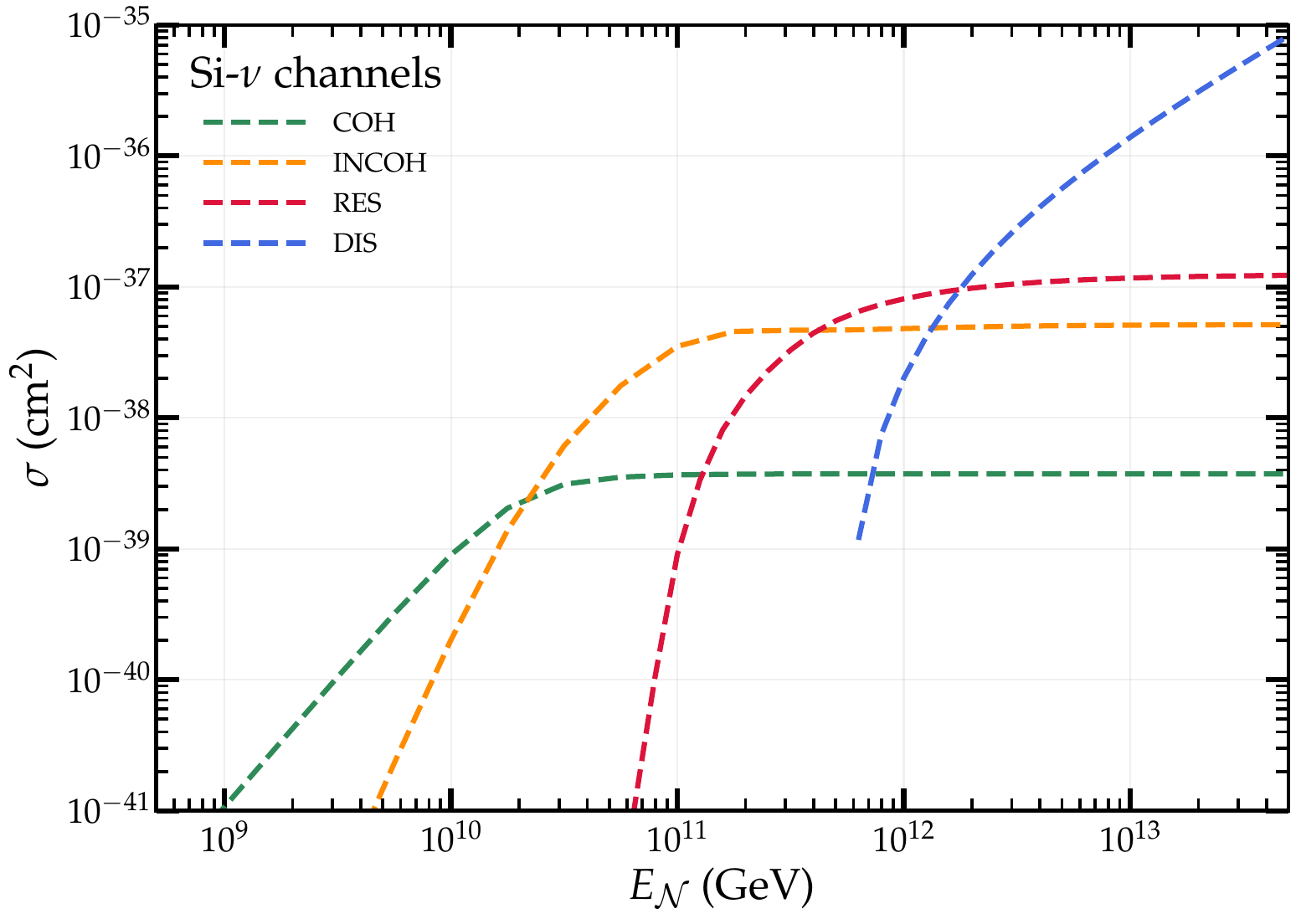}
    \end{minipage}

    \vspace{2mm}

    \begin{minipage}{0.48\textwidth}
        \centering
        \includegraphics[width=\textwidth]{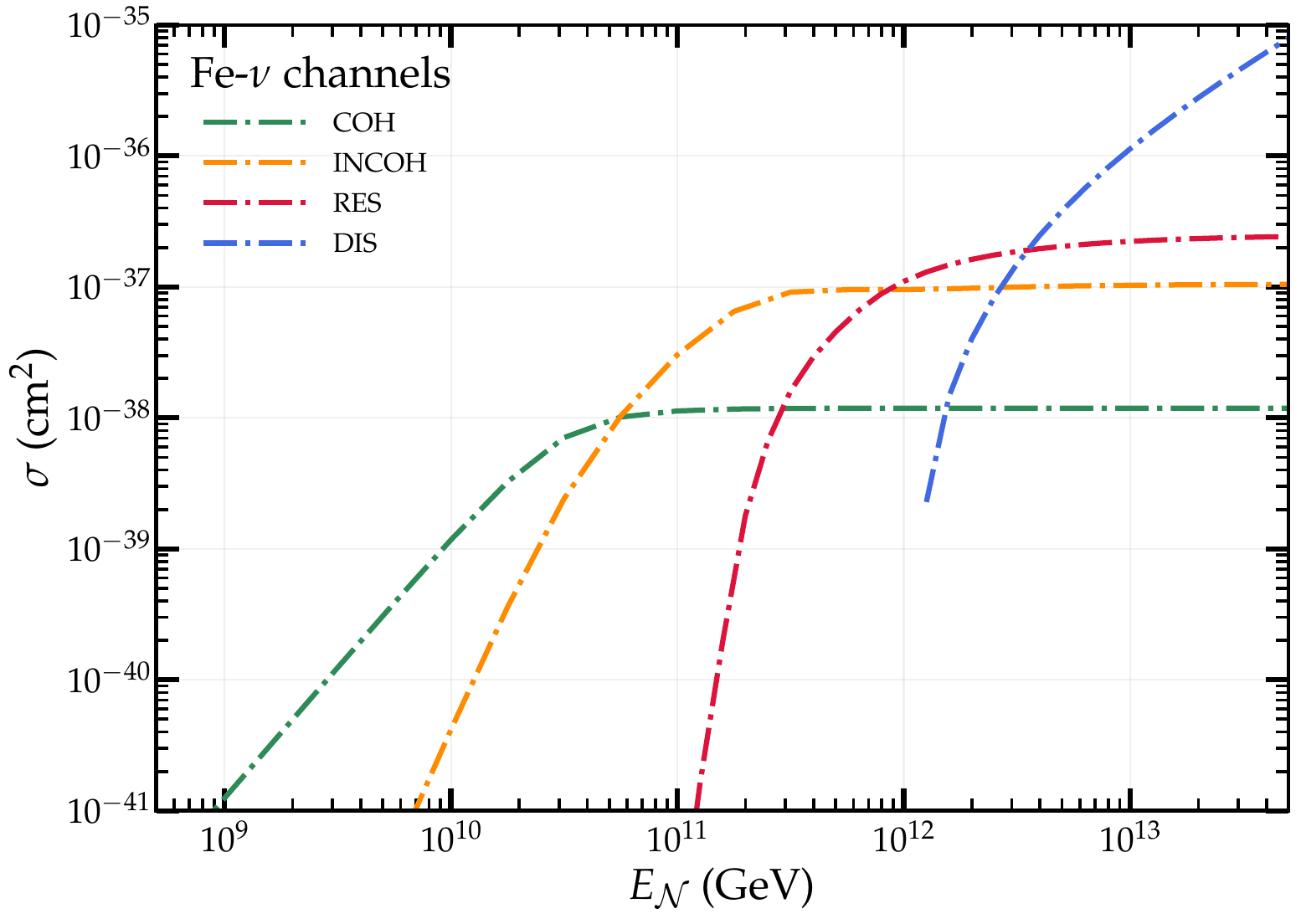}
    \end{minipage}
    \hfill
    \begin{minipage}{0.48\textwidth}
        \centering
        \includegraphics[width=\textwidth]{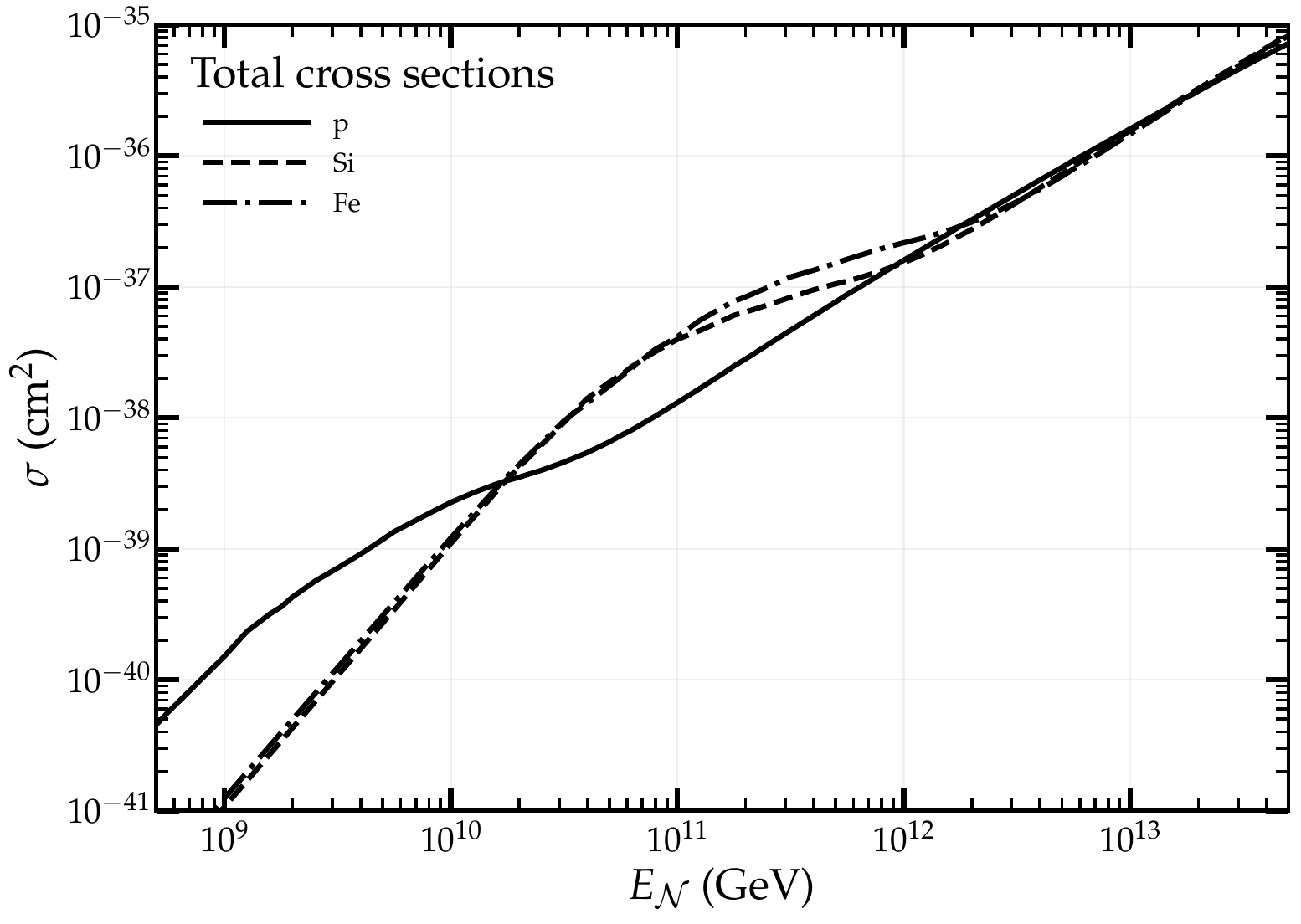}
    \end{minipage}

    \caption{
    Scattering cross sections as functions of the incoming UHECR energy $E_{\mathcal{N}}$ for $m_\nu=0.1~{\rm eV}$.
    The upper-left panel shows the proton case, including ES, RES, and DIS contributions.
    The upper-right and lower-left panels show the channel decomposition for silicon and iron nuclei, including COH, INCOH, RES, and DIS contributions.
    The lower-right panel compares the total cross sections for the proton, silicon, and iron UHECR components.
    }
    \label{fig:sigma_all_channels}
\end{figure*}

In Fig.~\ref{fig:sigma_all_channels}, we compare the scattering cross sections for the different channels considered in this work, taking $m_\nu=0.1~{\rm eV}$. 
For the proton case, shown in the upper-left panel, the ES contribution dominates at relatively low UHECR energies, approximately for $E_N\lesssim10^{10}~{\rm GeV}$. As the energy increases, the RES contribution becomes the leading contribution in the intermediate region, roughly from $10^{10}~{\rm GeV}$ to $6\times10^{10}~{\rm GeV}$. 
At higher energies, the DIS cross section grows rapidly and eventually dominates the proton cross section.

For nuclear UHECR components, illustrated by the silicon and iron examples in the upper-right and lower-left panels, the low-energy behavior is controlled by the COH contribution. 
At small momentum transfer, the nucleus is not resolved and the COH channel gives the largest cross section due to the coherent enhancement. 
As the UHECR energy increases, the nuclear form factor suppresses the COH contribution, and INCOH becomes dominant once the scattering energy is large enough to resolve individual nucleons. 
At higher energies, RES becomes sizable and can exceed INCOH before the DIS dominated regime is reached. 
The DIS curves start only above the perturbative PDF threshold, i.e., $Q^2>Q_{\rm DIS}^2=1.69~{\rm GeV}^2$, and then take over at the highest energies when the interaction probes the partonic structure of the nucleons.

The lower-right panel compares the total cross sections for proton, silicon, and iron UHECR components. 
For $E_{\mathcal{N}}\lesssim2\times10^{10}~{\rm GeV}$, the proton cross section is larger than the nuclear ones. 
In the intermediate energy range, the proton and nuclear components are dominated by different scattering regimes. 
For a proton UHECR, the full CR energy is carried by a single nucleon, so the scattering enters the RES dominated regime already around $E_{\mathcal{N}}\sim10^{10}~{\rm GeV}$. 
For a nuclear UHECR with mass number $A$, however, the relevant energy per nucleon is approximately $E_{\mathcal{N}}/A$. 
As a result, at the same total CR energy $E_{\mathcal{N}}$, the Si and Fe components are still mainly in the INCOH regime when the proton component has already entered the RES regime. 
At higher energies, the proton component reaches the DIS regime first, while the Si and Fe components are still dominated by RES. 
This delayed transition of nuclear components from INCOH to RES and then to DIS explains why the Si and Fe total cross sections exceed the proton one over the intermediate range, approximately from $2\times10^{10}~{\rm GeV}$ to $10^{12}~{\rm GeV}$.
For $E_{\mathcal{N}}\gtrsim5\times10^{12}~{\rm GeV}$, all components enter the DIS-dominated regime and their total cross sections become nearly identical. 
This can be understood from the approximate linear energy dependence of the nucleon DIS cross section, $\sigma_{\rm DIS}^{N}(E_N)\propto E_N$. 
For a nucleus with mass number $A$, the DIS cross section is approximated as a linear sum over its nucleons,$
\sigma_{\rm DIS}^{\mathcal{N}}(E_{\mathcal{N}})
\simeq
A\,\sigma_{\rm DIS}^{N}\left(\frac{E_{\mathcal{N}}}{A}\right)$.
Since $\sigma_{\rm DIS}^{N}$ is approximately proportional to the nucleon energy, this gives
$
A\,\sigma_{\rm DIS}^{N}\left(\frac{E_{\mathcal{N}}}{A}\right)
\propto
A\,\frac{E_{\mathcal{N}}}{A}
=
E_{\mathcal{N}}$.
Thus, the enhancement from the number of nucleons is compensated by the smaller energy carried by each nucleon, making the DIS-dominated total cross sections nearly independent of the nuclear mass number.

\medskip

%%%%%%%%%%%%%%%%%%%%%%%%%%%%%%%%%%%%%
\section{Ultra-High-Energy cosmic ray flux}
\label{sec:cr_flux}

The prediction of the boosted C$\nu$B flux is also sensitive to the flux of each UHECR nuclear component. 
Since the boosted C$\nu$B signal accumulates along the propagation path of UHECRs from their sources to Earth, it is mainly determined by the extragalactic CR flux over cosmological distances. 
In addition, only CR with sufficiently high energies can escape magnetic confinement in their host environments and propagate into intergalactic space~\cite{AlvesBatista:2019tlv}. 
Composition and anisotropy studies indicate that the transition from Galactic to extragalactic CR occurs around the second knee, at approximately $5\times10^{8}~{\rm GeV}$~\cite{Kachelriess:2019oqu,Zhang:2025rqh}. 
We therefore include only the contribution from CR with $E_{\mathcal{N}_i}>5\times10^{8}~{\rm GeV}$~\cite{Zhang:2025rqh}, and we have checked that the contribution below this energy is negligible for the boosted C$\nu$B flux. 

We adopt three different approaches for the UHECR flux in this work. 
The first one is obtained from propagation simulations with the PriNCe code~\cite{Heinze:2019jou}. 
PriNCe solves coupled transport equations for UHECR nuclei as functions of energy and redshift, including adiabatic energy losses from the cosmological expansion, electron--positron pair production, photohadronic interactions, and photonuclear interactions during propagation through cosmic photon backgrounds. 
It therefore provides redshift-dependent mixed-composition spectra after propagation from the sources to the Earth. 
The other two approaches are based on the Hillas model~\cite{Hillas:2005cs,Gaisser:2011klf}. 
We use both the H3a and H4a implementations: in H3a the ultra-high-energy part of the spectrum is dominated by heavy nuclei, whereas in H4a the corresponding high-energy part is dominated by protons. The detailed implementation of the three UHECR flux models is described below.

\subsection{PriNCe simulation}
\label{subsec:prince_flux}
The propagation of UHECR spectra is simulated using
the PriNCe code~\cite{Heinze:2019jou}. 
PriNCe solves the coupled transport equations for different nuclear species as functions of energy and redshift, and follows the evolution of the UHECR flux from the sources to Earth. 
The propagation includes adiabatic energy losses due to cosmic expansion, continuous energy losses from electron--positron pair production, photohadronic interactions, and photonuclear interactions~\cite{Heinze:2019jou}. 
The relevant photon backgrounds are the cosmic microwave background and the extragalactic background light, for which we use the model of Ref.~\cite{Gilmore2012}. 
Photonuclear interactions are evaluated with TALYS~\cite{Koning2005}, while photohadronic interactions are treated with SOPHIA~\cite{Mcke2000}. 
In this way, the PriNCe spectra include both the energy losses and the composition changes induced by propagation through cosmic photon backgrounds.

In the PriNCe calculation, the injected nuclear spectra are used as source terms in the coupled transport equations. 
The solver then evolves the injected nuclei from the source redshift to the target redshift, including energy losses and nuclear fragmentation during propagation~\cite{Heinze:2019jou}. For each injected nuclear species $\mathcal{N}_i$, the source spectrum is parametrized as~\cite{Gaisser:2013bla}
\begin{equation}
J_{\mathcal{N}_i}(E_{\mathcal{N}_i})
=
f_{\mathcal{N}_i}
\left(
\frac{E_{\mathcal{N}_i}}{10^9~{\rm GeV}}
\right)^{-\gamma}
\times
\begin{cases}
1,
&
E_{\mathcal{N}_i}<Z_i R_{\rm max},
\\[1ex]
\exp\left(1-\dfrac{E_{\mathcal{N}_i}}{Z_iR_{\rm max}}\right),
&
E_{\mathcal{N}_i}>Z_i R_{\rm max}\,,
\end{cases}
\label{eq:prince_injection}
\end{equation}
where $R_{\rm max}$ is the maximum rigidity, $\gamma$ is the injection spectral index, and $f_{\mathcal{N}_i}$ gives the normalization of each injected nuclear component.

\begin{table}[t]
\centering
\begin{tabular}{c|cc|ccccc}
\hline
Source 
& $R_{\rm max}$ (GV) 
& $\gamma$ 
& $f_{\rm H}$ 
& $f_{\rm He}$ 
& $f_{\rm N}$ 
& $f_{\rm Si}$ 
& $f_{\rm Fe}$ \\
\hline
SFR 
& $10^{9.25}$ 
& $-0.8$ 
& $1.0$ 
& $0.45$ 
& $0.062$ 
& $0.0041$ 
& $9.0\times10^{-5}$ \\
QSO 
& $10^{9.20}$ 
& $-1.0$ 
& $2.2$ 
& $0.77$ 
& $0.061$ 
& $0.0023$ 
& $6.6\times10^{-5}$ \\
GRB 
& $10^{9.25}$ 
& $-0.8$ 
& $0.17$ 
& $0.046$ 
& $0.055$ 
& $0.0039$ 
& $9.5\times10^{-5}$ \\
\hline
\end{tabular}
\caption{
Source parameters used for the PriNCe UHECR propagation calculations with SFR, QSO, and GRB source evolution models~\cite{Sandrock:2025nzb}. 
The abundances $f_i$ are given in units of $10^{-45}$.
The parameters are obtained from fits to Pierre Auger Observatory spectrum and composition data~\cite{Veberic:2017hwu}.
}
\label{tab:prince_params}
\end{table}

In this work, we consider three representative source evolution models. 
The detailed forms of these source evolution functions are provided in Appendix~\ref{app:source_distribution}. The source parameters used in the PriNCe simulations are summarized in Table~\ref{tab:prince_params}. 
They were obtained in Ref.~\cite{Sandrock:2025nzb} by fitting the propagated UHECR spectra and composition to Pierre Auger Observatory measurements of the all-particle energy spectrum, the mean shower maximum $X_{\rm max}$, and the shower-maximum variance $\sigma(X_{\rm max})$~\cite{Veberic:2017hwu}. 
The fitting procedure uses the companion package \texttt{PriNCe-analysis-tools}\footnote{The code is available at \url{https://github.com/joheinze/PriNCe-analysis-tools/}.}, which was developed for PriNCe.

In the boosted C$\nu$B calculation, we use the redshift-dependent propagated UHECR spectra produced by PriNCe for each source evolution model. The attenuation of UHECRs due to scattering with the C$\nu$B is not included in the PriNCe propagation simulation.

\subsection{Hillas H3a and H4a models}
\label{subsec:hillas_flux}

The Hillas model has also been widely used to describe the present-day CR spectrum~\cite{Hillas:2005cs,Gaisser:2011klf,Gaisser:2013bla}. 
In this model, the observed CR spectrum is written as the sum of rigidity-dependent populations (Pop.). 
Pop.~1 is associated with Galactic CR and mainly accounts for the knee region. 
At the highest energies, the spectrum is described by Pop.~3, which represents the extragalactic component. 
Between these two regimes, Pop.~2 is introduced as an additional component of uncertain origin to connect the Galactic and extragalactic parts of the spectrum between the knee and the ankle. 
Each population contains representative nuclear groups and has an exponential cutoff controlled by the particle rigidity.

For each nuclear species $\mathcal{N}_i$, the present-day differential flux is written as~\cite{Gaisser:2011klf,Gaisser:2013bla}
\begin{equation}
\left.
\frac{d\phi_{\mathcal{N}_i}}{dE_{\mathcal{N}_i}}
\right|_{z=0}
=
k_{\rm H}
\sum_{j\in \mathcal{J}}
a_{i,j}\,
E_{\mathcal{N}_i}^{-\gamma_{i,j}-1}
\exp\left(
-\frac{E_{\mathcal{N}_i}}{Z_iR_{c,j}}
\right)\,,
\label{eq:hillas_flux}
\end{equation}
where $j$ labels the populations listed in Table~\ref{tab:hillas_params}, $R_{c,j}$ is the cutoff rigidity of population $j$, and $a_{i,j}$ and $\gamma_{i,j}$ are the normalization and spectral index of the component $\mathcal{N}_i$ in that population. 
The factor $k_{\rm H}$ is an overall normalization used to match the resulting spectrum to the UHECR flux measured by the Pierre Auger Observatory~\cite{Veberic:2017hwu}.

\begin{table}[t]
\centering
\renewcommand{\arraystretch}{1.15}
\begin{tabular}{c c c c c c c}
\hline
Population & Parameter & $p$ & He & N & Si & Fe \\
\hline
Pop.~1 
& $a_{i,j}$, $R_c=4~\mathrm{PV}$ 
& 7860 
& 3550 
& 2200 
& 1430 
& 2120 \\
& $\gamma_{i,j}$ 
& 1.66 
& 1.58 
& 1.63 
& 1.67 
& 1.63 \\
\hline
Pop.~2 
& $a_{i,j}$, $R_c=30~\mathrm{PV}$ 
& 20 
& 20 
& 13.4 
& 13.4 
& 13.4 \\
& $\gamma_{i,j}$ 
& 1.4 
& 1.4 
& 1.4 
& 1.4 
& 1.4 \\
\hline
Pop.~3 
& $a_{i,j}$, $R_c=2~\mathrm{EV}$ 
& 1.7 
& 1.7 
& 1.14 
& 1.14 
& 1.14 \\
& $\gamma_{i,j}$ 
& 1.4 
& 1.4 
& 1.4 
& 1.4 
& 1.4 \\
\hline
Pop.~3$^\ast$ 
& $a_{i,j}$, $R_c=60~\mathrm{EV}$ 
& 200 
& 0 
& 0 
& 0 
& 0 \\
& $\gamma_{i,j}$ 
& 1.6 
& -- 
& -- 
& -- 
& -- \\
\hline
\end{tabular}
\caption{
Parameters of the Hillas model used for the H3a and H4a implementations~\cite{Hillas:2005cs,Gaisser:2013bla}. 
For each population, the first row gives the normalization $a_{i,j}$ together with the cutoff rigidity $R_c$, and the second row gives the spectral index $\gamma_{i,j}$. 
Pop.~1, Pop.~2, and Pop.~3 define H3a, while Pop.~3$^\ast$ replaces Pop.~3 in H4a.
}
\label{tab:hillas_params}
\end{table}

We use two implementations of the Hillas model, H3a and H4a, which differ in the composition of Pop.~3~\cite{Gaisser:2011klf,Gaisser:2013bla}.
In H3a, Pop.~3 is mixed in composition and contains $p$, He, N, Si, and Fe, while in H4a it is replaced by a proton-only component, denoted as Pop.~3$^\ast$. 
Thus, $\mathcal{J}=\{1,2,3\}$ for H3a and $\mathcal{J}=\{1,2,3^\ast\}$ for H4a. 
The normalization factors are $k_{\rm H3a}=0.601$~\cite{Zhang:2025rqh} and $k_{\rm H4a}=0.933$, respectively.~\footnote{
The original H3a and H4a parameters were not obtained by fitting the Pierre Auger data~\cite{Veberic:2017hwu} used in this work. 
They were introduced as phenomenological Hillas model implementations based on earlier cosmic ray data~\cite{Gaisser:2011klf}. 
In the present calculation, the Pierre Auger data of Ref.~\cite{Veberic:2017hwu} are used only to fix the overall normalization of the H3a and H4a spectra in the ultra-high-energy region.
}
The parameters used for the two implementations are summarized in Table~\ref{tab:hillas_params}. 
In our calculation, we include only CRs with
$E_{N_i}>5\times10^8~{\rm GeV}$~\cite{Kachelriess:2019oqu}.
Since the Galactic Pop.~1 has a cutoff rigidity of only $R_c=4~{\rm PV}$, its contribution is already strongly suppressed in this energy range and can be safely neglected.
Therefore, the relevant high-energy CR flux is dominated by the higher-rigidity, predominantly extragalactic populations, in particular
Pop.~3 in H3a and Pop.~3$^\ast$ in H4a at the highest energies.
Although the proton-dominated H4a case is not favored by current Pierre Auger composition measurements at the highest energies, we include it as a useful benchmark model to illustrate how the assumed UHECR composition and maximum rigidity modify the boosted C$\nu$B flux at the highest energies.

The Hillas model describes only the UHECR flux at $z=0$. Following Refs.~\cite{Herrera:2024upj, Herrera:2026pzj}, we rescale the $z=0$ spectrum by the normalized source evolution function $f(z)$ to obtain the CR flux at redshift $z$,
\begin{equation}
\frac{d\phi_{\mathcal{N}_i}}{dE_{\mathcal{N}_i}}(z)
=
f(z)
\left.
\frac{d\phi_{\mathcal{N}_i}}{dE_{\mathcal{N}_i}}
\right|_{z=0}\,,
\label{eq:hillas_redshift}
\end{equation}
where $f(z)$ is a normalized source evolution function given in Appendix~\ref{app:source_distribution}.
Note that this prescription does not explicitly follow UHECR energy losses or composition changes during propagation. 
For this reason, the PriNCe simulation can be regarded as the propagation-based benchmark model, while the Hillas implementation is a phenomenological comparison.

\subsection{Comparison of UHECR fluxes}
\label{subsec:cr_flux_redshift}

\begin{figure*}[t]
    \centering

    % Row 1: z = 0
    \begin{minipage}{0.32\textwidth}
        \centering
        \includegraphics[width=\textwidth]{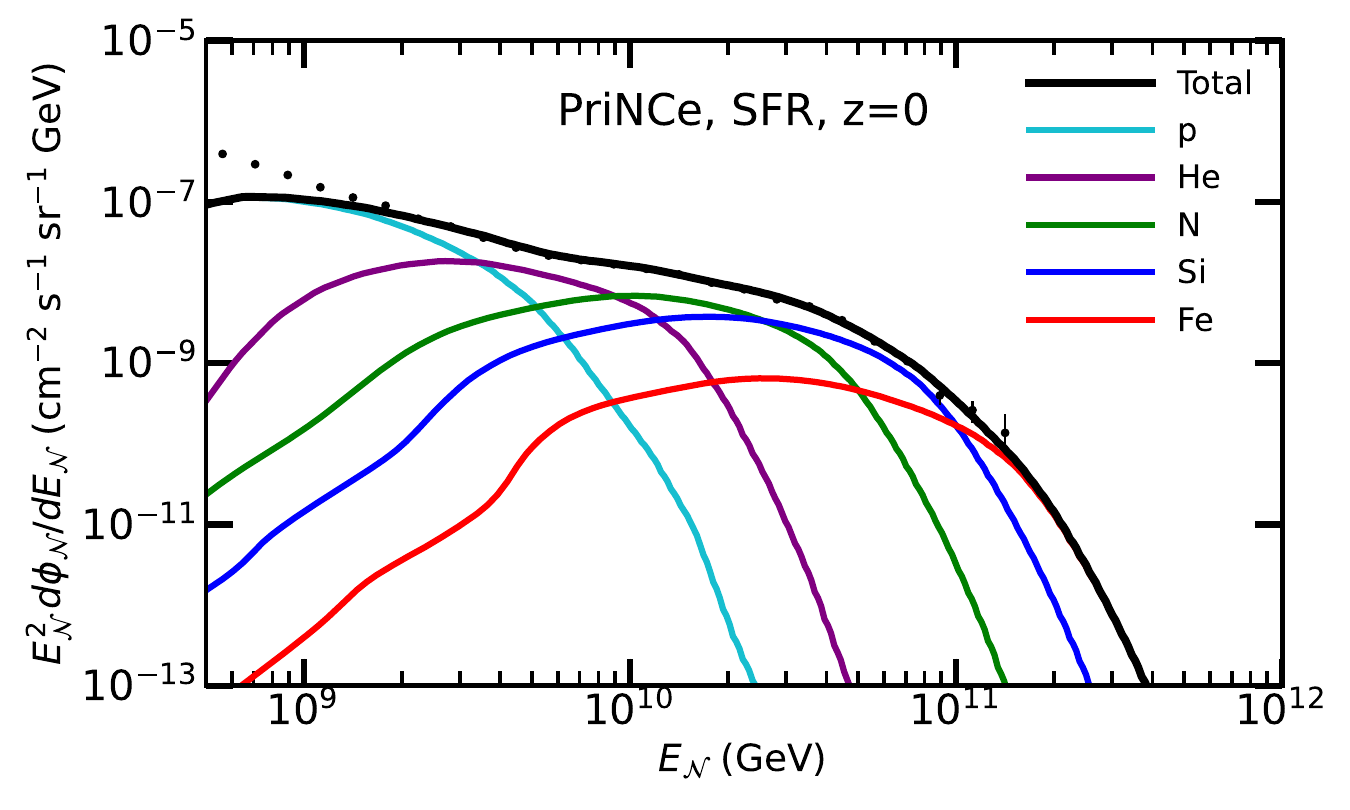}
    \end{minipage}
    \hfill
    \begin{minipage}{0.32\textwidth}
        \centering
        \includegraphics[width=\textwidth]{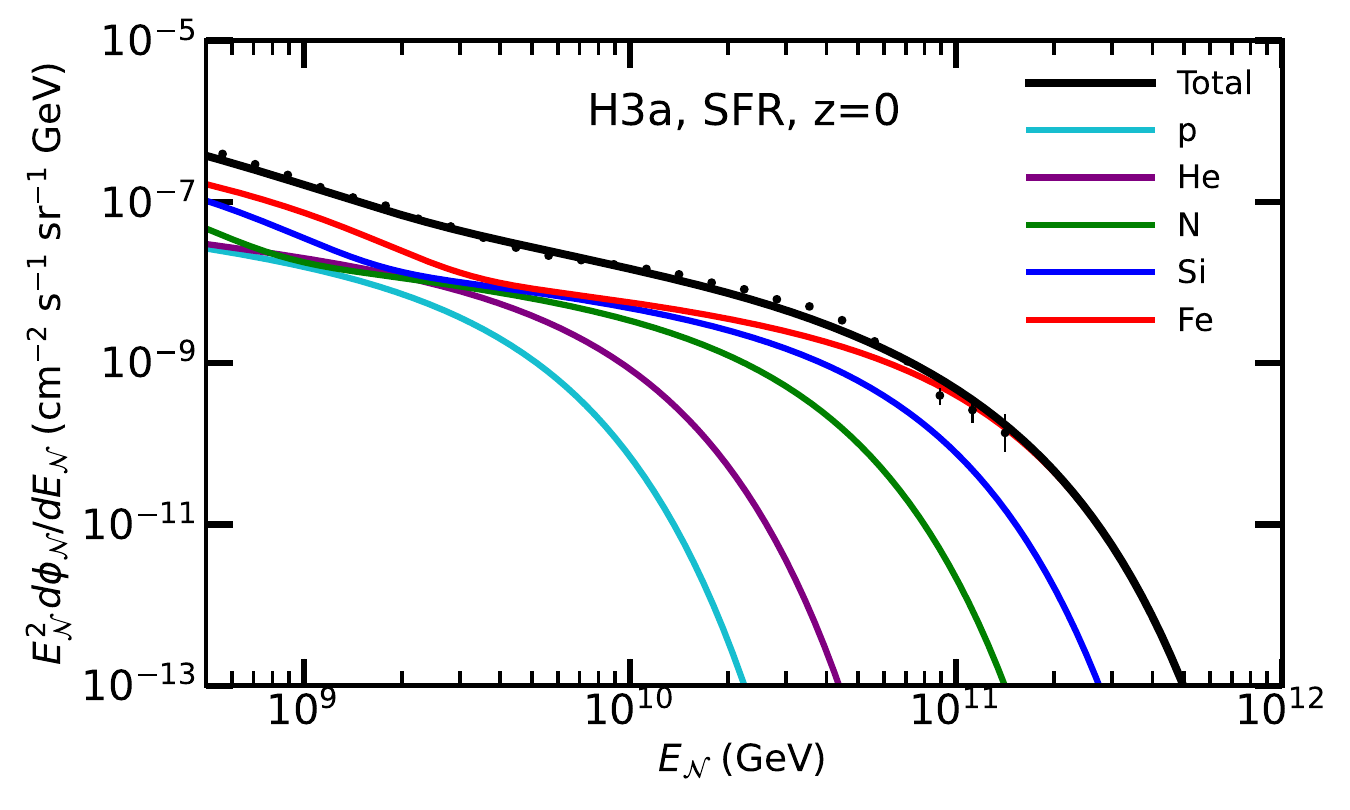}
    \end{minipage}
    \hfill
    \begin{minipage}{0.32\textwidth}
        \centering
        \includegraphics[width=\textwidth]{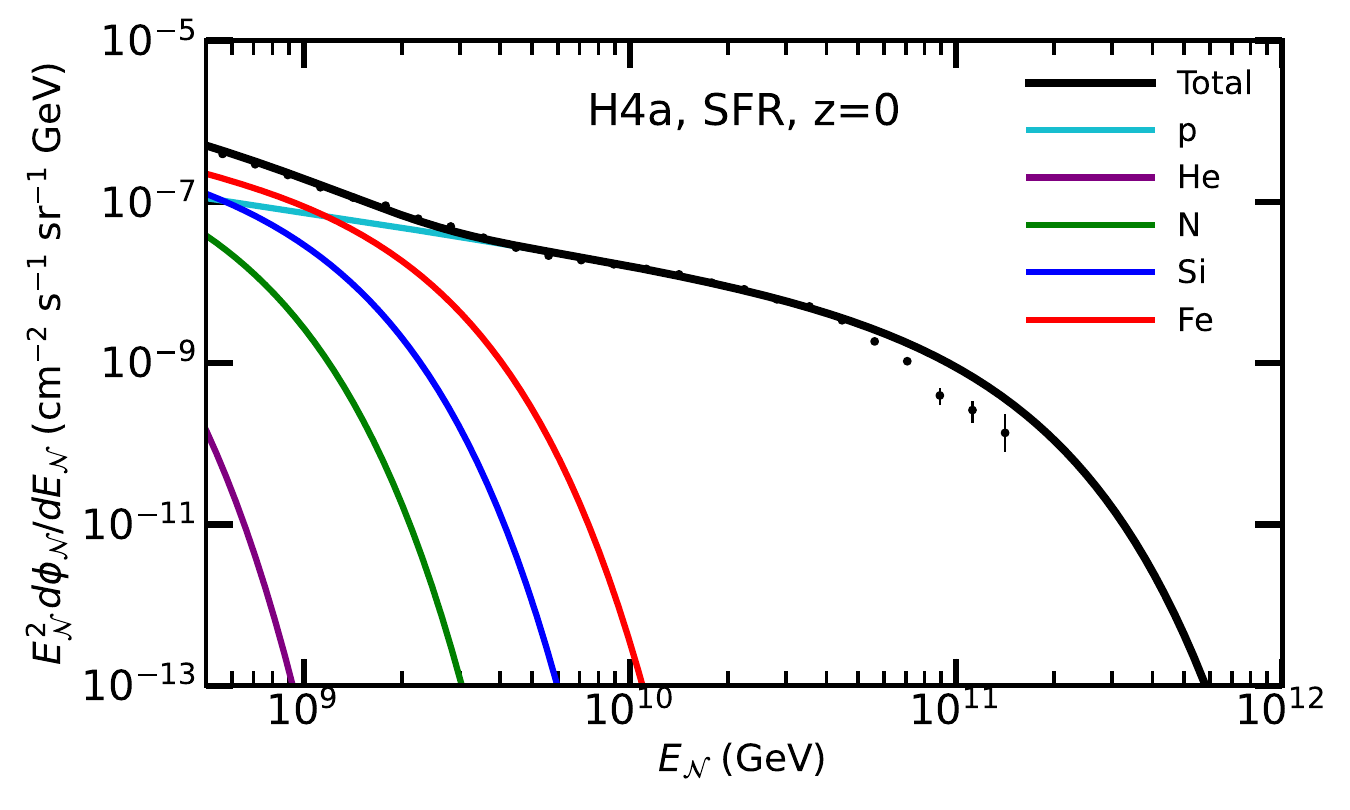}
    \end{minipage}

    \vspace{0.6em}

    % Row 2: z = 1
    \begin{minipage}{0.32\textwidth}
        \centering
        \includegraphics[width=\textwidth]{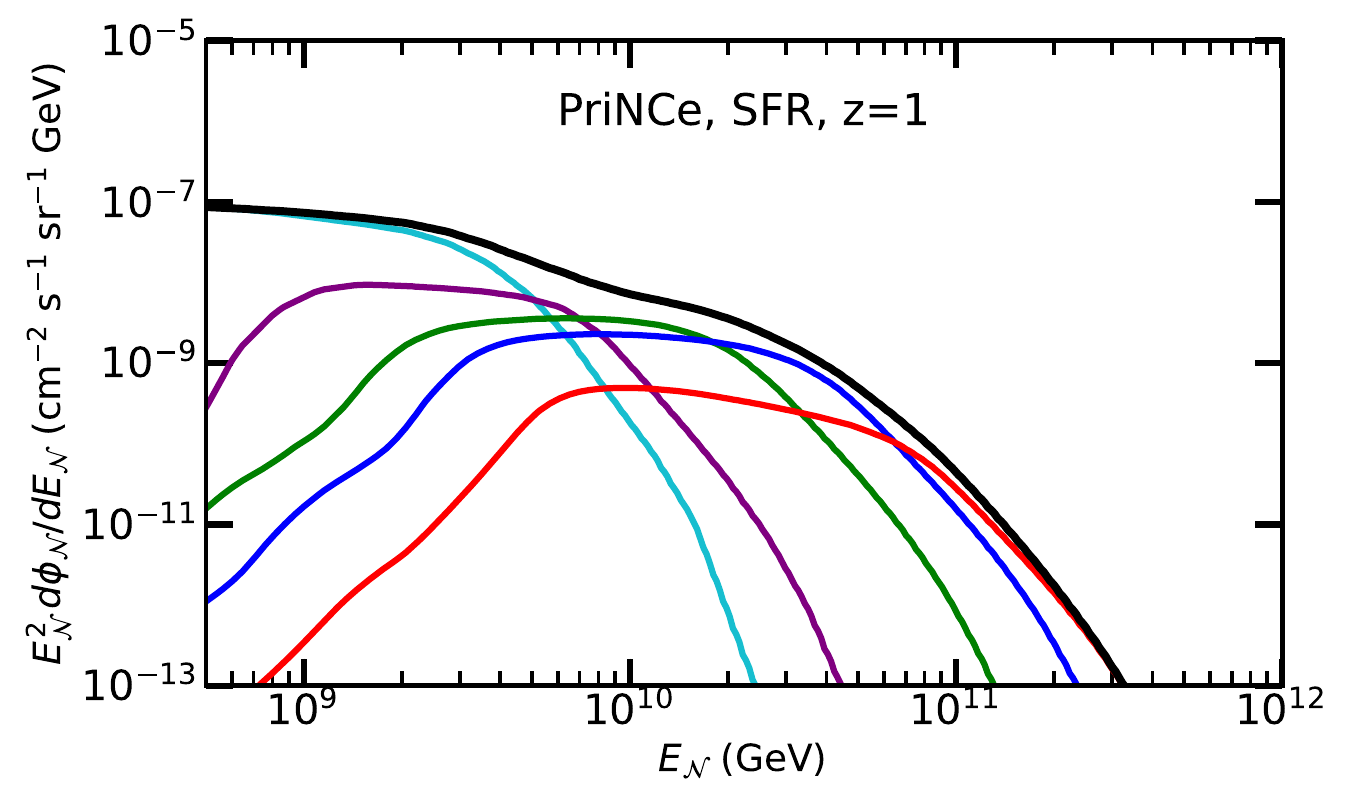}
    \end{minipage}
    \hfill
    \begin{minipage}{0.32\textwidth}
        \centering
        \includegraphics[width=\textwidth]{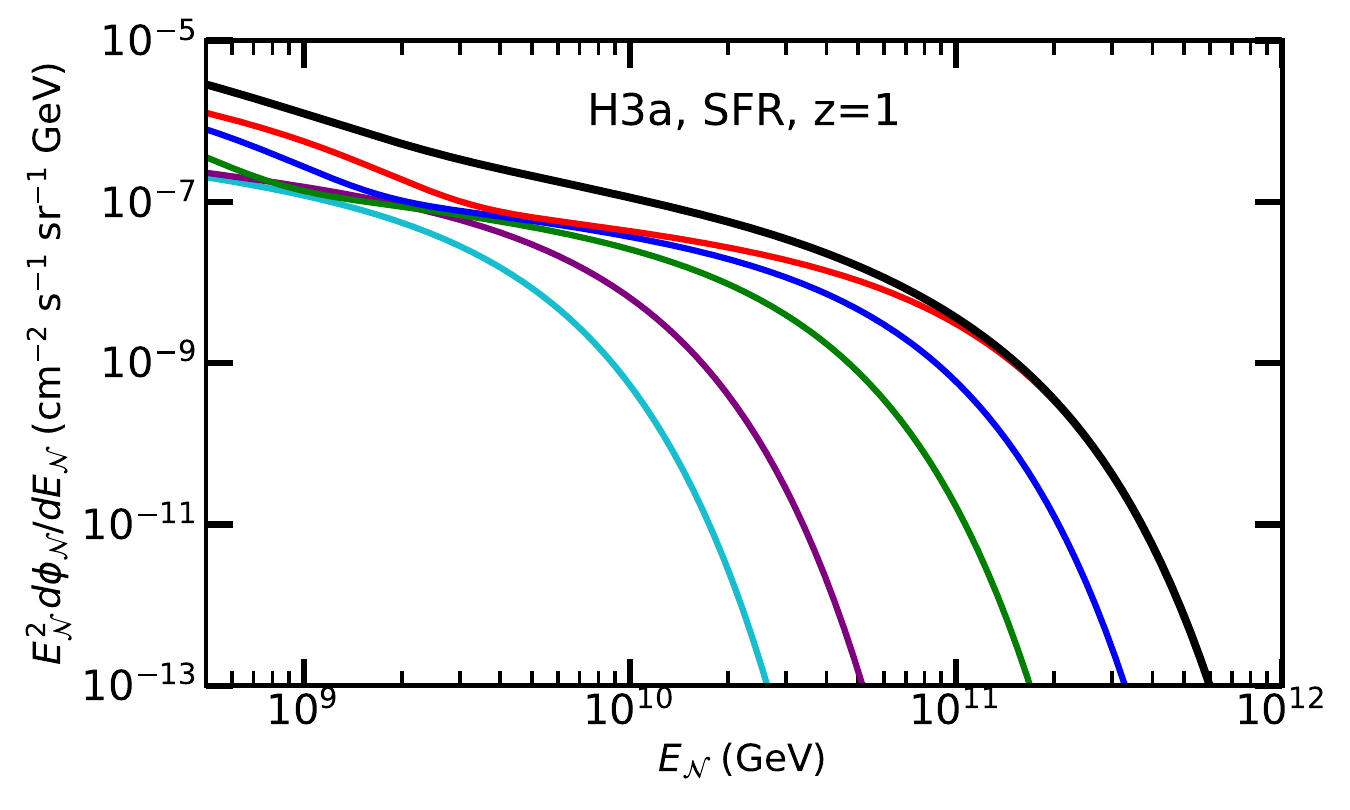}
    \end{minipage}
    \hfill
    \begin{minipage}{0.32\textwidth}
        \centering
        \includegraphics[width=\textwidth]{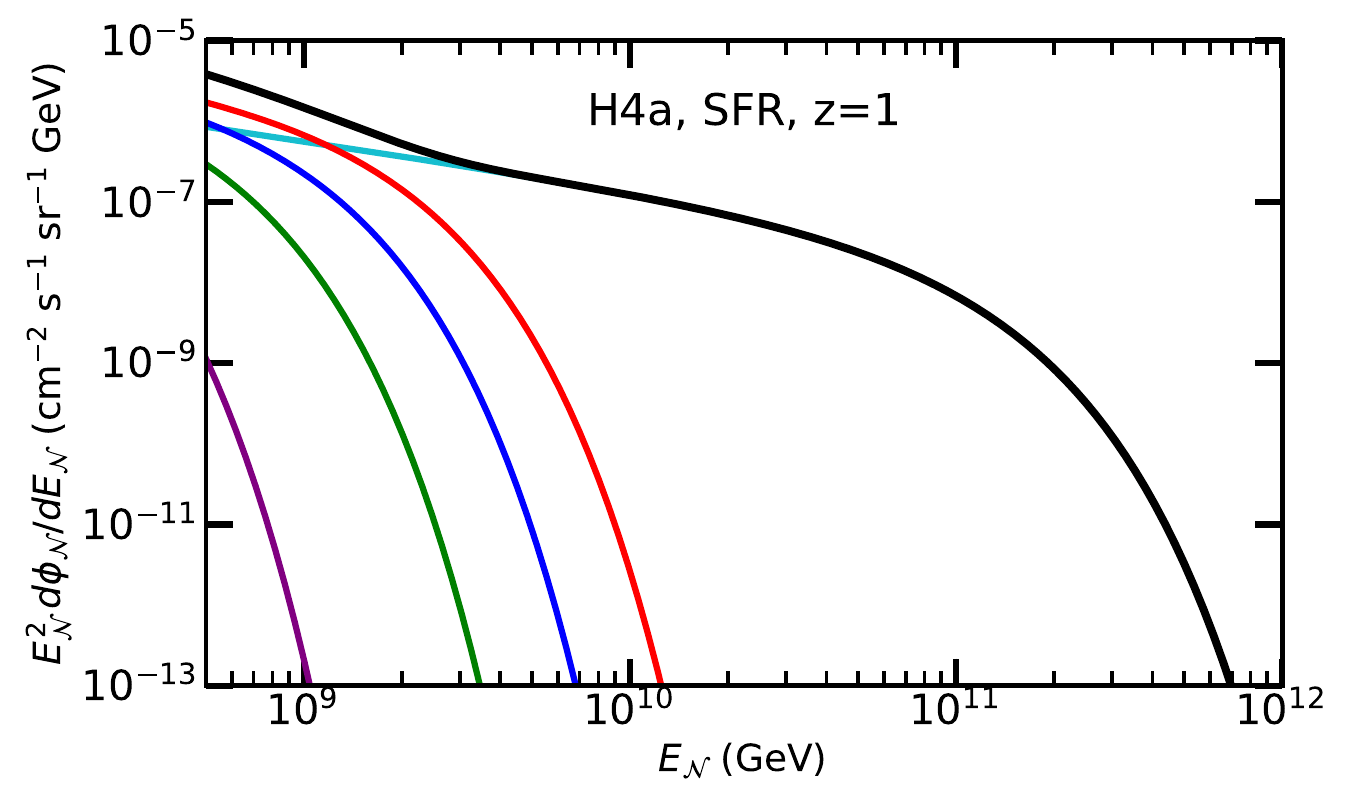}
    \end{minipage}

    \vspace{0.6em}

    % Row 3: z = 5
    \begin{minipage}{0.32\textwidth}
        \centering
        \includegraphics[width=\textwidth]{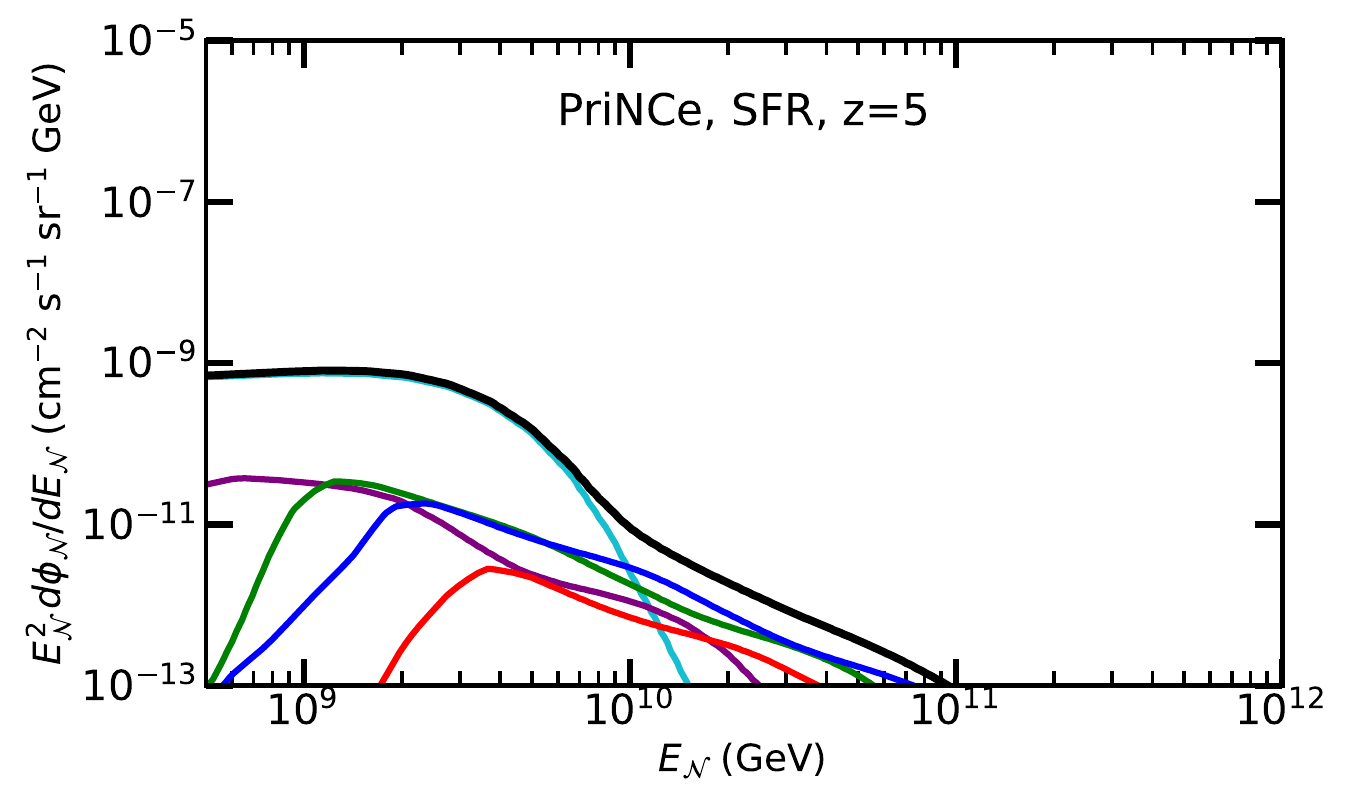}
    \end{minipage}
    \hfill
    \begin{minipage}{0.32\textwidth}
        \centering
        \includegraphics[width=\textwidth]{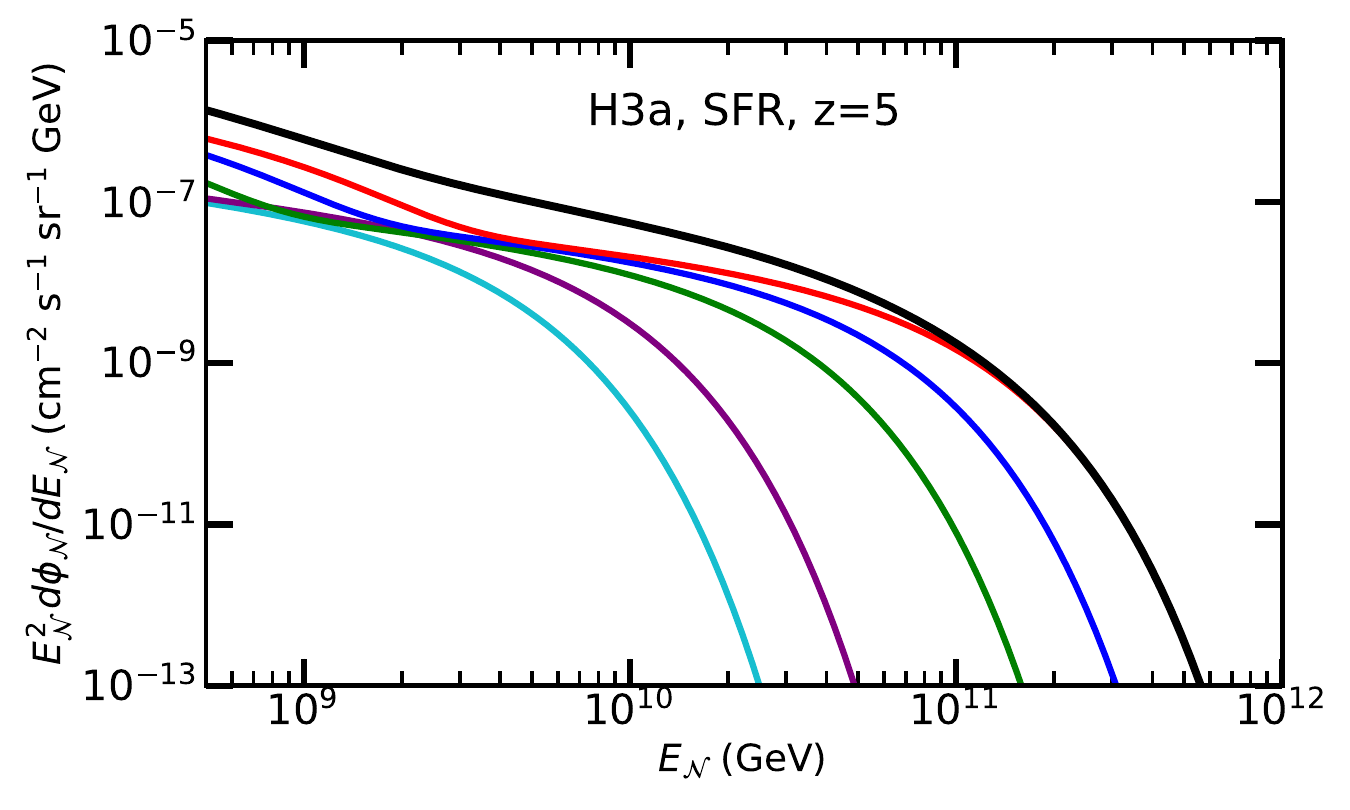}
    \end{minipage}
    \hfill
    \begin{minipage}{0.32\textwidth}
        \centering
        \includegraphics[width=\textwidth]{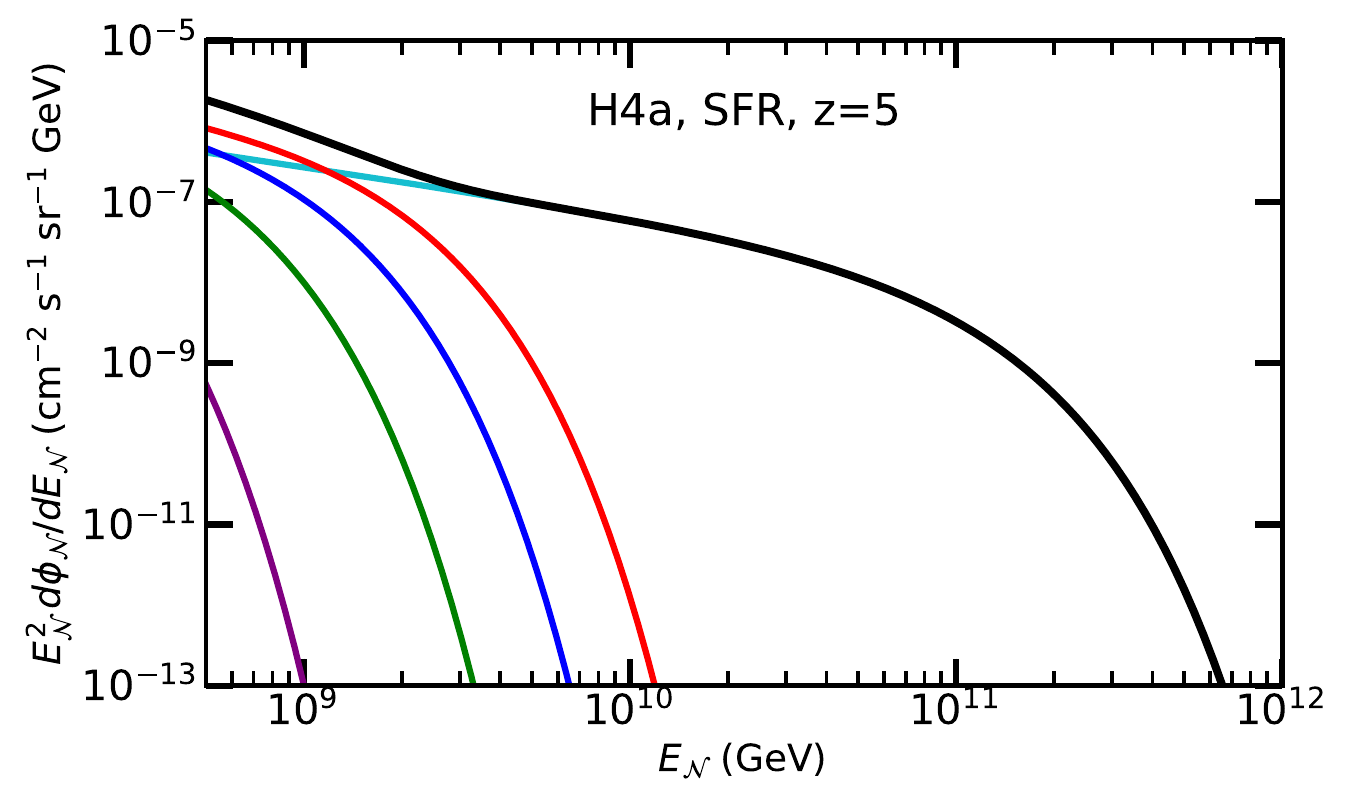}
    \end{minipage}

    \caption{
    Redshift dependence of the UHECR flux for the SFR source evolution model. From left to right, the columns show the PriNCe, H3a, and H4a spectra, respectively. From top to bottom, the rows correspond to $z=0$, $z=1$, and $z=5$. The colored curves denote different nuclear components, and the black curve gives the total flux. The data points at $z=0$ are taken from the Pierre Auger Observatory measurement~\cite{Veberic:2017hwu}.
    }
    \label{fig:cr_flux_sfr_redshift}
\end{figure*}

Figure~\ref{fig:cr_flux_sfr_redshift} compares the redshift dependence of the three UHECR flux descriptions used in this work. 
For illustration, we show the spectra for the SFR source evolution model at $z=0$, $z=1$, and $z=5$. 
The left column shows the PriNCe propagated spectra. 
In this case, the UHECR nuclei are propagated through cosmological photon backgrounds before reaching the Earth. 
As the redshift increases, the CMB and the extragalactic background light (EBL) become denser and more energetic, so pair production, photohadronic interactions, and photodisintegration become more efficient. 
These propagation effects suppress the high energy part of the UHECR spectrum and shift the composition dependent cutoffs to lower observed energies. 
As a result, the increase in the SFR source density at high redshift is partly offset by stronger propagation losses, leading to a suppressed propagated PriNCe flux, especially at $z=5$.

The middle and right columns show the Hillas-based H3a and H4a spectra. 
In these two cases, the local present-day Hillas spectra are not propagated through photon backgrounds, and they are only rescaled by the source-evolution function.
Therefore, their redshift dependence directly follows the assumed source evolution, rather than propagation-induced energy losses or composition changes. 
The main difference between H3a and H4a appears at the highest UHECR energies. 
For $E_{\mathcal{N}}\gtrsim10^{10}~{\rm GeV}$, H3a is dominated by heavy nuclear components, whereas H4a is dominated by protons. 
This difference is important for the prediction of boosted C$\nu$B flux, because heavy nuclei mainly enhance the low energy coherent contribution, whereas a high energy proton component plays a more important role in the ES, RES, and DIS regimes.

%%%%%%%%%%%%%%%%%%%%%%%% 
\section{UHECR-boosted C$\nu$B flux}
\label{sec:boosted_flux}
UHECRs propagate over cosmological distances from their sources to Earth and can scatter off the C$\nu$B along their trajectories. 
These interactions transfer part of the CR energy to the C$\nu$B and generate a diffuse flux of boosted neutrinos at Earth. In this section, we calculate the boosted C$\nu$B flux for different UHECR flux models.
%\subsection{Comparison of boosted C$\nu$B flux of different CR models}
%\label{subsec:Comparison_of_boosted_C$\nu$B_flux}
For a given UHECR flux and C$\nu$B overdensity, the differential boosted C$\nu$B flux can be written as~\cite{Herrera:2024upj,Zhang:2025rqh,Herrera:2026pzj}
\begin{equation}
\frac{d\phi_\nu}{dE_\nu}
=
\sum_{i,j}
\int_{z_{\min}}^{z_{\max}}
dz\,
\frac{c}{H(z)}\,
\eta\,n_{\nu_j}\,(1+z)^3
\int_0^\infty
dE_{\mathcal{N}_i}\,
\frac{d\sigma^{\nu \mathcal{N}_i}}{dE_\nu'}
\,
\frac{d\phi_{\mathcal{N}_i}}{dE_{\mathcal{N}_i}}(z)\,
\Theta\!\left[E_\nu^{\max}-E_\nu'\right] \,,
\label{eq:boosted_flux}
\end{equation}
The relic-neutrino density at redshift $z$ is scaled by the factor $(1+z)^3$, and $\eta$ is the relic neutrino overdensity. 
The differential cross section is evaluated at the production redshift, where the observed boosted-neutrino energy is related to the production energy by $E_\nu'=E_\nu(1+z)$.
The kinematic upper limits for the outgoing boosted relic neutrino energy $E_\nu^{\max}$ depend on the scattering channel and are given in Sec.~\ref{sec:cross_sections}.
We adopt a flat $\Lambda$CDM cosmology,
$H(z)=H_0\sqrt{\Omega_m(1+z)^3+\Omega_\Lambda}$
with $c/H_0=1.1267\times 10^{28}~{\rm cm}$~\cite{ParticleDataGroup:2024cfk}, $\Omega_m=0.315$, $\Omega_\Lambda=0.685$, and perform the redshift integral over $0\le z\le 6$~\cite{hopkins2006normalization,Herrera:2024upj,Zhang:2025rqh}. For the mass splitting parameters, we use the best-fit values in Ref.~\cite{Esteban:2020cvm} for both the normal ordering (NO) and inverted ordering (IO).
Note that Eq.~\eqref{eq:boosted_flux} assumes that C$\nu$B scattering does not appreciably modify the UHECR spectrum during propagation.
We have examined the validity of this approximation in Appendix~\ref{app:cr_energy_loss} and find that C$\nu$B-induced energy losses have only a small impact on the UHECR spectrum in the parameter region relevant to the present analysis. \footnote{This is also supported by a recent propagation study that incorporated UHECR--C$\nu$B scattering consistently into the full UHECR propagation framework~\cite{Azeredo:2026qnc}.
For $m_\nu=0.1~{\rm eV}$, relic-neutrino overdensities below $\eta\sim10^9$ produce only small modifications to the UHECR spectrum, and UHECR--C$\nu$B interactions become comparable to interactions with diffuse photon backgrounds only at $\eta\gtrsim10^{10}$~\cite{Azeredo:2026qnc}.}

\begin{figure*}[t]
    \centering

    % Row 1: PriNCe
    \begin{minipage}{0.48\textwidth}
        \centering
        \includegraphics[width=\textwidth]{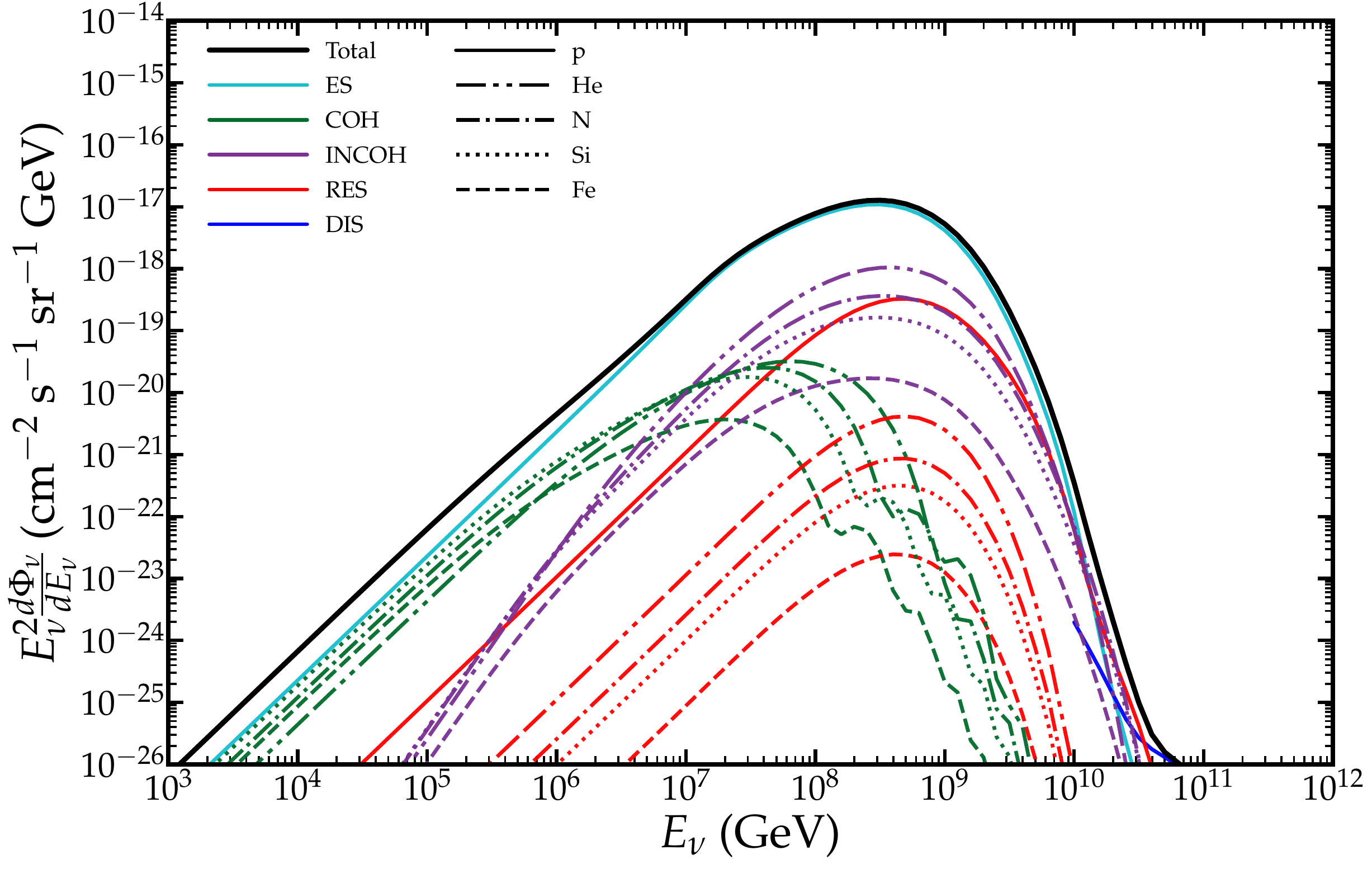}
    \end{minipage}
    \hfill
    \begin{minipage}{0.48\textwidth}
        \centering
        \includegraphics[width=\textwidth]{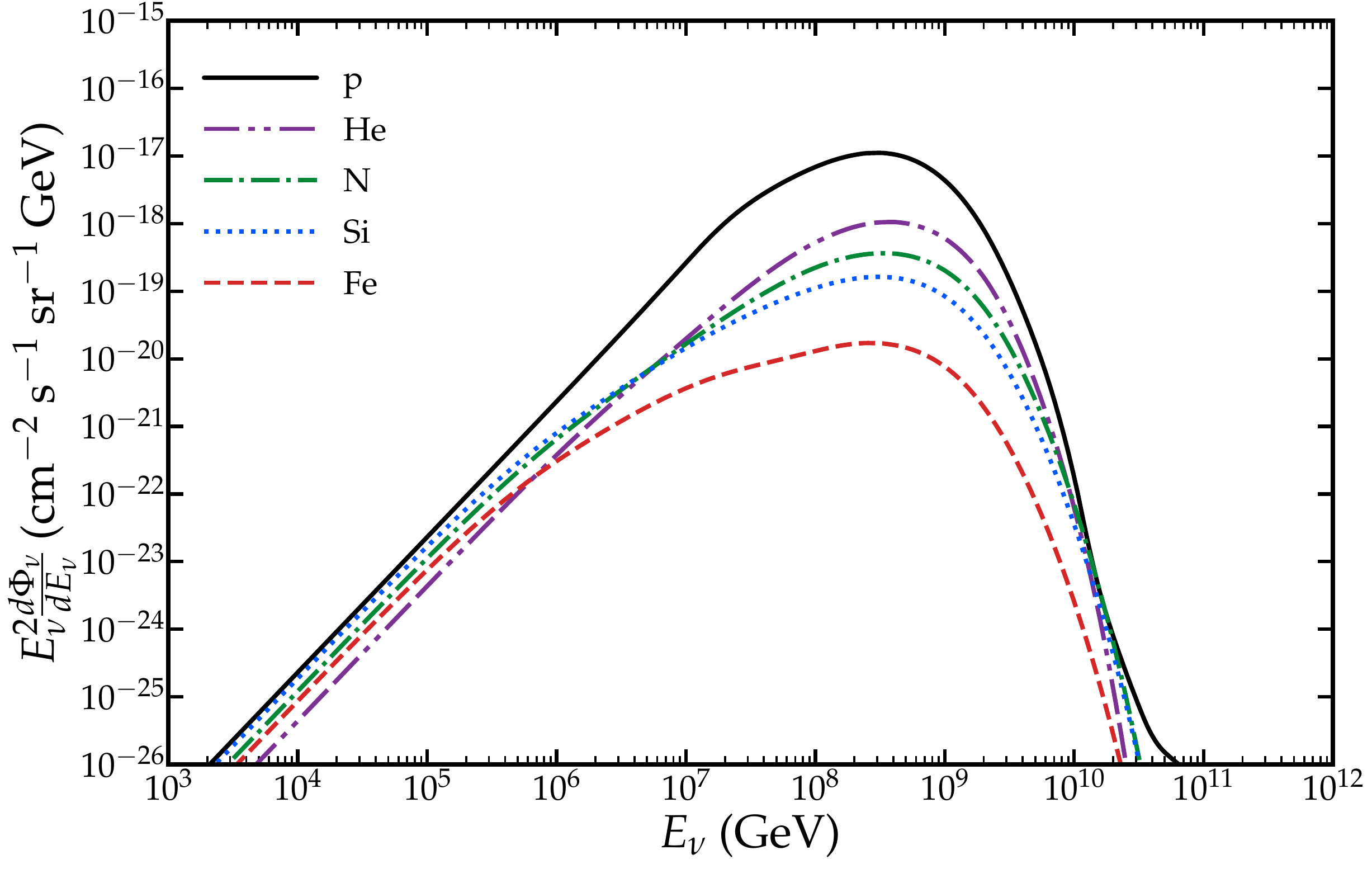}
    \end{minipage}

    \vspace{0.8em}

    % Row 2: H3a
    \begin{minipage}{0.48\textwidth}
        \centering
        \includegraphics[width=\textwidth]{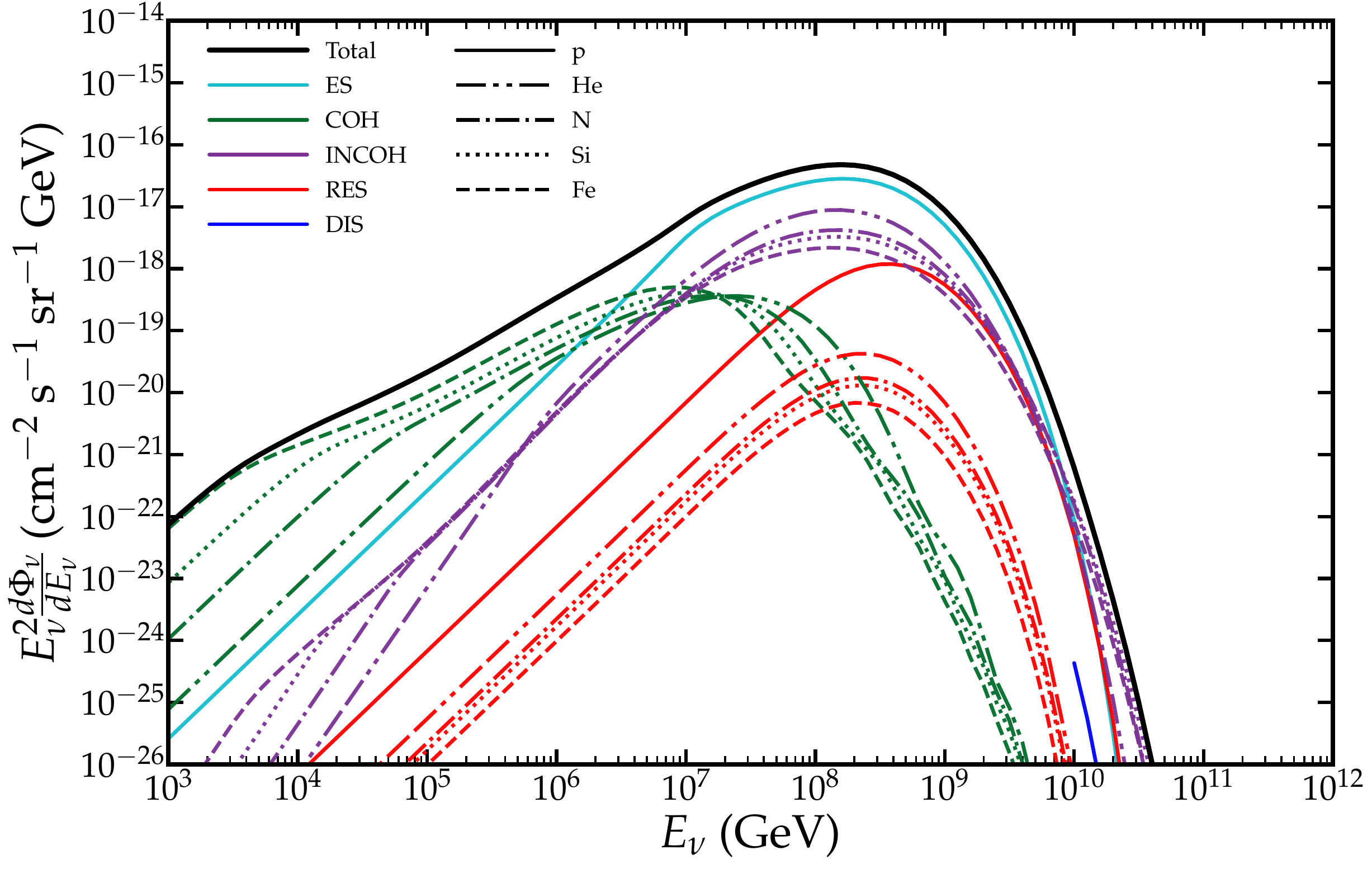}
    \end{minipage}
    \hfill
    \begin{minipage}{0.48\textwidth}
        \centering
        \includegraphics[width=\textwidth]{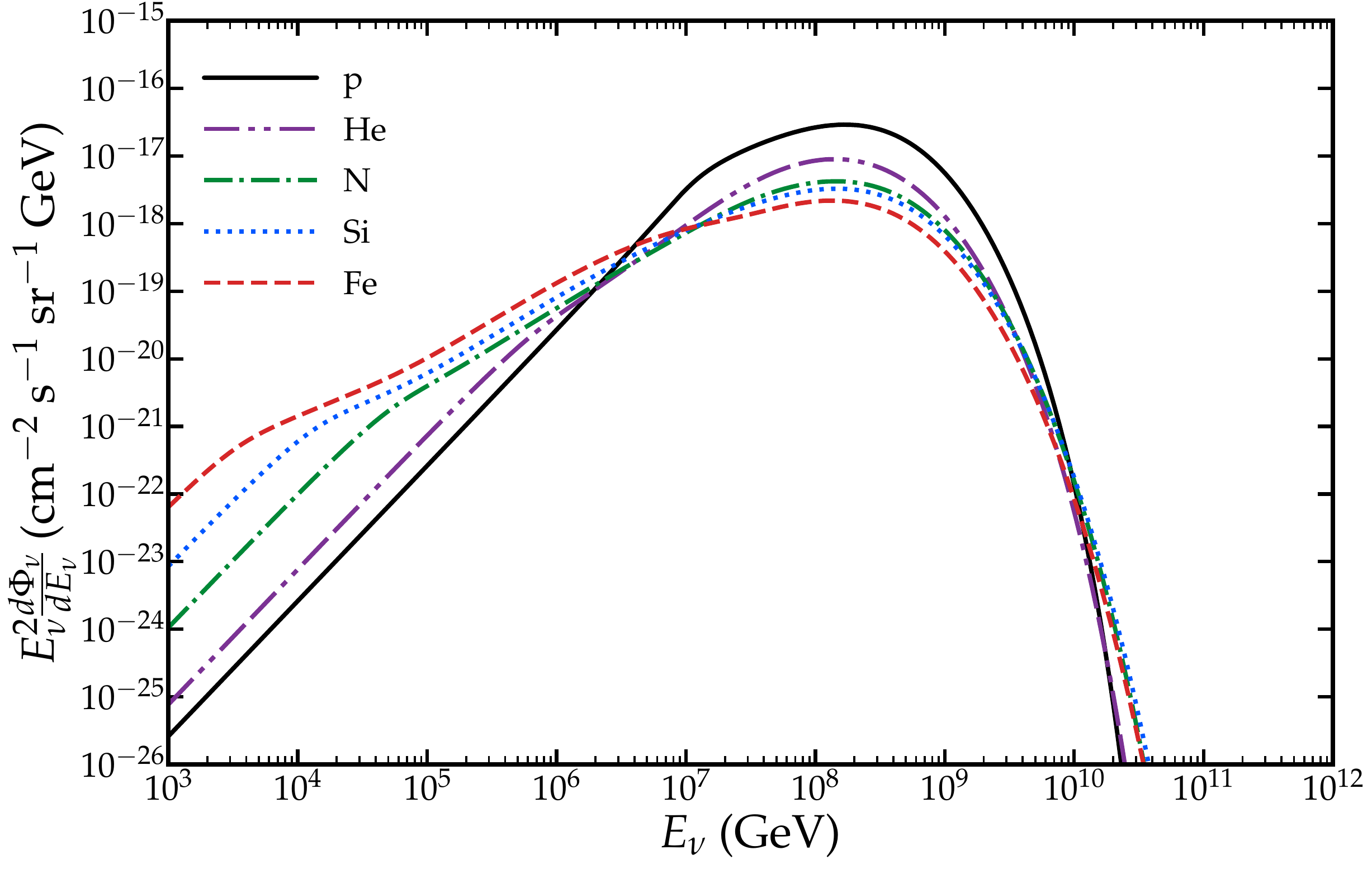}
    \end{minipage}

    \vspace{0.8em}

    % Row 3: H4a
    \begin{minipage}{0.48\textwidth}
        \centering
        \includegraphics[width=\textwidth]{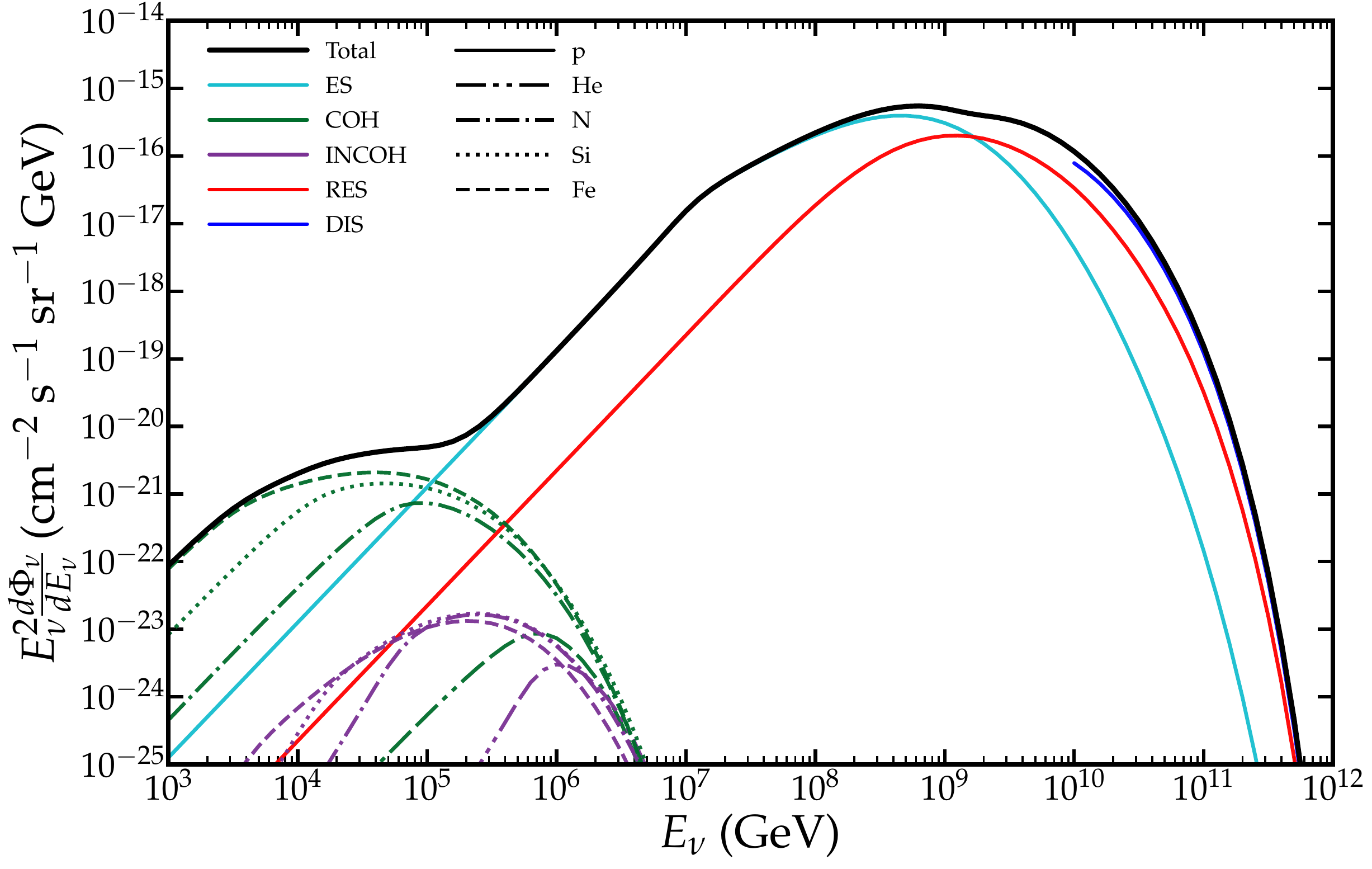}
    \end{minipage}
    \hfill
    \begin{minipage}{0.48\textwidth}
        \centering
        \includegraphics[width=\textwidth]{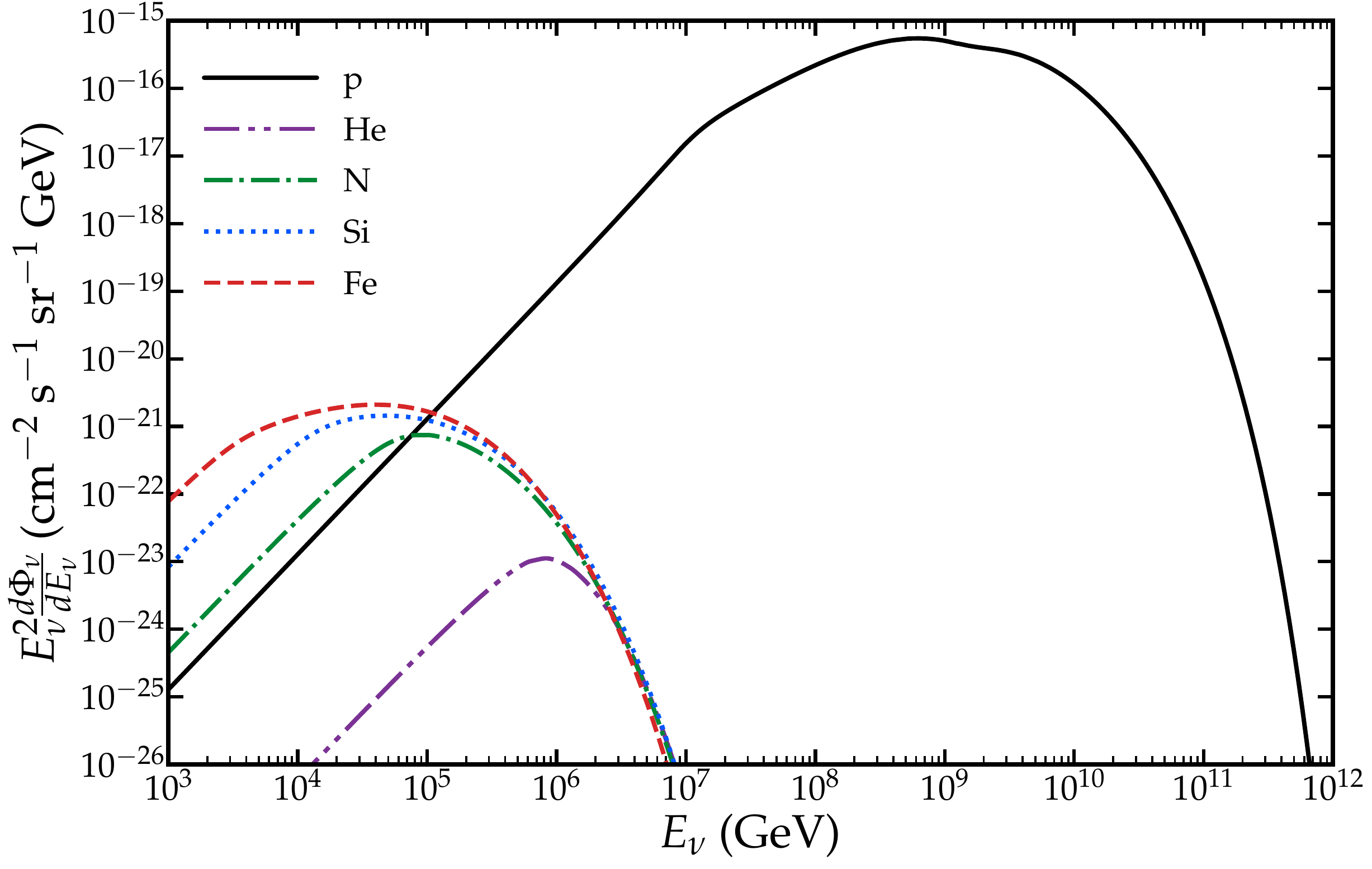}
    \end{minipage}

    \caption{
    Boosted C$\nu$B fluxes for $\eta=1$ and $m_1=0.1~{\rm eV}$ obtained with the PriNCe, H3a, and H4a UHECR flux models for the SFR source evolution model.
    From top to bottom, the rows correspond to PriNCe, H3a, and H4a, respectively.
    The left column shows the decomposition by scattering channel and UHECR nuclear component, while the right column shows the contributions of individual UHECR nuclear component to the total boosted C$\nu$B flux.
    In the left column, colors denote scattering channels, line styles denote UHECR nuclear components, and the black solid curve gives the total flux.
    In the right column, different curves show the contributions from different UHECR nuclear components.
    }
    \label{fig:flux_all_models}
\end{figure*}

We show the boosted C$\nu$B flux obtained with the PriNCe, H3a, and H4a UHECR flux models for the SFR source evolution model in Fig.~\ref{fig:flux_all_models}. Here we fix $\eta=1$ and assume the NO with $m_1=0.1~{\rm eV}$. 
The left column shows the decomposition into scattering channels and UHECR nuclear components, while the right column shows the contributions of each UHECR nuclear component to the total boosted C$\nu$B flux.
The first row of Fig.~\ref{fig:flux_all_models} corresponds to the PriNCe propagated spectra. 
In this case, the proton component gives the dominant contribution over nearly the entire boosted-neutrino energy range, as shown in the right panel. 
This behavior is consistent with the UHECR spectra shown in Fig.~\ref{fig:cr_flux_sfr_redshift}, where the proton flux is much larger than the heavy-nucleus fluxes in the energy range $E_{\mathcal{N}}\sim 5\times10^8$ -- $ 3\times10^{9}~{\rm GeV}$ that mainly contributes to the boosted C$\nu$B signal. 
The heavier nuclear components therefore remain subdominant, even though they can receive coherent enhancement at low energies. 
As shown in the left column, the ES channel associated with UHECR protons gives the leading contribution, while the nuclear COH, INCOH, RES, and DIS channels are subdominant.

The second row of Fig.~\ref{fig:flux_all_models} shows the results obtained with the H3a implementation of the Hillas model with three mixed-composition populations. 
Unlike the PriNCe case, the low-energy part of the boosted C$\nu$B spectrum receives a sizable contribution from nuclear UHECR components. 
At $E_\nu\lesssim5\times10^6~{\rm GeV}$, heavier nuclei, especially Si and Fe, give the largest contributions because the scattering is still in the coherent regime.
As $E_\nu$ increases, the nuclear form factor suppresses the coherent contribution, and the flux from heavy components drops rapidly. 
At higher energies, lighter components become more important because, for a fixed nuclear energy, they carry a larger energy per nucleon and can more efficiently produce high-energy boosted C$\nu$B neutrinos. 
In the channel decomposition, COH dominates the low-energy region, while ES and INCOH become more important after coherence is suppressed. 
The RES channel becomes visible for $E_\nu\gtrsim10^9~{\rm GeV}$ and contributes to the high-energy tail, whereas DIS appears only at the highest energies and remains much smaller than the total flux in the H3a case.

The third row of Fig.~\ref{fig:flux_all_models} shows the results with the H4a implementation of the Hillas model with an additional high-rigidity proton-only population, Pop.~3$^\ast$. 
Compared with H3a, H4a gives a larger boosted C$\nu$B flux and extends the spectrum to higher neutrino energies. 
This behavior is mainly driven by the high-rigidity proton component in H4a. 
Since this component can reach larger UHECR energies, the scattering can access larger momentum transfers more efficiently, making the proton-level ES, RES, and DIS channels more important. 
Consequently, the proton component dominates the boosted C$\nu$B flux across nearly the entire energy range, while the nuclear components contribute mainly at lower energies. 
The ES channel gives the leading contribution at lower boosted-neutrino energies, whereas RES becomes important in the high-energy region and DIS appears at the highest energies. 
This channel hierarchy differs from that in H3a, where the flux is mainly controlled by nuclear components and drops rapidly for $E_\nu\gtrsim10^9~{\rm GeV}$. 
By contrast, in H4a the decrease becomes pronounced only for $E_\nu\gtrsim10^{10}~{\rm GeV}$. 
Therefore, the comparison between H3a and H4a shows that the predicted boosted C$\nu$B spectrum is sensitive not only to the UHECR composition, but also to the maximum rigidity of the dominant high-energy component.

\section{Constraints on the C$\nu$B overdensity}
\label{sec:eta_constraints}

\begin{figure*}[t]
    \centering

    % Row 1: SFR
    \begin{minipage}{0.32\textwidth}
        \centering
        \includegraphics[width=\textwidth]{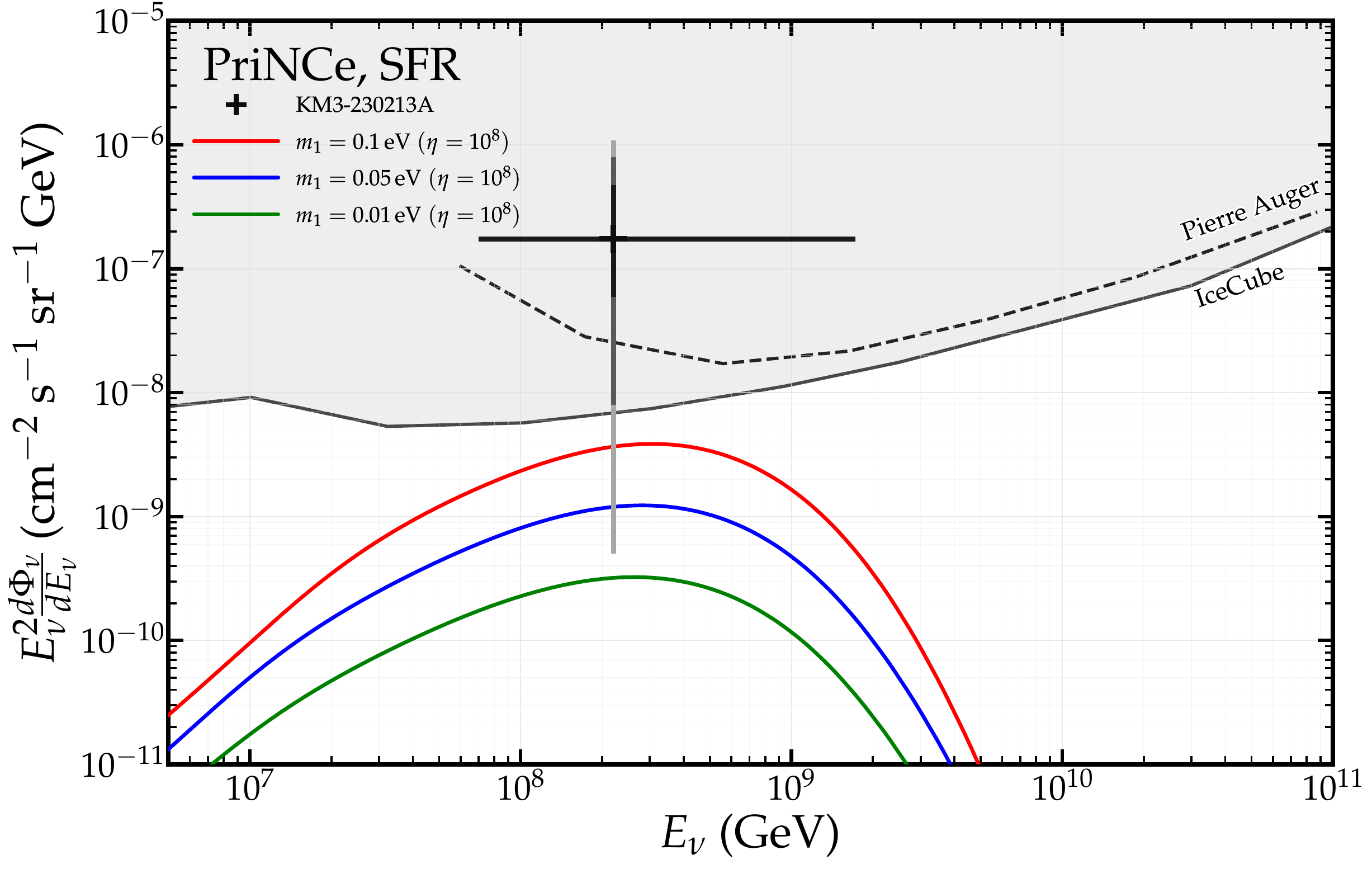}
    \end{minipage}
    \hfill
    \begin{minipage}{0.32\textwidth}
        \centering
        \includegraphics[width=\textwidth]{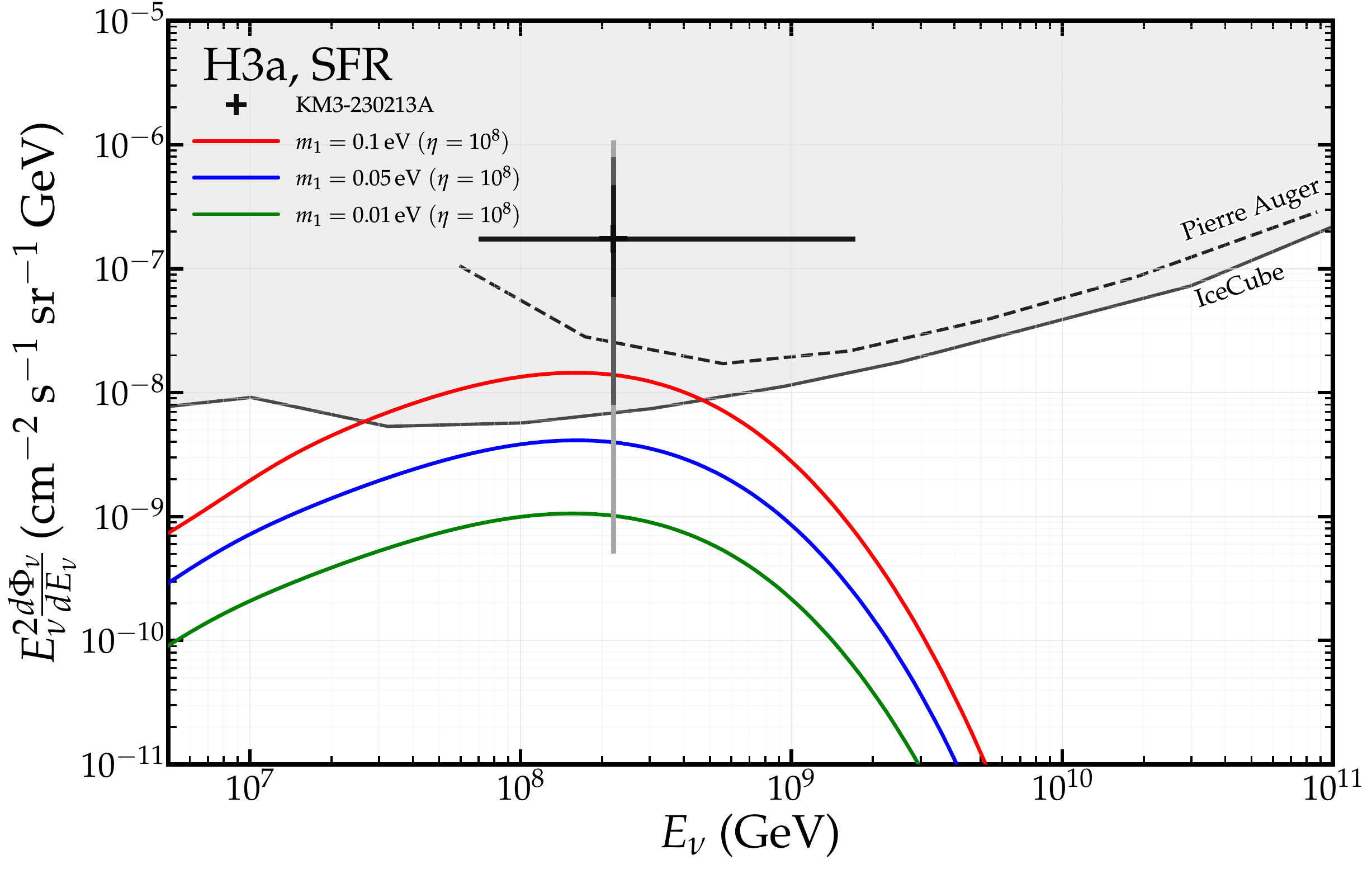}
    \end{minipage}
    \hfill
    \begin{minipage}{0.32\textwidth}
        \centering
        \includegraphics[width=\textwidth]{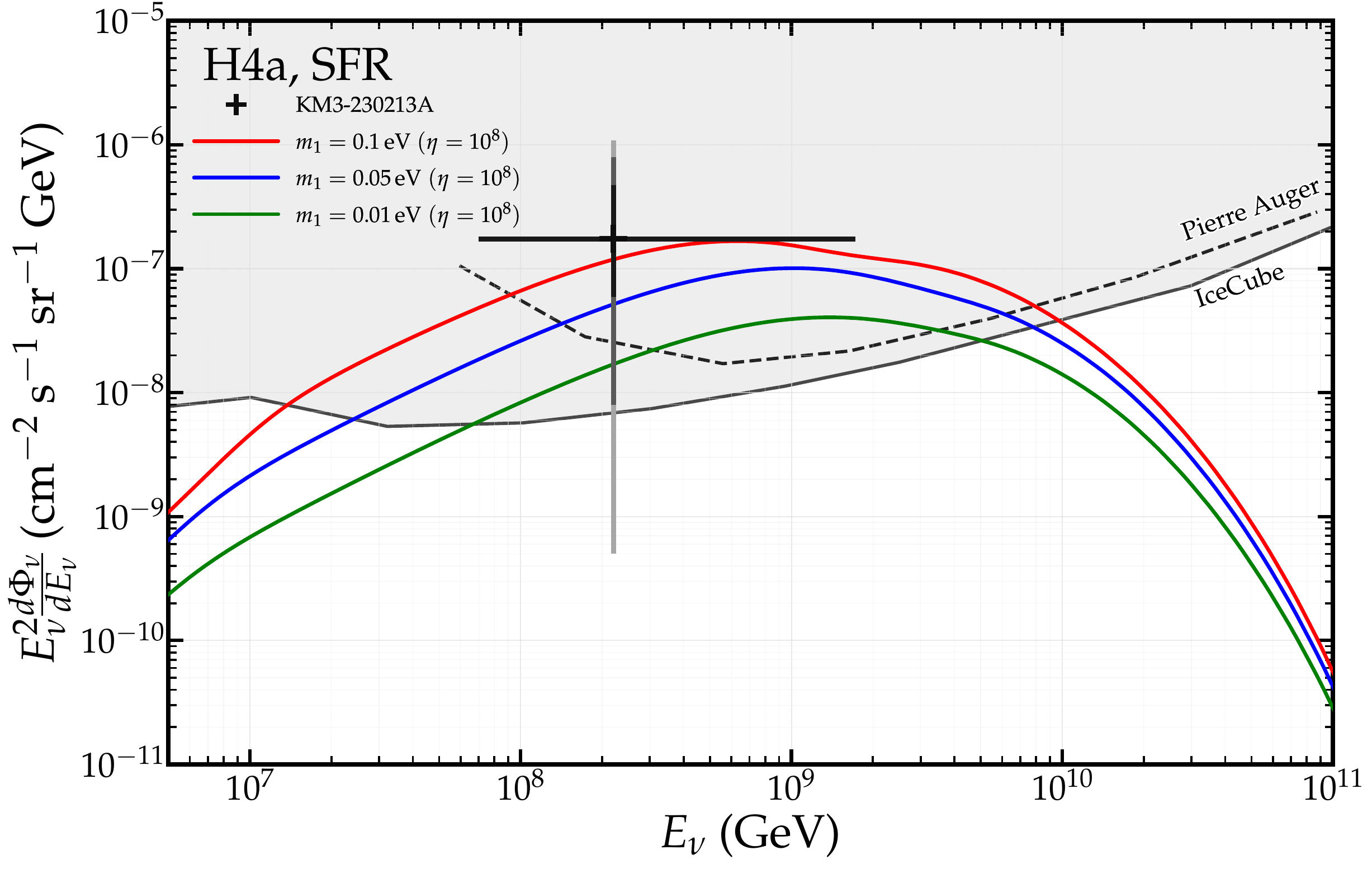}
    \end{minipage}

    \vspace{0.6em}

    % Row 2: QSO
    \begin{minipage}{0.32\textwidth}
        \centering
        \includegraphics[width=\textwidth]{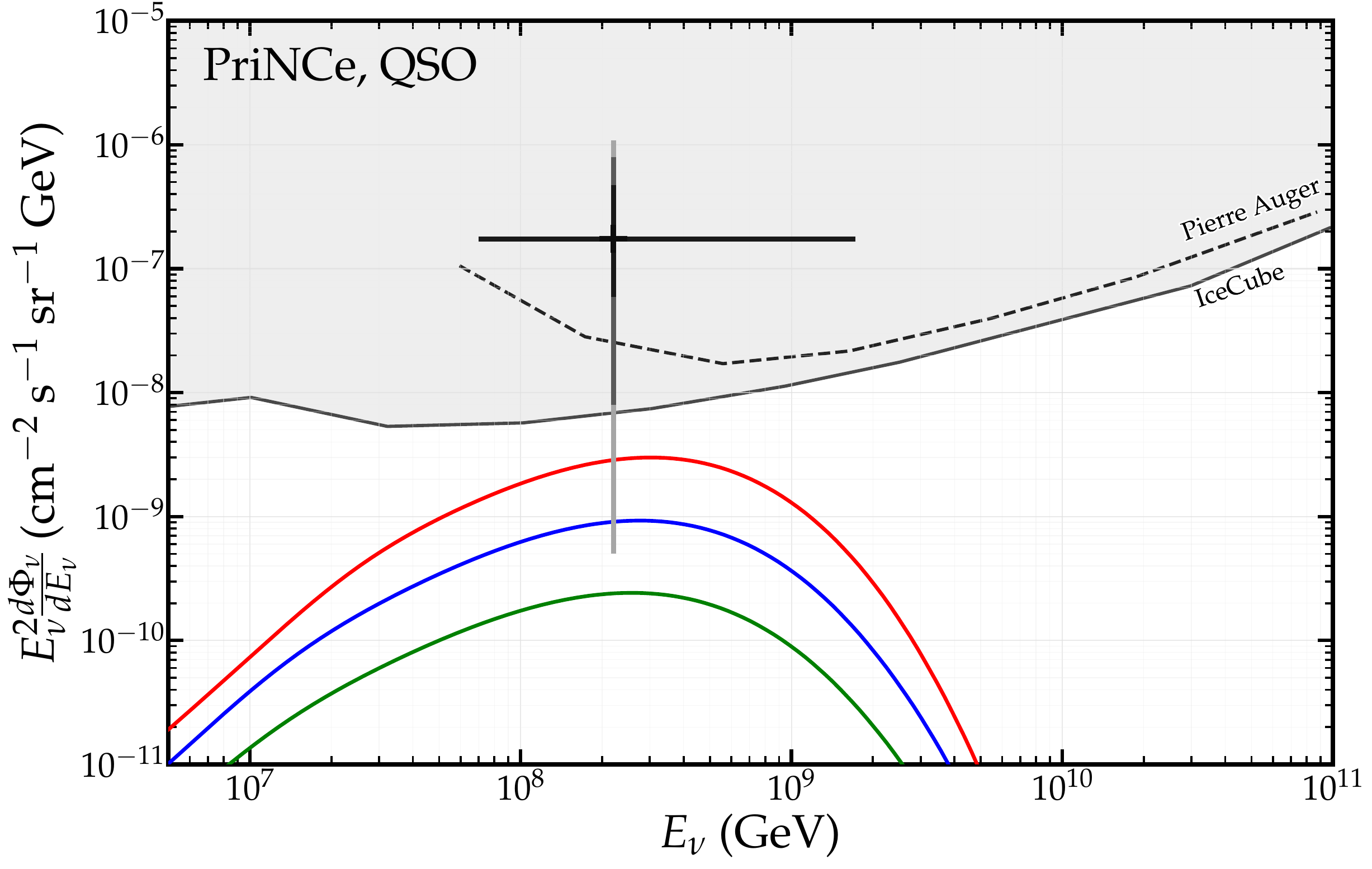}
    \end{minipage}
    \hfill
    \begin{minipage}{0.32\textwidth}
        \centering
        \includegraphics[width=\textwidth]{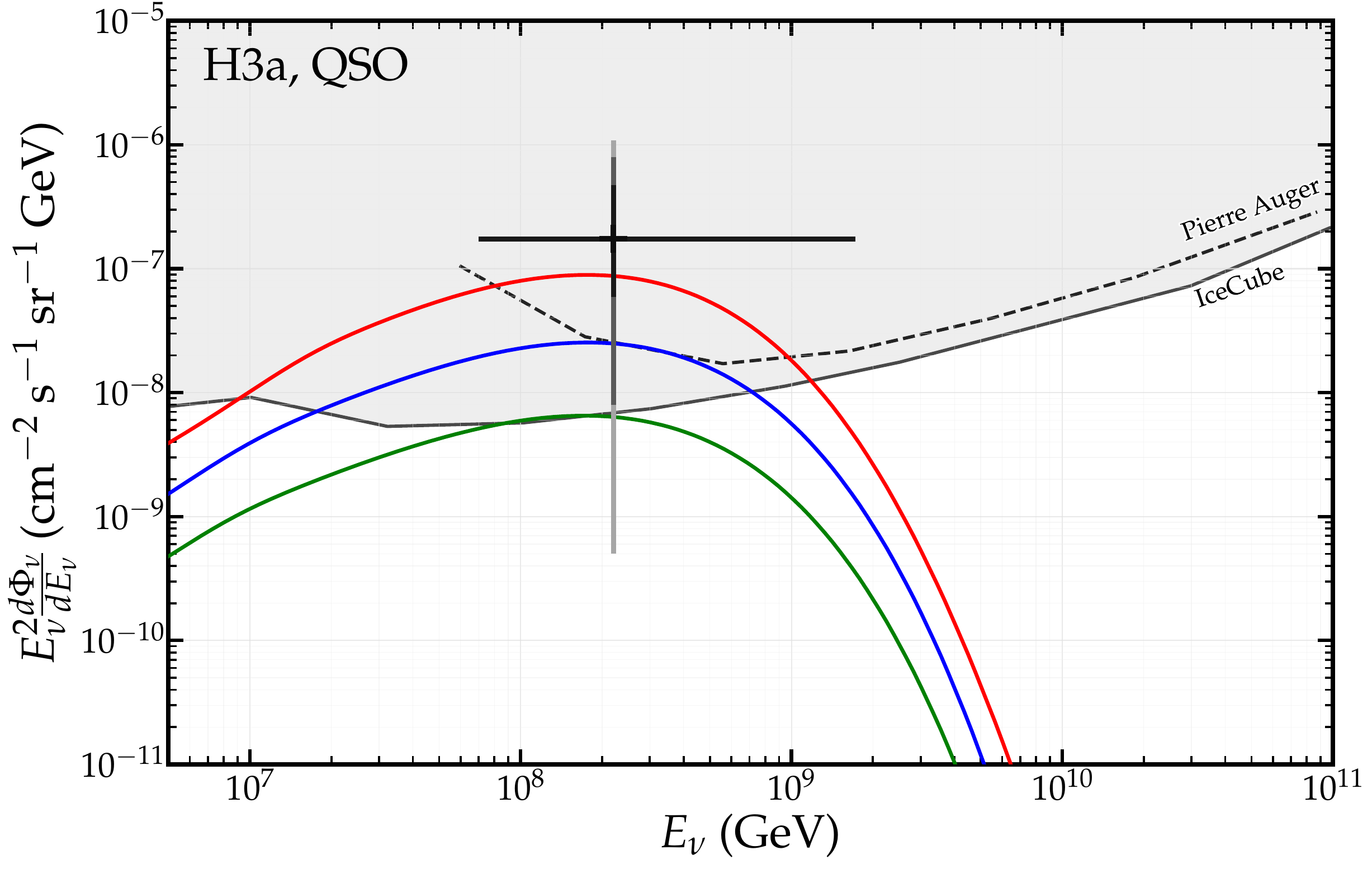}
    \end{minipage}
    \hfill
    \begin{minipage}{0.32\textwidth}
        \centering
        \includegraphics[width=\textwidth]{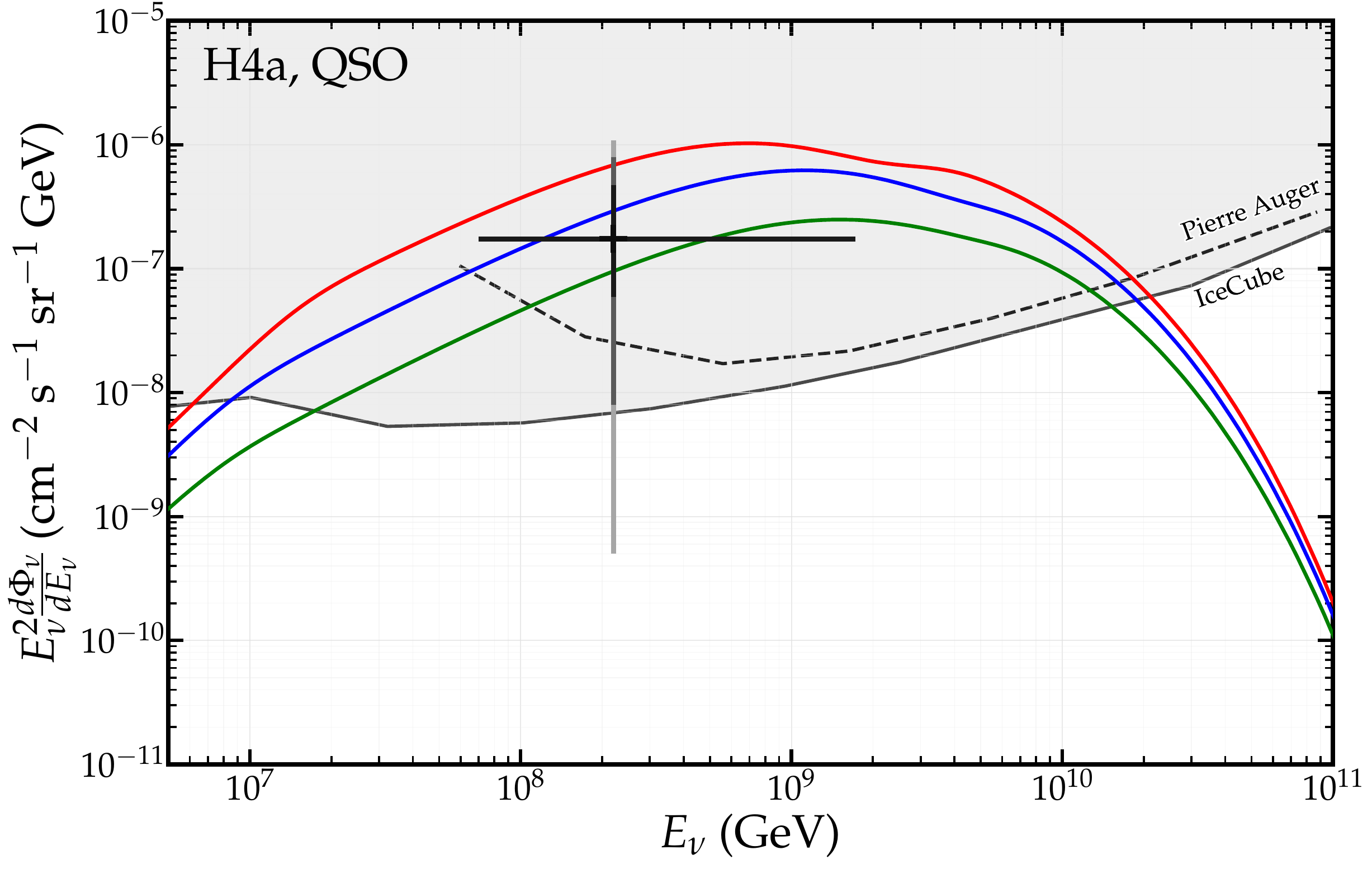}
    \end{minipage}

    \vspace{0.6em}

    % Row 3: GRB
    \begin{minipage}{0.32\textwidth}
        \centering
        \includegraphics[width=\textwidth]{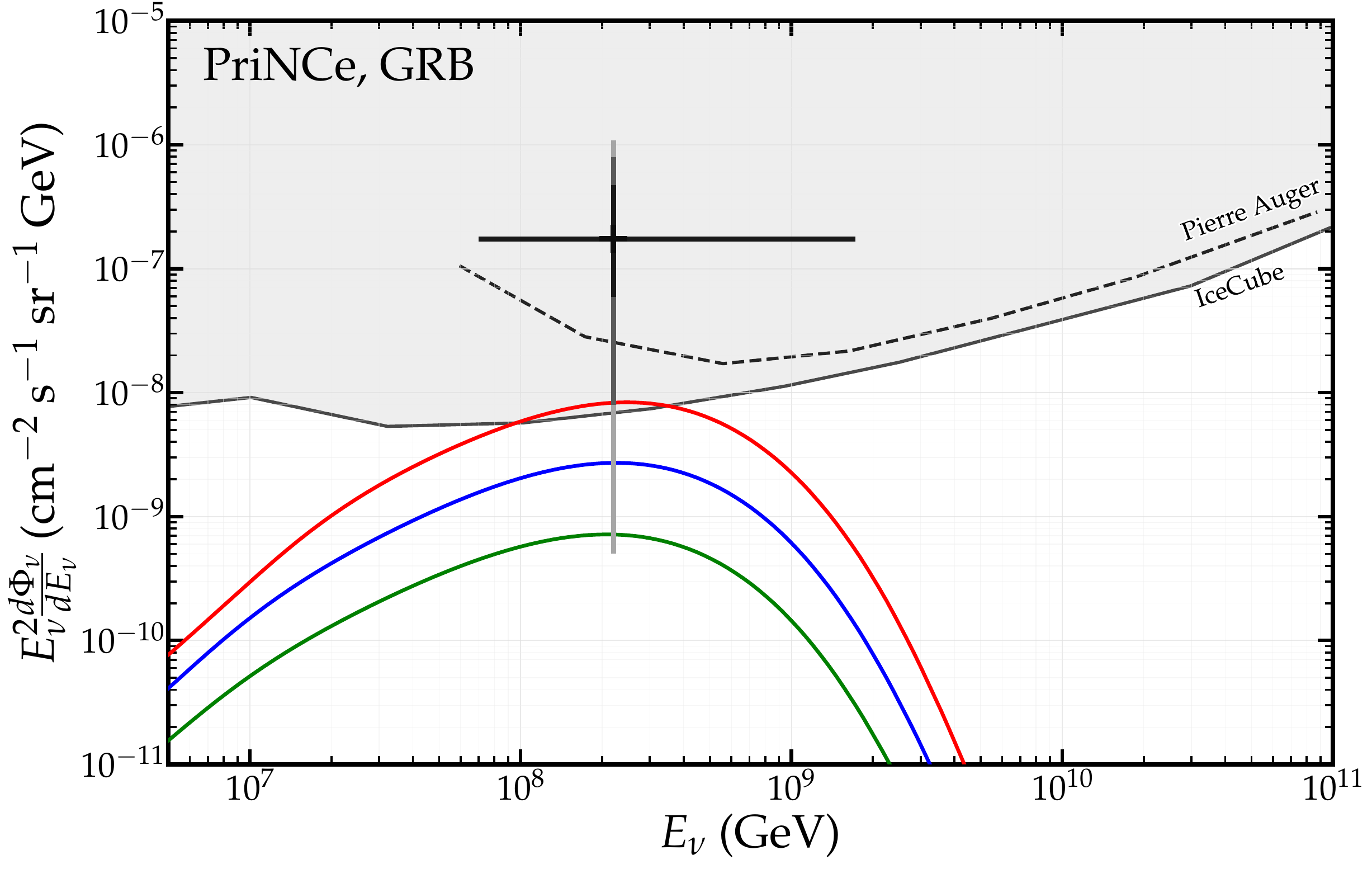}
    \end{minipage}
    \hfill
    \begin{minipage}{0.32\textwidth}
        \centering
        \includegraphics[width=\textwidth]{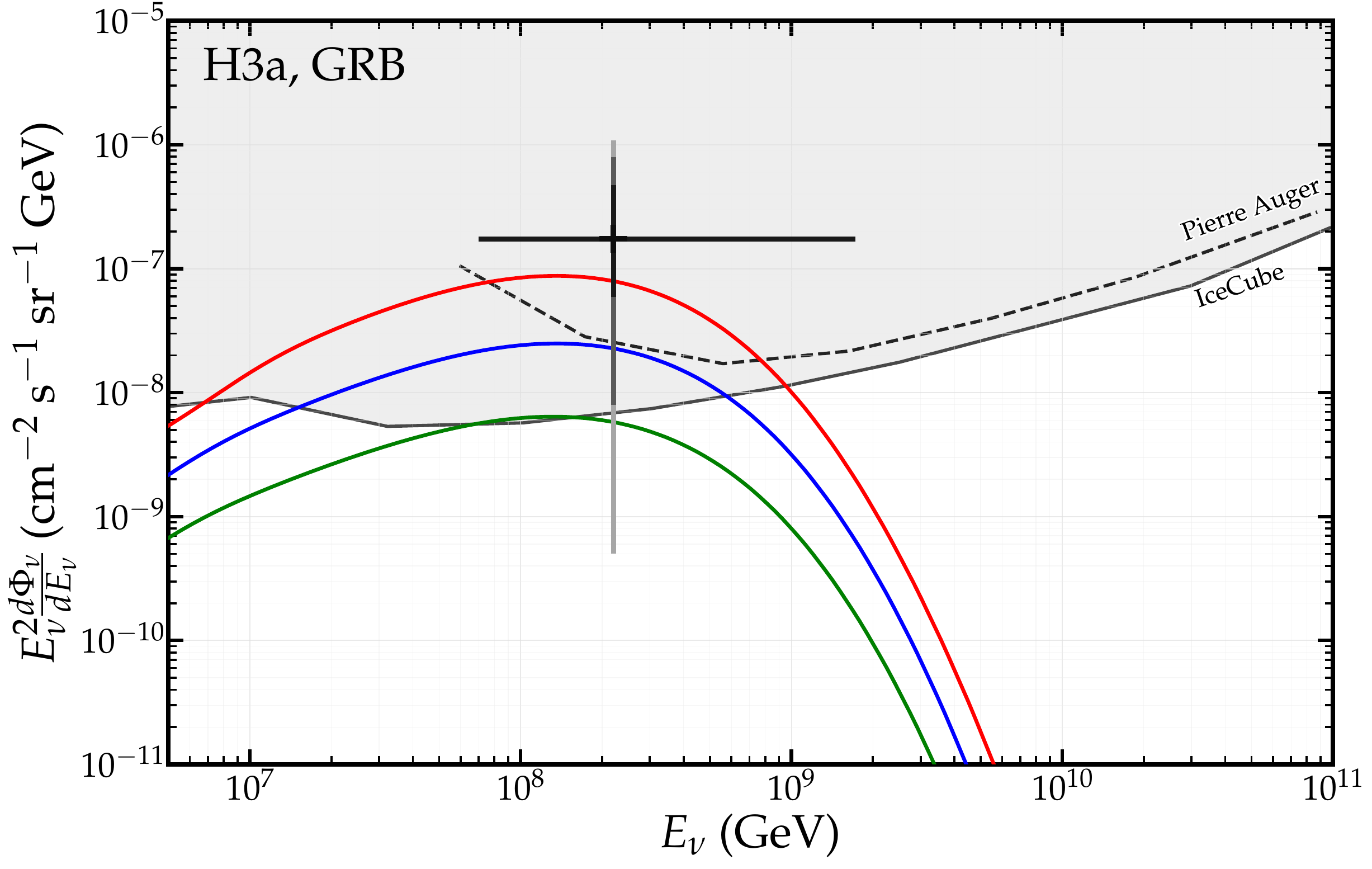}
    \end{minipage}
    \hfill
    \begin{minipage}{0.32\textwidth}
        \centering
        \includegraphics[width=\textwidth]{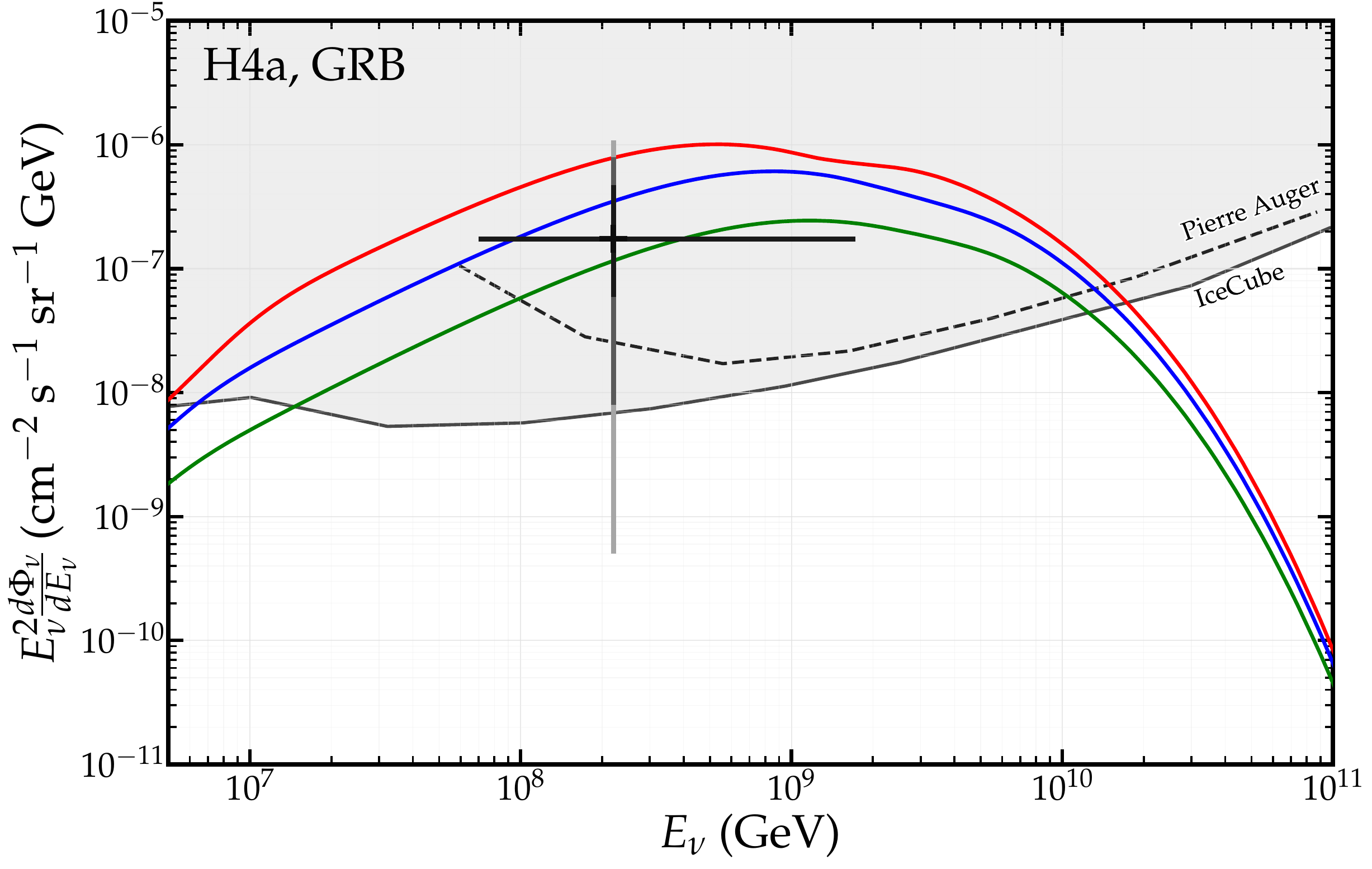}
    \end{minipage}

    \caption{
    Boosted C$\nu$B fluxes for three lightest-neutrino masses, $m_1=0.1$, $0.05$, and $0.01~{\rm eV}$, with $\eta=10^8$.
    From top to bottom, the rows correspond to the SFR, QSO, and GRB source evolution models.
    From left to right, the columns show the results obtained with the PriNCe, H3a, and H4a UHECR flux models.
    The gray solid and dashed curves show the current constraints from IceCube~\cite{IceCube:2025ezc,Meier:2024flg} and the Pierre Auger Observatory~\cite{PierreAuger:2019ens}, respectively, and the black cross marks the KM3-230213A event~\cite{KM3NeT:2025npi}.
    } 
    \label{fig:all_flux_constraints}
\end{figure*}

Using the boosted C$\nu$B fluxes obtained in Sec.~\ref{sec:boosted_flux}, we now derive constraints on the relic-neutrino overdensity $\eta$ from current high-energy neutrino observations.
We first show in Fig.~\ref{fig:all_flux_constraints} the predicted spectra for three representative lightest-neutrino masses, $m_1=0.1$, $0.05$, and $0.01~{\rm eV}$, with $\eta=10^8$. 
From left to right, the columns correspond to the PriNCe, H3a, and H4a UHECR flux models, while from top to bottom the rows correspond to the SFR, QSO, and GRB source evolution models. 
% For all UHECR flux models, the boosted C$\nu$B flux decreases as $m_1$ becomes smaller. 
% This is because the relevant scattering cross sections are approximately proportional to the relic-neutrino mass, so a smaller $m_1$ leads to a smaller boosted C$\nu$B flux for fixed UHECR flux and fixed $\eta$.

For the PriNCe and H3a cases, the predicted spectra peak at energies of several hundred PeV range, overlapping the reconstructed energy range of KM3-230213A~\cite{KM3NeT:2025npi}. This spectral overlap motivates us to examine the parameter region compatible with the KM3NeT observation.
In contrast, the H4a spectra peak at higher neutrino energies than the PriNCe and H3a spectra. 
This shift is due to the proton-dominated high-energy component in H4a, which makes the RES and DIS channels more important at large momentum transfer. 

We then use current high-energy neutrino data to constrain $\eta$. 
For a given experiment, the expected number of boosted C$\nu$B events is computed as
\begin{equation}
N_{\rm ev}
=
4\pi T
\int_{E_{\min}}^{E_{\max}}
dE_\nu\,
\frac{d\phi_\nu}{dE_\nu}(\eta,m_1)\,
A_{\rm eff}(E_\nu)\,,
\label{eq:expected_events}
\end{equation}
where $T$ is the data-taking time, $A_{\rm eff}(E_\nu)$ is the all-flavor effective area, and $d\phi_\nu/dE_\nu$ is the boosted C$\nu$B flux calculated from Eq.~\eqref{eq:boosted_flux}.
For IceCube, we use the public all-flavor effective area from the 12.6yr EHE search~\cite{Meier:2024flg}. 
To avoid the lower-energy region containing PeV-scale neutrino candidates~\cite{IceCube:2025ezc,IceCube:2021rpz,2019GCN.24028....1I,IceCube:2016umi} and the associated background uncertainties, we follow the treatment of the IceCube EHE data adopted by the KM3NeT Collaboration in Ref.~\cite{KM3NeT:2025ccp}, and 
restrict the analysis to
$20~{\rm PeV}\leq E_\nu\leq50~{\rm EeV}$.
No EHE candidate is observed in this energy range, and the expected background is negligible. Taking both the observed event count and the expected background to be zero, the Feldman--Cousins 90\% C.L. upper limit is $N^{90}_{\rm IC}=2.44$~\cite{Feldman:1997qc}.
For the Pierre Auger Observatory, we use the 14.7yr data set, in which no neutrino candidate was identified~\cite{PierreAuger:2019ens}.
The energy-dependent exposure is taken from Fig.~5 of Ref.~\cite{PierreAuger:2019ens}. We adopt the corresponding Feldman--Cousins upper limit
$N_{\rm ev}^{\rm PAO}<2.39$, including the uncertainty in the exposure~\cite{PierreAuger:2015ihf}.
For KM3NeT, we use a data-taking time of
$T=335~{\rm days}$ and the corresponding sky-averaged effective area reported in Ref.~\cite{KM3NeT:2025npi}.
Taking the observed event count to be one and the expected background to be negligible, we use the Feldman--Cousins 90\% C.L. interval
$N_{\rm ev}^{\rm KM3NeT}\in[0.11,\,4.36]$~\cite{Feldman:1997qc}. 
The corresponding constraints on the overdensity are obtained from the predicted event numbers and the experimental limits described above.

\begin{figure*}[t]
    \centering

    % Row 1: SFR
    \begin{minipage}{0.48\textwidth}
        \centering
        \includegraphics[width=\textwidth]{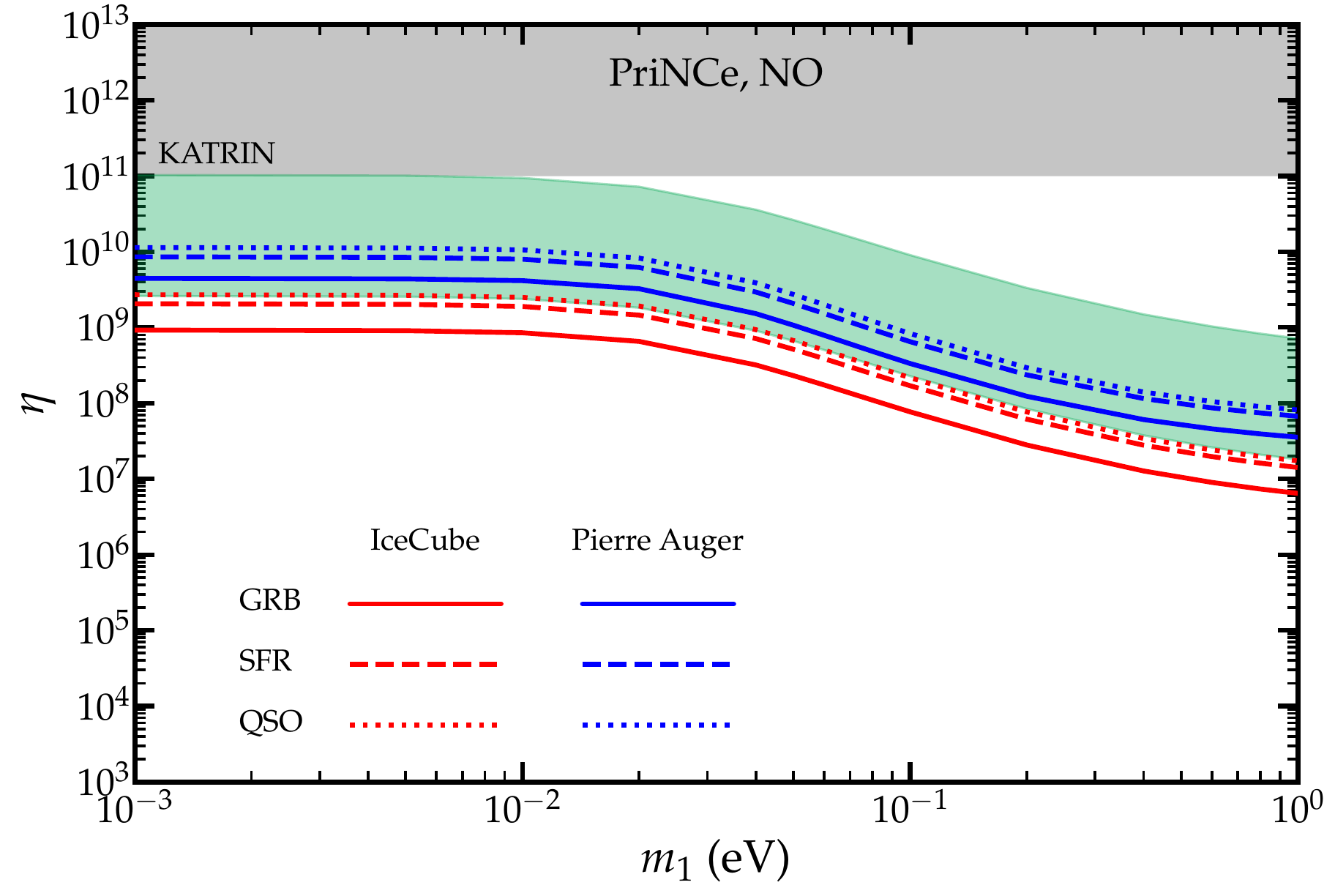}
    \end{minipage}
    \hfill
    \begin{minipage}{0.48\textwidth}
        \centering
        \includegraphics[width=\textwidth]{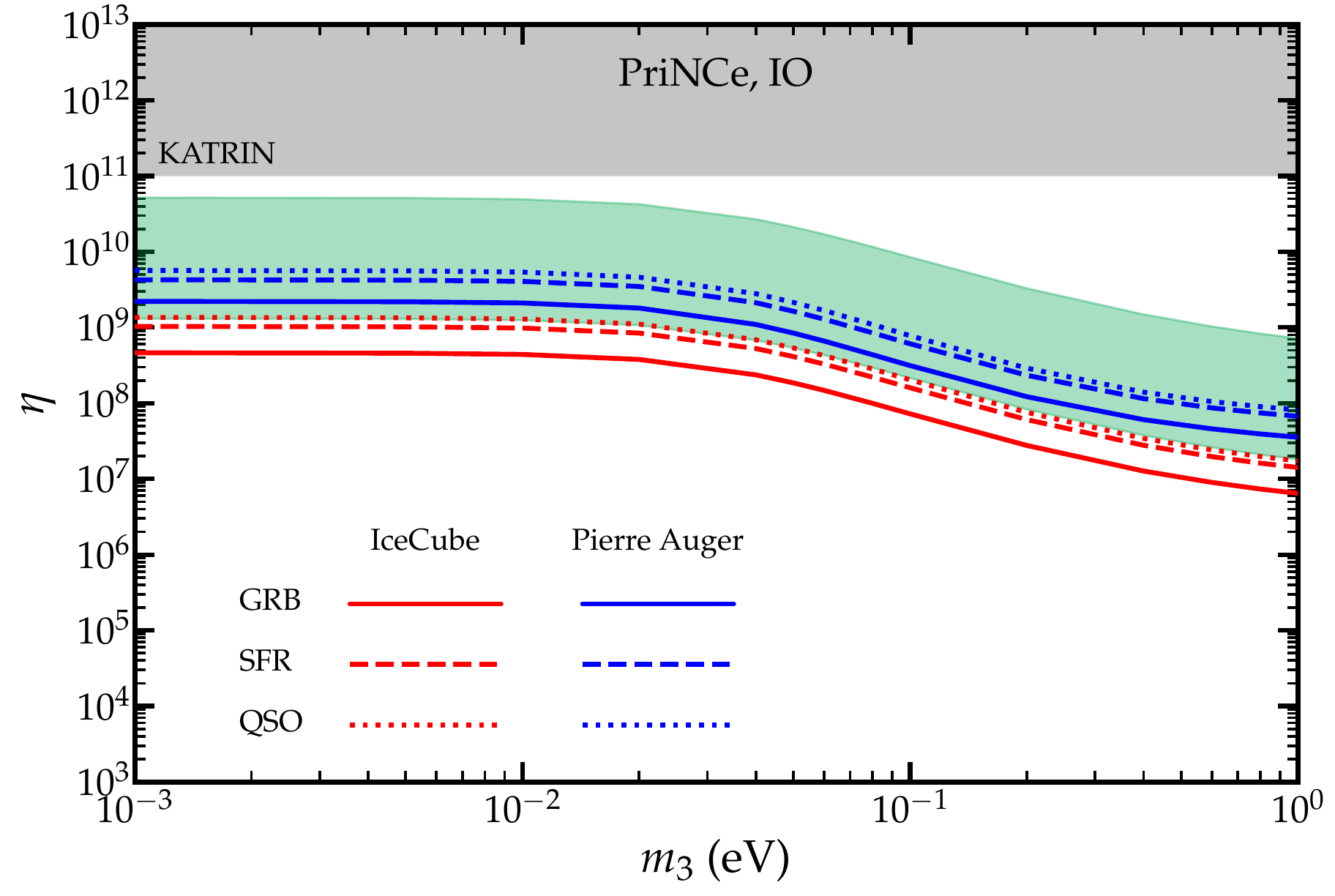}
    \end{minipage}

    \vspace{0.6em}

    % Row 2: QSO
    \begin{minipage}{0.48\textwidth}
        \centering
        \includegraphics[width=\textwidth]{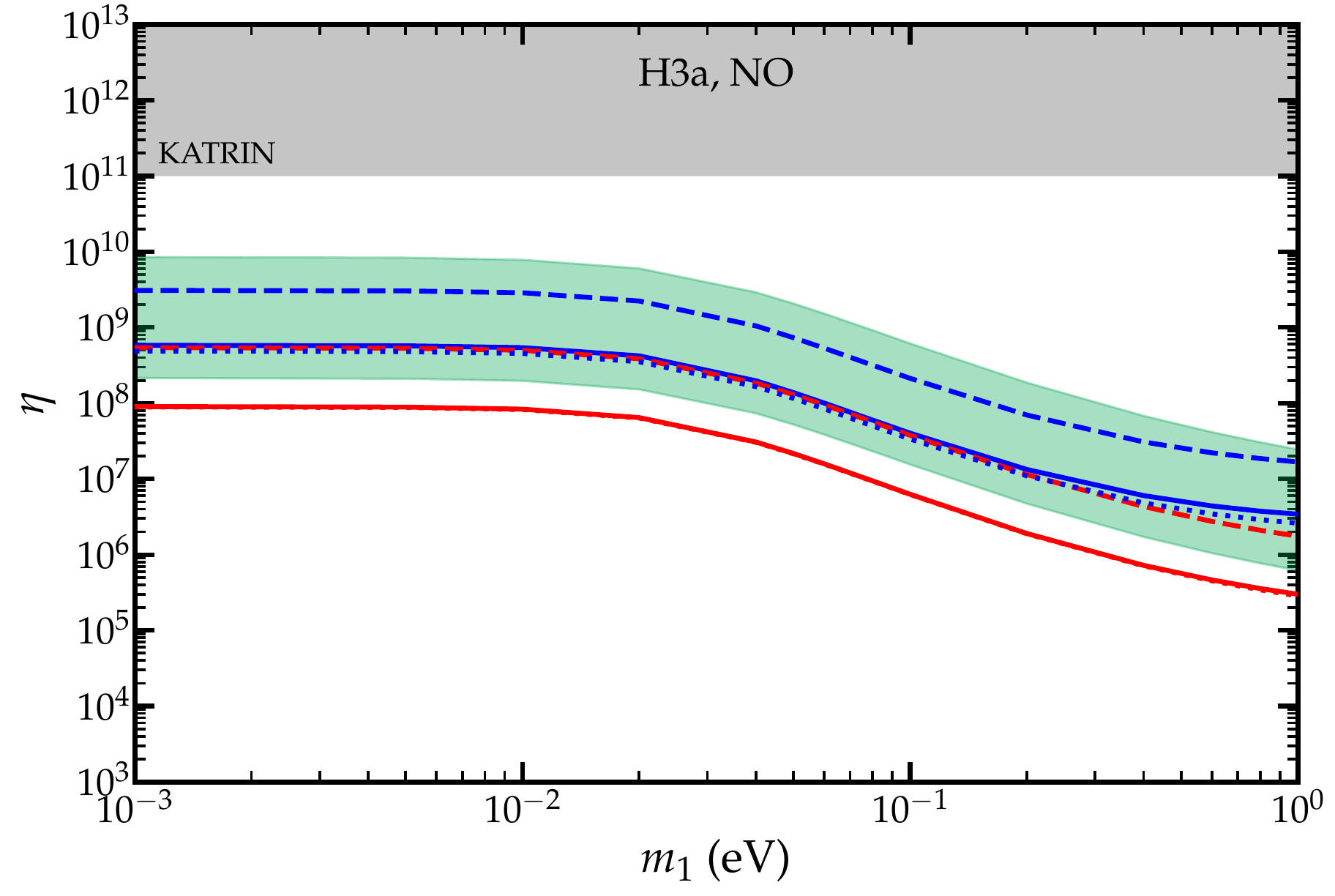}
    \end{minipage}
    \hfill
    \begin{minipage}{0.48\textwidth}
        \centering
        \includegraphics[width=\textwidth]{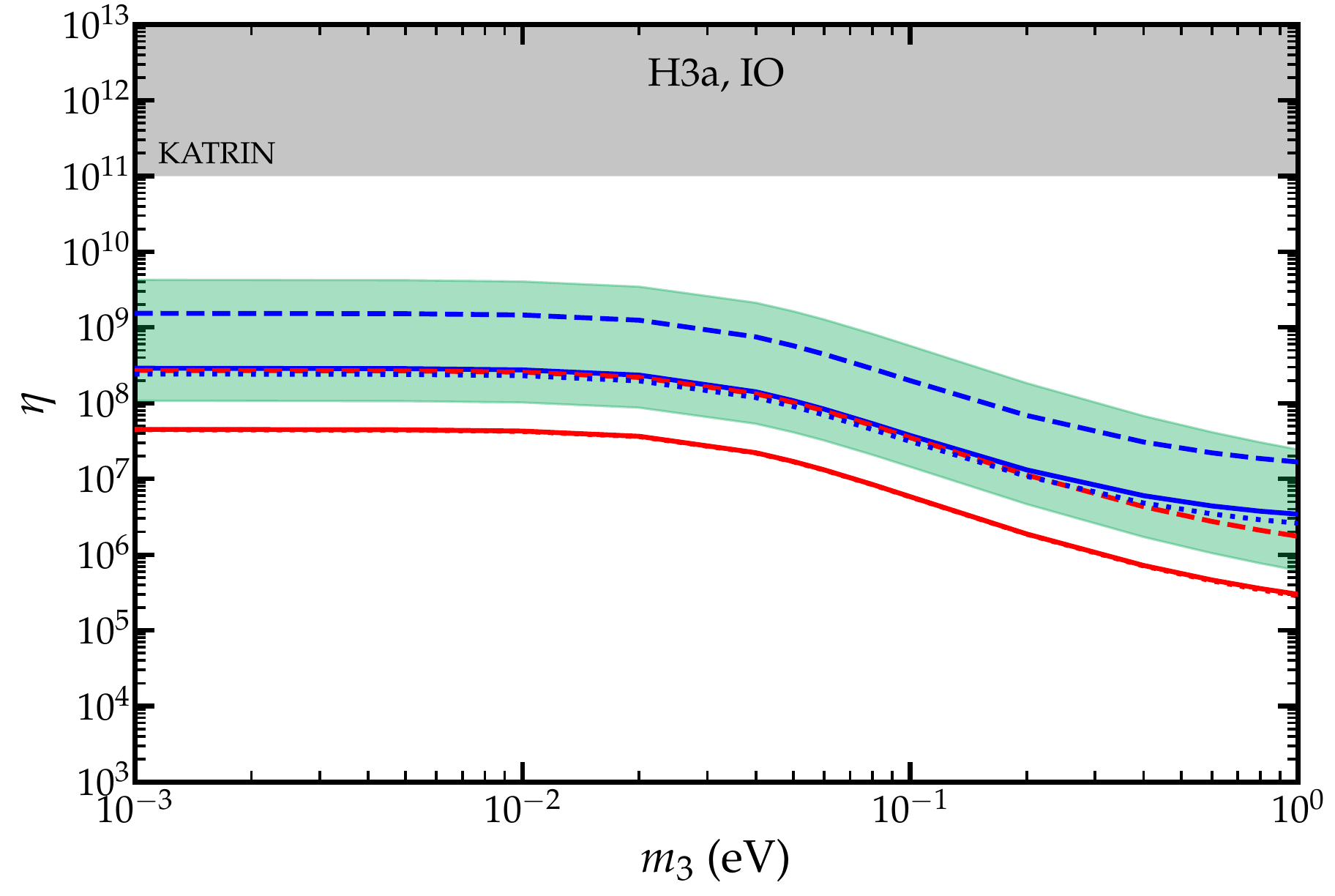}
    \end{minipage}

    \vspace{0.6em}

    % Row 3: GRB
    \begin{minipage}{0.48\textwidth}
        \centering
        \includegraphics[width=\textwidth]{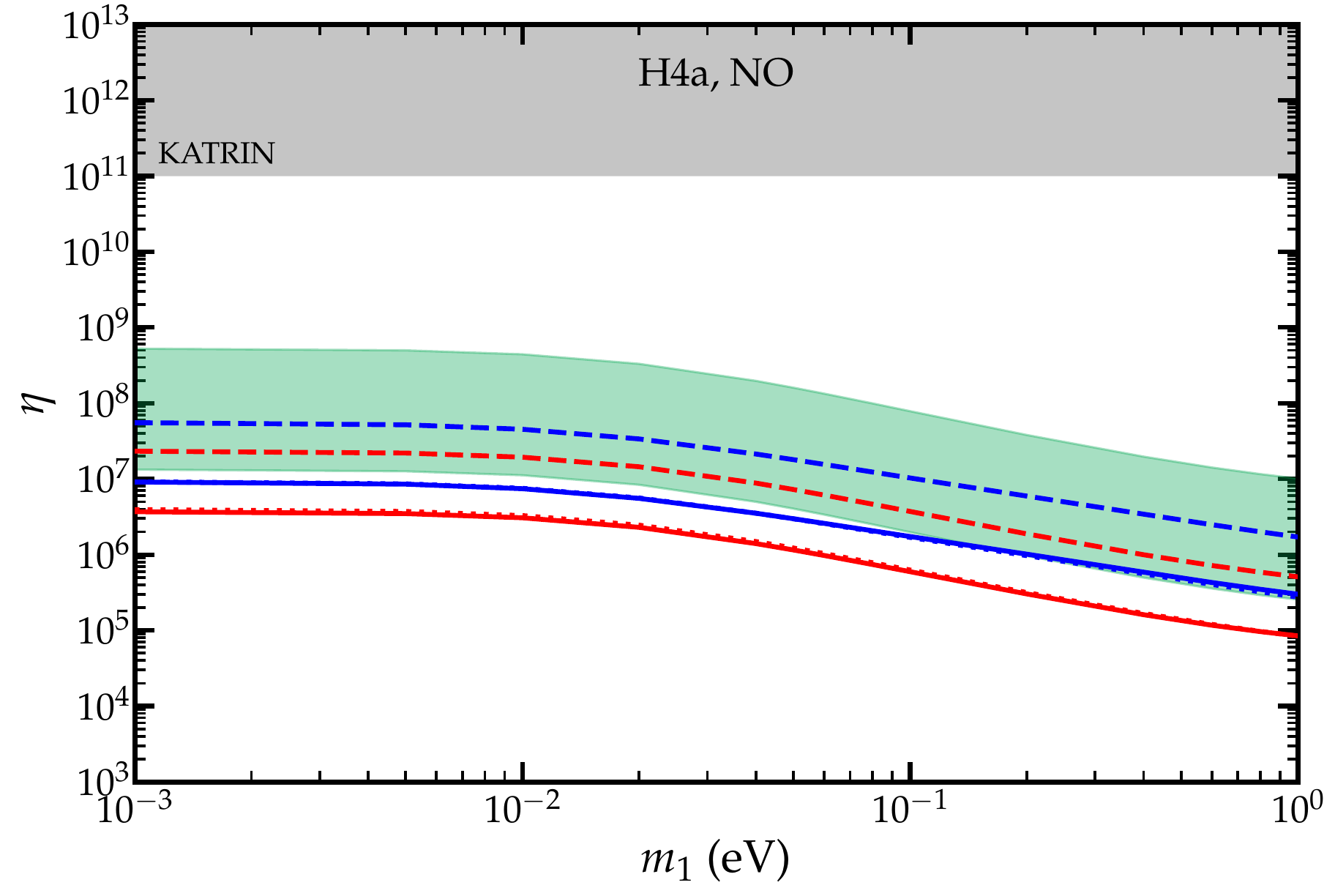}
    \end{minipage}
    \hfill
    \begin{minipage}{0.48\textwidth}
        \centering
        \includegraphics[width=\textwidth]{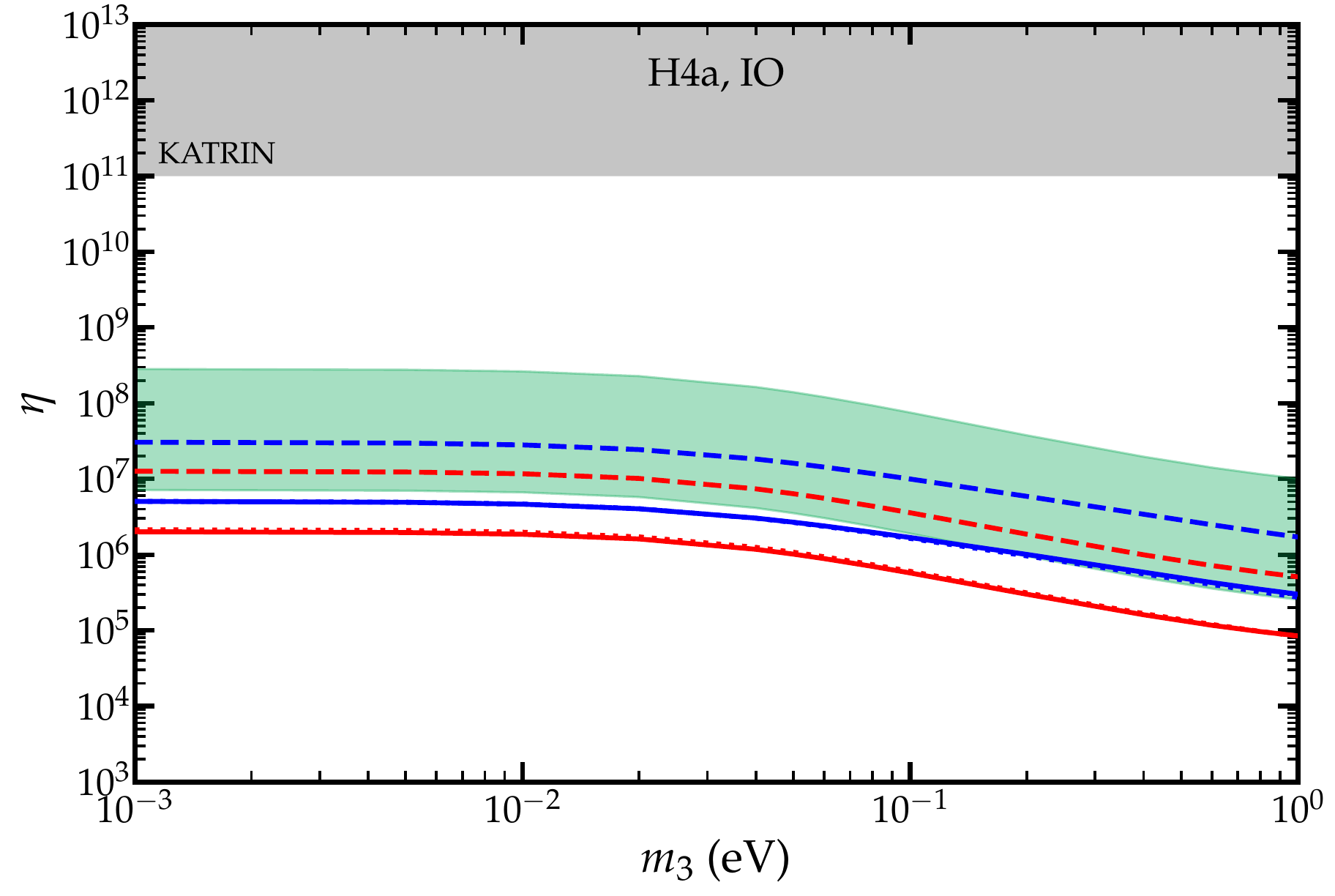}
    \end{minipage}

    \caption{
    Constraints on the C$\nu$B overdensity $\eta$ as a function of the lightest neutrino mass.
    From top to bottom, the rows correspond to the PriNCe, H3a, and H4a UHECR flux models, respectively.
    The left and right columns show the results for NO and IO, respectively.
    The red and blue curves show the 90\% C.L. constraints obtained from IceCube and the Pierre Auger Observatory.
    Different line styles denote the GRB, SFR, and QSO source evolution models.
    The green shaded regions show the KM3NeT allowed regions for the UHECR boosted relic neutrinos with the GRB source evolution.
    The gray shaded region shows the current direct KATRIN bound~\cite{KATRIN:2022kkv}.
    }
\label{fig:eta_limits_all_models}
\end{figure*}
The resulting limits are shown in Fig.~\ref{fig:eta_limits_all_models}. 
For each UHECR flux model, we show the constraints obtained with the SFR, GRB, and QSO source evolution models in the same panel. 
The left and right columns correspond to NO and IO, respectively. 
For sufficiently small lightest-neutrino mass $m_{\rm lightest}$, the constraint curves become nearly flat. 
This is because the lightest mass eigenstate gives only a small contribution to the total boosted C$\nu$B flux, while the signal is dominated by the heavier eigenstates. 
For larger $m_{\rm lightest}$, the boosted C$\nu$B flux increases because the relevant scattering cross sections scale approximately with $m_\nu$. 
Therefore, the same observational upper limit requires a smaller allowed overdensity $\eta$, leading to stronger constraints.

The remaining differences among the constraint curves are driven by the assumed UHECR flux model and source evolution.
Among the three UHECR flux models, H4a gives the strongest constraints, H3a gives intermediate constraints, and PriNCe gives the weakest constraints. 
For a given UHECR flux model and mass ordering, the spread among the SFR, GRB, and QSO curves quantifies this source evolution dependence.
For the NO, the strongest PriNCe constraint is obtained for the GRB source evolution. 
At $m_1=0.1~{\rm eV}$, the limits are $\eta<7.69\times10^7$($3.34\times10^8$) at IceCube (PAO), while at $m_1=0.01~{\rm eV}$ they become $\eta<8.51\times10^8$($4.15\times10^9$). 
The corresponding KM3NeT allowed regions are $\eta\in[2.28\times10^8,\,9.04\times10^9]$ for
$m_1=0.1~{\rm eV}$ and
$\eta\in[2.38\times10^9,\,9.44\times10^{10}]$ for $m_1=0.01~{\rm eV}$. As shown in Fig.~\ref{fig:eta_limits_all_models}, the KM3NeT allowed regions are in tension with the IceCube constraints.
\footnote{A quantitative assessment of this tension would require a joint statistical analysis of the KM3NeT, IceCube, and Pierre Auger data.
Since such an analysis is not performed here, we present the KM3NeT allowed regions only as a phenomenological comparison.}
For H3a, the GRB and QSO evolutions give nearly comparable constraints, and both are stronger than those obtained for the SFR source evolution. 
Using the GRB evolution, the limits at $m_1=0.1~{\rm eV}$ are $\eta<6.27\times10^6$ from IceCube and $\eta<4.02\times10^7$ from the Pierre Auger Observatory, while at $m_1=0.01~{\rm eV}$ the corresponding limits are $\eta<8.35\times10^7$ and $\eta<5.42\times10^8$.
For H4a, using the GRB source evolution, the limits at $m_1=0.1~{\rm eV}$ are $\eta<5.96\times10^5$ from IceCube and $\eta<1.72\times10^6$ from the Pierre Auger Observatory, while at $m_1=0.01~{\rm eV}$ they are $\eta<3.01\times10^6$ and $\eta<7.41\times10^6$, respectively.
In all cases, the high-energy neutrino constraints probe overdensities well below the current direct bound from KATRIN~\cite{KATRIN:2022kkv}.
Note that for the SM interaction channels and the UHECR models considered here, the constraints from current high-energy neutrino data are still weaker than the cosmological upper bound $\eta\lesssim2\times10^4$~\cite{Bondarenko:2023ukx}. Nevertheless, our results show that UHECR-boosted relic neutrinos can provide a complementary astrophysical probe of the C$\nu$B and broaden the range of observational strategies for relic neutrinos.
Future experiments, such as IceCube-Gen2~\cite{Meier:2024flg}, Trinity-18~\cite{Otte:2019knb}, RNO-G~\cite{RNO-G:2023jlv}, TAMBO~\cite{Romero-Wolf:2020pzh}, POEMMA~\cite{POEMMA:2020ykm}, and GRAND~\cite{GRAND:2017pux}, will substantially improve the sensitivity to UHECR-boosted relic neutrinos.
A detailed analysis of the projected sensitivity of these experiments is left for future work.

The difference between the two mass orderings is most visible when the lightest neutrino mass is small. 
For $m_{\rm lightest}\lesssim 10^{-2}~{\rm eV}$, the IO constraints are stronger by roughly a factor of two. 
In the NO, the dominant contribution in this limit mainly comes from the heaviest state $m_3$. 
For the IO, however, both $m_1$ and $m_2$ are relatively heavy and contribute efficiently to the boosted C$\nu$B flux. 
The IO therefore gives a larger boosted flux and hence stronger overdensity limits. 
For $m_{\rm lightest}\gtrsim 0.1~{\rm eV}$, the three mass eigenstates become nearly degenerate in both mass orderings, so the predicted boosted C$\nu$B flux and the resulting constraints are almost identical.

%%%%%%%%%%%%%%%%%%%%%%%%%%%%%%%% 

\section{Conclusions}
\label{sec:Conclusions}
In this work, we have performed a systematic study of the diffuse flux of the cosmic neutrino background boosted by UHECRs. 
To cover a broad range of momentum transfers, we constructed a common SM neutral-current framework that includes ES, COH, INCOH, RES, and DIS contributions to the boosted relic neutrino flux.
A key extension of this work is the inclusion of the RES contribution together with a consistent treatment of the RES and DIS regions.
The RES cross section was evaluated with the GENIE resonance configuration including the full set of 17 baryon resonances, while the DIS contribution was computed for $W>W_{\rm cut}$ with PDF inputs in the high $W$ region to avoid double counting with resonance production. 
This framework allows us to follow the boosted C$\nu$B production continuously from the coherent nuclear regime to the nucleon-resonance and partonic regimes.

We combined the cross sections with three descriptions of the UHECR flux: the redshift-dependent propagated spectra obtained with PriNCe, the H3a implementation of the Hillas model with three mixed-composition populations, and the H4a implementation in which the highest-energy Pop.~3 consists of only protons. 
For each UHECR description, we included the representative nuclear components $p$, He, N, Si, and Fe, and considered the SFR, QSO, and GRB source evolution models. 
The main results can be summarized as follows.

\begin{itemize}

\item 
For UHECR protons, the relevant SM channels are ES, RES, and DIS. 
For $m_\nu=0.1~{\rm eV}$, ES dominates at low UHECR energies, approximately for $E_N\lesssim10^{10}~{\rm GeV}$. 
The RES contribution becomes dominant in the intermediate region, roughly from $E_N\simeq10^{10}~{\rm GeV}$ to $E_N\simeq6\times10^{10}~{\rm GeV}$. 
At higher energies, the DIS cross section grows rapidly and eventually dominates. 
This demonstrates that RES provides an essential bridge between the ES- and DIS-dominated regimes and should be included in reliable boosted C$\nu$B flux predictions.

\item 
For nuclear UHECR components, the cross sections for scattering with relic neutrinos exhibit a characteristic hierarchy as the momentum transfer increases. 
At low energies, the cross section is dominated by COH, since the momentum transfer is small enough for the nucleus to be probed coherently. 
As the UHECR energy increases, the nuclear form factor suppresses the coherent contribution, and INCOH becomes important once individual nucleons are resolved. 
At still higher energies, RES becomes sizable because resonance production can occur in neutrino scattering with the nucleons inside the nucleus.
DIS appears only after the perturbative PDF threshold is reached and eventually dominates at the highest energies. 
This channel hierarchy in the cross sections is directly reflected in the boosted C$\nu$B flux: COH controls the low-energy flux for nuclear components, ES and INCOH become important after coherence is suppressed, RES contributes to the high-energy tail, and DIS appears only at the highest boosted-neutrino energies.

\item 
The boosted C$\nu$B flux is sensitive to the assumed UHECR flux model with different composition and maximum rigidity. 
For the PriNCe propagated spectra, the proton component dominates over most of the boosted-neutrino energy range. 
As a result, ES gives the leading contribution and the nuclear channels remain subdominant. 
For H3a, heavy nuclear components enhance the low-energy COH contribution, while lighter components become more important after coherence is suppressed. 
For H4a, the high-rigidity proton component extends the boosted spectrum to higher neutrino energies, enhancing the RES contribution and allowing DIS to appear at the highest energies.

\item 
Source evolution affects the predicted boosted C$\nu$B flux through the redshift weighting of the UHECR flux. 
The SFR, QSO, and GRB source evolution models lead to different boosted C$\nu$B fluxes and therefore different overdensity limits. 
For the benchmark models considered here, the strongest PriNCe constraint is obtained for the GRB source evolution. 
For H3a and H4a, the GRB and QSO source evolution models give nearly comparable constraints, and both are stronger than those obtained for the SFR source evolution.

\item 
%\sout{Using current high-energy neutrino data from IceCube and the Pierre Auger Observatory, we obtained constraints on the C$\nu$B overdensity factor $\eta$.}
Using current high-energy neutrino data, we obtained upper limits on the C$\nu$B overdensity from IceCube and the Pierre Auger Observatory. We also show the allowed regions corresponding to the observed KM3-230213A event for comparison.
The strength of the constraint is mainly determined by the predicted boosted C$\nu$B flux: a larger flux gives a stronger upper limit on $\eta$. 
Among the three UHECR descriptions, H4a gives the strongest limits, H3a gives intermediate limits, and PriNCe gives the weakest limits. 

\item 
The dependence on the lightest neutrino mass and the mass ordering is governed by which mass eigenstates dominate the boosted C$\nu$B flux. 
For $m_{\rm lightest}\lesssim10^{-2}~{\rm eV}$, the flux becomes nearly insensitive to $m_{\rm lightest}$, and the overdensity limits are nearly flat in the small mass region. 
%In this regime, IO gives stronger constraints than NO, typically by about a factor of two. 
As $m_{\rm lightest}$ increases, the boosted flux increases and the corresponding overdensity constraints become stronger. 
For $m_{\rm lightest}\gtrsim0.1~{\rm eV}$, the three mass eigenstates become nearly degenerate in both orderings, and the predicted boosted C$\nu$B flux and the resulting constraints become almost identical.

\end{itemize}

\acknowledgments
We thank G. Herrera, D. Marfatia, X. Qi, A. Sandrock, and B. Yue for helpful discussions. J. L. and J. Z. are supported by the National Natural Science Foundation of China under Grant No. 12275368.

\newpage

%\begin{appendix}
\appendix
	%%%%%%%%%%%%%%%%%%%%%%%%%%%%%%%%%%%%%%%%%%%%%%%%

\section{Elastic neutrino--nucleon cross section}
\label{app:ES_coefficients}
    
Here we summarize the coefficient and form factor inputs used in the elastic neutrino--nucleon cross section. 
Elastic neutrino--nucleon scattering in the standard high-energy neutrino-scattering kinematics has been discussed extensively in Refs.~\cite{Giunti:2007ry,Formaggio:2012cpf,DeMarchi:2024zer}. 
For the UHECR--C$\nu$B process considered here, where the incoming relic neutrino is approximately at rest in the laboratory frame, we follow the corresponding result derived in Ref.~\cite{Zhang:2025rqh}. 
Since all cross sections in this work are defined as the sum of neutrino and antineutrino contributions, the elastic differential cross section is written as
\begin{equation}
\frac{d\sigma_{\nu N}^{\rm ES}}{dE_\nu}
=
\frac{G_F^2 m_\nu m_N^4}{\pi (s-m_N^2)^2}
\left[
A_N(Q^2)
+
C_N(Q^2)\frac{(s-u)^2}{m_N^4}
\right] \,,
\label{eq:app_ES_cross_section}
\end{equation}
The Mandelstam variables are approximated as
$s\simeq m_N^2+2m_\nu E_N$,
$t\equiv -Q^2$,
and
$u\simeq 2m_\nu^2+2m_N^2-s+Q^2$.
For the boosted-neutrino energies considered here, the momentum transfer is $Q^2\simeq 2m_\nu(E_\nu-m_\nu)\simeq 2m_\nu E_\nu$.
The maximum energy of the outgoing boosted C$\nu$B neutrino is
$E_\nu^{\max}(E_N)
=
{E_N^2}/[{E_N+m_N^2/(2m_\nu)}]$.

The coefficient functions are
\begin{align}
A_N(Q^2)
=
\frac{Q^2}{m_N^2}
\Bigg[
&
\left(1+\frac{Q^2}{4m_N^2}\right)
\left(G_A^{ZN}\right)^2
-
\left(1-\frac{Q^2}{4m_N^2}\right)
\left(
\left(F_1^{ZN}\right)^2
-
\frac{Q^2}{4m_N^2}
\left(F_2^{ZN}\right)^2
\right)
\nonumber\\
&
+
\frac{Q^2}{m_N^2}
F_1^{ZN}F_2^{ZN}
\Bigg],
\label{eq:A_ES}
\end{align}
and
\begin{equation}
C_N(Q^2)
=
\frac{1}{4}
\left[
\left(G_A^{ZN}\right)^2
+
\left(F_1^{ZN}\right)^2
+
\frac{Q^2}{4m_N^2}
\left(F_2^{ZN}\right)^2
\right]\,,
\label{eq:C_ES}
\end{equation}

The weak neutral-current vector form factors are obtained from the electromagnetic form factors as
$F_i^{Zp}=\frac{1}{2}(F_i^p-F_i^n)-2\sin^2\theta_W F_i^p$ and
$F_i^{Zn}=-\frac{1}{2}(F_i^p-F_i^n)-2\sin^2\theta_W F_i^n$, with $i=1,2$. 
The corresponding axial neutral-current form factors are
$G_A^{Zp}(Q^2)=\frac{1}{2}G_A(Q^2)$ and
$G_A^{Zn}(Q^2)=-\frac{1}{2}G_A(Q^2)$.
The electromagnetic Dirac and Pauli form factors are related to the Sachs electric and magnetic form factors by~\cite{Giunti:2007ry,Formaggio:2012cpf}
\begin{equation}
F_1^N(Q^2)
=
\frac{G_E^N(Q^2)+\tau G_M^N(Q^2)}{1+\tau},
\qquad
F_2^N(Q^2)
=
\frac{G_M^N(Q^2)-G_E^N(Q^2)}{1+\tau},
\qquad
\tau=\frac{Q^2}{4m_N^2}\,.
\label{eq:F1F2_from_Sachs}
\end{equation}
For the numerical evaluation, we use the standard dipole parametrization for the electromagnetic and axial form factors~\cite{Giunti:2007ry,Formaggio:2012cpf,Zhang:2025rqh}. 
At zero momentum transfer, we take
$G_E^p(0)=1$, $G_E^n(0)=0$, $G_M^p(0)=\mu_p/\mu_N=2.79$, and $G_M^n(0)=\mu_n/\mu_N=-1.91$, where the magnetic moments are taken from Ref.~\cite{ParticleDataGroup:2024cfk}. 
The axial form factor is normalized as $G_A(0)\simeq 1.245$ and parametrized with an axial mass $m_A\simeq 1.17~{\rm GeV}$~\cite{Gao:2021sml,Alexandrou:2023qbg}. 
These inputs determine the functions $A_N(Q^2)$ and $C_N(Q^2)$ used in Eq.~\eqref{eq:ES_general}.

\section{Neutral-current \texorpdfstring{$P_{33}(1232)$}{P33(1232)} resonance cross section}
\label{app:RES_delta}

This appendix gives the neutral-current resonance-production formula used in the calculation, with the $P_{33}(1232)$ channel written explicitly as a representative example. 
The calculation adopts the GENIE RES implementation of the Rein--Sehgal resonance framework for free-nucleon neutral-current scattering~\cite{Berger:2007rq,Andreopoulos:2009rq,GENIE:2021npt}. The relevant free-nucleon neutral-current processes are
\begin{equation}
\nu+N\to \nu+\mathcal{R}_N,
\qquad
\bar{\nu}+N\to \bar{\nu}+\mathcal{R}_N,
\qquad
N=p,n\,,
\end{equation} 
For the $P_{33}(1232)$ state, the double-differential cross sections used in the calculation are~\cite{Kuzmin:2003ji,Berger:2007rq,Rein:1980wg,Andreopoulos:2009rq}
\begin{align}
\frac{d^2\sigma_{\nu N\to\nu P_{33}}}
{dW\,dQ^2}
&=
\frac{1}{2}\,
\sigma_0
\left[
V^2\sigma_R^N
+U^2\sigma_L^N
+2UV\sigma_S^N
\right]
\mathcal{B}_{P_{33}}(W)\,
\Theta(W_{\rm cut}-W),
\label{eq:app_p33_nu}
\\
\frac{d^2\sigma_{\bar{\nu}N\to\bar{\nu}P_{33}}}
{dW\,dQ^2}
&=
\frac{1}{2}\,
\sigma_0
\left[
U^2\sigma_R^N
+V^2\sigma_L^N
+2UV\sigma_S^N
\right]
\mathcal{B}_{P_{33}}(W)\,
\Theta(W_{\rm cut}-W)\,,
\label{eq:app_p33_nubar}
\end{align}
Here $W$ is the invariant mass of the final-state hadronic system.
The factors $\sigma_0$, $U$ and $V$ are the leptonic kinematic factors, and $\sigma_L^N$, $\sigma_R^N$, and $\sigma_S^N$ denote the left-transverse, right-transverse, and scalar helicity cross sections for a target nucleon $N$. 
The function $\mathcal{B}_{P_{33}}(W)$ describes the resonance line shape, while $\Theta(W_{\rm cut}-W)$ restricts the RES contribution to the low-$W$ region in order to avoid double counting with DIS. The neutrino and antineutrino expressions differ by the interchange of the two transverse contributions, while the scalar term enters in the same way. The factor $1/2$ accounts for the fact that present-day C$\nu$B neutrinos are nonrelativistic helicity eigenstates, containing approximately equal left- and right-chiral components~\cite{Long:2014zva}, while only the active chiral component participates in the weak neutral-current interaction~\cite{Zhang:2025rqh}.

The kinematic quantities are evaluated in the nucleon rest frame. 
The incoming relic-neutrino energy in this frame is $E_\nu^\ast$. 
For a free nucleon,
$s=M_N^2+2M_NE_\nu^\ast$.
In the UHECR--C$\nu$B kinematics, the relic neutrino is approximately at rest in the laboratory frame, so $E_\nu^\ast\simeq m_\nu E_N/M_N$, where $E_N$ is the energy of the incident CR nucleon.
For a given hadronic invariant mass $W$, the energy transfer is
$w=(W^2-M_N^2+Q^2)/(2M_N)$,
the outgoing-neutrino energy is
$E_\nu^{\prime\ast}=E_\nu^\ast-w$,
and
$|\mathbf q|=\sqrt{w^2+Q^2}$.
The leptonic kinematic factors appearing in Eqs.~\eqref{eq:app_p33_nu} and \eqref{eq:app_p33_nubar} are
\begin{equation}
U=
\frac{E_\nu^\ast+E_\nu^{\prime\ast}+|\mathbf q|}
{2E_\nu^\ast},
\qquad
V=
\frac{E_\nu^\ast+E_\nu^{\prime\ast}-|\mathbf q|}
{2E_\nu^\ast}\,,
\end{equation}

The helicity cross sections are~\cite{Andreopoulos:2009rq}
\begin{align}
\sigma_0
&=
\frac{G_F^2}{8\pi}
\frac{Q^2}{|\mathbf q|^2}
\frac{W}{M_N},
\\
\sigma_L^N
&=
\frac{W}{M_N}
\left(
|f_{+3}^{N}|^2+|f_{+1}^{N}|^2
\right),
\\
\sigma_R^N
&=
\frac{W}{M_N}
\left(
|f_{-3}^{N}|^2+|f_{-1}^{N}|^2
\right),
\\
\sigma_S^N
&=
\frac{M_N}{W}
\frac{|\mathbf q|^2}{Q^2}
\left(
|f_{0+}^{N}|^2+|f_{0-}^{N}|^2
\right)\,,
\end{align}
where $f_{\lambda}^{N}$ are the neutral-current helicity amplitudes for the transition
$N\to P_{33}(1232)$, with $\lambda=\pm1,\pm3,0\pm$ labeling the transverse and scalar helicity components in the Rein--Sehgal convention. 
For the $P_{33}(1232)$ channel, the explicit GENIE amplitudes are the same for proton and neutron targets. 
With $\xi=\sin^2\theta_W$, they are~\cite{Rein:1980wg}
\begin{align}
f_{-1}^{N}
&=
-\sqrt{2}\left(R_-+2\xi R_V\right),
\\
f_{+1}^{N}
&=
\sqrt{2}\left(R_++2\xi R_V\right),
\\
f_{-3}^{N}
&=
-\sqrt{6}\left(R_-+2\xi R_V\right),
\\
f_{+3}^{N}
&=
\sqrt{6}\left(R_++2\xi R_V\right),
\\
f_{0-}^{N}
&=
f_{0+}^{N}
=
2\sqrt{2}\,C \,,
\end{align}
The transverse combinations are defined as
$R_+=-(R_V+R_A)\,,
R_-=-(R_V-R_A).$
The quantities $R_V$, $R_A$, and $C$ are the FKR resonance-model amplitudes used in the GENIE Berger--Sehgal implementation~\cite{Berger:2007rq}. 
Defining $D=(W+M_N)^2-q^2$, they are
\begin{align}
R_V
&=
\sqrt{2}\,
\frac{M_N|\mathbf q|}{W}
\frac{W+M_N}{D}
G_V,
\\
R_A
&=
\frac{\sqrt{2}}{6}
\frac{\zeta}{W}
\left(W+M_N\right)G_A,
\\
C
&=
\frac{\zeta}{6|\mathbf q|}
\left(W^2-M_N^2\right)
\frac{G_A}{M_N}\,,
\end{align}
The parameter $\zeta=0.76338$ is the GENIE default value used in the 
Berger--Sehgal RES implementation~\cite{Andreopoulos:2009rq,Berger:2007rq}.
The vector and axial transition factors entering the amplitudes are
\begin{align}
G_V(q^2,W)
&=
\frac{1}{2}
\left(1-\frac{q^2}{(M_N+W)^2}\right)^{1/2}
\sqrt{3(g_3^V)^2+(g_1^V)^2},
\\
G_A(q^2,W)
&=
\frac{\sqrt{3}}{2}
\left(1-\frac{q^2}{(M_N+W)^2}\right)^{1/2}
\left[
1-\frac{W^2+q^2-M_N^2}{8M_N^2}
\right]
C_5^A(q^2)\,,
\end{align}
The vector combinations appearing in $G_V$ are
\begin{align}
g_3^V
&=
\frac{1}{2\sqrt{3}}
\left[
C_3^V\frac{W+M_N}{M_N}
+C_4^V\frac{W^2+q^2-M_N^2}{2M_N^2}
+C_5^V\frac{W^2-q^2-M_N^2}{2M_N^2}
\right],
\\
g_1^V
&=
-\frac{1}{2\sqrt{3}}
\left[
C_3^V\frac{M_N^2-q^2+M_NW}{WM_N}
+C_4^V\frac{W^2+q^2-M_N^2}{2M_N^2}
+C_5^V\frac{W^2-q^2-M_N^2}{2M_N^2}
\right]\,,
\end{align}
The axial form factor is parametrized as
\begin{equation}
C_5^A(q^2)
=
1.2
\left(1-\frac{q^2}{M_A^2}\right)^{-2}\,,
\end{equation}
with $M_A=1.120~{\rm GeV}$~\cite{Andreopoulos:2009rq}. 
For the vector form factors, we define
$
C_0^V(q^2)=(1-q^2/4M_V^2)^{-1}
$,
with $M_V=0.840~{\rm GeV}$, and use
\begin{align}
C_3^V(q^2)
&=
2.13\,C_0^V(q^2)
\left(1-\frac{q^2}{M_V^2}\right)^{-2},
\\
C_4^V(q^2)
&=
-1.51\,C_0^V(q^2)
\left(1-\frac{q^2}{M_V^2}\right)^{-2},
\\
C_5^V(q^2)
&=
0.48\,C_0^V(q^2)
\left(1-\frac{q^2}{0.766\,M_V^2}\right)^{-2}\,,
\end{align}

The function $\mathcal{B}_{P_{33}}(W)$ in Eqs.~\eqref{eq:app_p33_nu} and \eqref{eq:app_p33_nubar} is the normalized Breit--Wigner line shape used in GENIE. 
For $P_{33}(1232)$, it is written as~\cite{Andreopoulos:2009rq}
\begin{equation}
\mathcal{B}_{P_{33}}(W)
=
\frac{1}{2\pi}
\frac{\Gamma(W)/\mathcal{N}_{P_{33}}}
{(W-M_\Delta)^2+\Gamma^2(W)/4}\,,
\end{equation}
Here $M_\Delta$ is the mass of the $\Delta(1232)$ resonance, and $\Gamma(W)$ is the resonance width. 
For the $P_{33}(1232)$ state, GENIE uses
\begin{equation}
\Gamma(W)
=
\Gamma_\Delta^0
\left[
\frac{p_\pi(W)}{p_\pi(M_\Delta)}
\right]^3,
\qquad
p_\pi(W)
=
\frac{
\sqrt{\left[W^2-(M_N+m_{\pi^0})^2\right]
\left[W^2-(M_N-m_{\pi^0})^2\right]}
}{2W}\,,
\end{equation}
where $\Gamma_\Delta^0=0.120~{\rm GeV}$~\cite{Andreopoulos:2009rq} is the on-shell width of the $\Delta(1232)$ resonance, $p_\pi(W)$ is the pion momentum in the $N\pi$ center-of-mass frame, and $\mathcal{N}_{P_{33}}$ is a normalization constant.

To obtain the RES differential cross section, we integrate over the allowed hadronic invariant mass. The physical lower limit is the one-pion threshold, $W_{\rm min}=M_N+m_{\pi^0}$.
The phase-space upper limit is $W_{\rm max}^{\rm phys}=\sqrt{M_N^2+2M_NE_\nu^\ast}$.
For the $P_{33}(1232)$ state, GENIE also restricts the resonance window to $M_\Delta+6\Gamma_\Delta^0$. 
In addition, the RES contribution is restricted to $W<W_{\rm cut}$ with $W_{\rm cut}=1.7~{\rm GeV}$ to avoid overlap with the DIS contribution.
Therefore, the effective upper limit is
\begin{equation}
W_{\rm max}^{P_{33}}
=
\min\left(
W_{\rm max}^{\rm phys},
M_\Delta+6\Gamma_\Delta^0,
W_{\rm cut}
\right)\,,
\end{equation}
The single-differential cross section is then
\begin{equation}
\frac{d\sigma_{\nu N\to\nu P_{33}}}{dQ^2}
=
\int_{W_{\rm min}}^{W_{\rm max}^{P_{33}}}
dW\,
\frac{d^2\sigma_{\nu N\to\nu P_{33}}}
{dW\,dQ^2}\,.
\end{equation}
and analogously for antineutrinos.
The formulae above display only the $P_{33}(1232)$ contribution. 
For the results presented in the main text, we evaluate the cross sections for all 17 default resonance states following the GENIE implementation and then sum the individual resonance contributions.

\section{Distribution of cosmic ray sources}
\label{app:source_distribution}
For the redshift distribution of CR sources, we consider three benchmark cases: star formation rate (SFR), quasi-stellar objects (QSO), and gamma-ray bursts (GRB). 
We describe the redshift dependence by a normalized source evolution function,
\begin{equation}
f(z)
\equiv
\frac{\rho(z)}{\rho(z_{\rm min})}\,,
\end{equation}
where $\rho(z)$ denotes the comoving source density and $z_{\rm min}=0$ is used as the reference redshift. 
This normalization fixes $f(0)=1$ for all three source evolution models.

For the SFR case, we use~\cite{hopkins2006normalization}
\begin{equation}
\rho_{\rm SFR}(z)
=
\frac{(a+bz)h}{1+(z/c)^d}
\;M_\odot\,{\rm yr}^{-1}\,{\rm Mpc}^{-3}\,,
\label{eq:sfr_density}
\end{equation}
with $a=0.0170$, $b=0.13$, $c=3.3$, $d=5.3$, and $h\equiv H_0/(100~{\rm km\,s^{-1}\,Mpc^{-1}})$.

For the QSO case, we adopt the logarithmic density parametrization~\cite{Wall:2004tg}
\begin{equation}
\log\left[\frac{\rho_{\rm QSO}(z)}{{\rm Mpc}^{-3}}\right]
=
-a_0+a_1z-a_2z^2+a_3z^3-a_4z^4\,,
\label{eq:qso_density}
\end{equation}
with $a_0=12.49$, $a_1=2.704$, $a_2=1.145$, $a_3=0.1796$, and $a_4=0.01019$.

For the GRB case, we assume that the GRB rate follows the SFR density with an additional redshift-dependent factor~\cite{Lan:2021uuf},
\begin{equation}
\rho_{\rm GRB}(z)
=
\kappa\,\rho_{\rm SFR}(z)(1+z)^\delta\,,
\label{eq:grb_density}
\end{equation}
where $\kappa=8.5$ and $\delta=1.26$. 
For each benchmark case, the corresponding source distribution function $f(z)$ is obtained by normalizing $\rho(z)$ to its value at $z=0$.

% \begin{figure}[t]
%     \centering
%     \includegraphics[width=0.78\textwidth]{{CR source}.pdf}
%     \caption{
%     Redshift evolution functions of the cosmic-ray source distributions considered in this work. 
%     The three curves correspond to the SFR, QSO, and GRB source evolution models, normalized at $z_{\rm min}=0$.
%     }
%     \label{fig:cr_source_evolution}
% \end{figure}

% Figure~\ref{fig:cr_source_evolution} shows the three source evolution models used to describe the redshift dependence of the cosmic-ray source density. 
% The SFR evolution increases moderately with redshift and peaks around $z\simeq 2$--$3$, while the QSO evolution rises more rapidly and gives a larger source density at intermediate redshifts. 
% The GRB evolution remains sizable over a broader redshift range and has a relatively stronger high-redshift tail. 
% These different redshift dependences lead to different weights for UHECR production at earlier cosmic times and therefore affect the predicted boosted C$\nu$B flux.

%%%%%%%%%%%%%%%%%%%%%%%%%%%%%%%%%%%%%%%%%%%%%%%%

\section{Estimate of UHECR energy loss induced by C$\nu$B scattering}
\label{app:cr_energy_loss}

To estimate the possible impact of C$\nu$B scattering on the parent UHECR spectrum, we calculate the corresponding UHECR energy loss.
In each UHECR--C$\nu$B scattering, an energy $E_\nu$ is transferred from the UHECR to the relic neutrino.
In the continuous limit, the corresponding energy-loss rate can be written as~\cite{Zhang:2025rqh, Cappiello:2018hsu}
\begin{equation}
\frac{d\mathcal{Q}}{dt}
=
\int_0^{E_\nu^{\rm max}}
c\,\eta n_\nu\,
E_\nu
\frac{d\sigma^{\nu\mathcal{N}_i}}{dE_\nu}
\,dE_\nu ,
\label{eq:cr_energy_loss_rate}
\end{equation}
where $\mathcal{Q}\equiv\Delta E_{\mathcal{N}_i}$ denotes the energy loss of a UHECR nucleus $\mathcal{N}_i$, and $E_\nu^{\rm max}$ is the maximum kinematically allowed boosted-neutrino energy.
The relation between cosmic time and redshift is given by
$dt=dz[(1+z)H_0\sqrt{\Omega_m(1+z)^3+\Omega_\Lambda}]^{-1}$.
In evaluating the accumulated loss, we treat the energy degradation continuously, so that the UHECR energy entering Eq.~\eqref{eq:cr_energy_loss_rate} is updated during propagation.

\begin{figure*}[t]
    \centering

    \begin{minipage}{0.48\textwidth}
        \centering
        \includegraphics[width=\textwidth]{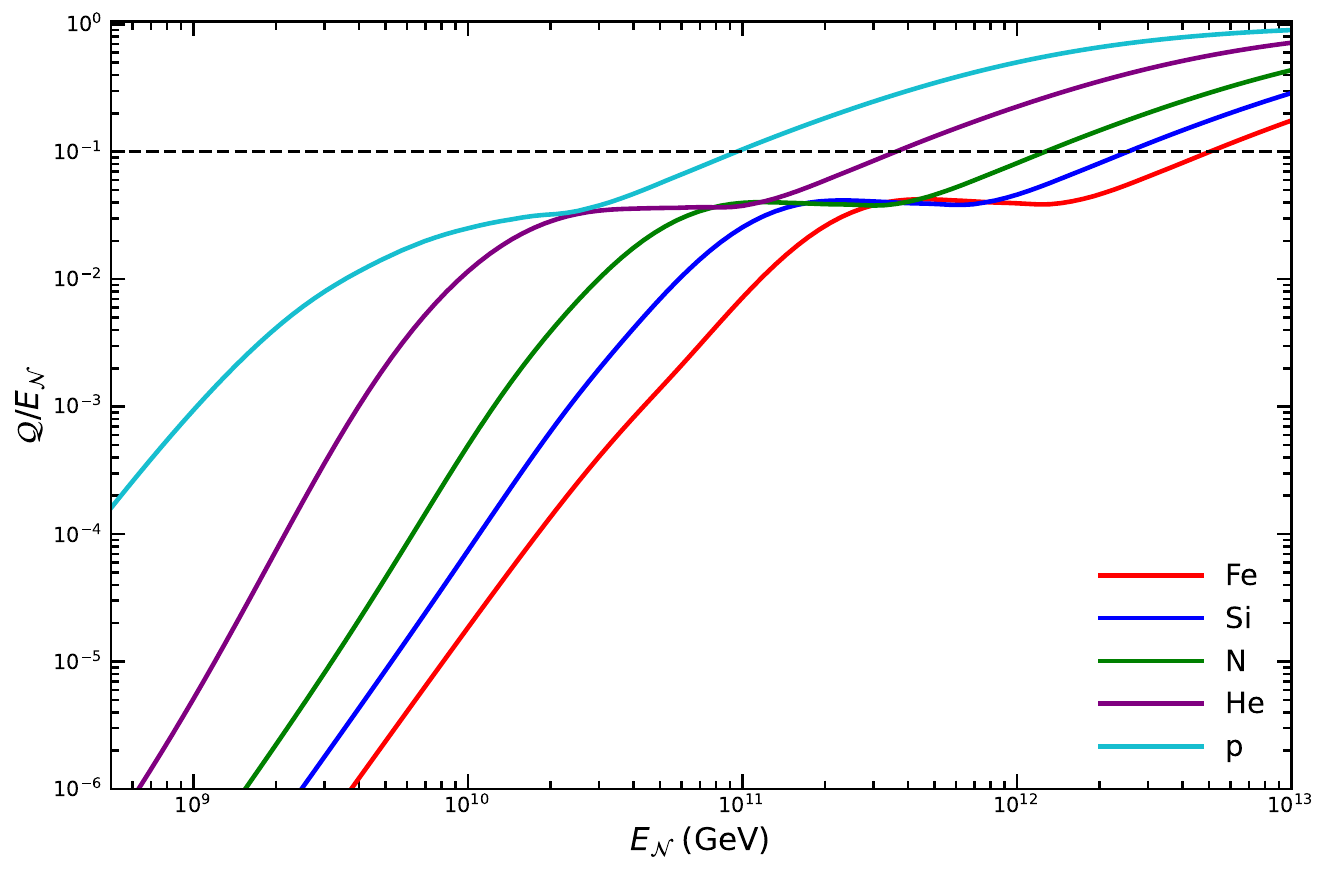}
    \end{minipage}
    \hfill
    \begin{minipage}{0.48\textwidth}
        \centering
        \includegraphics[width=\textwidth]{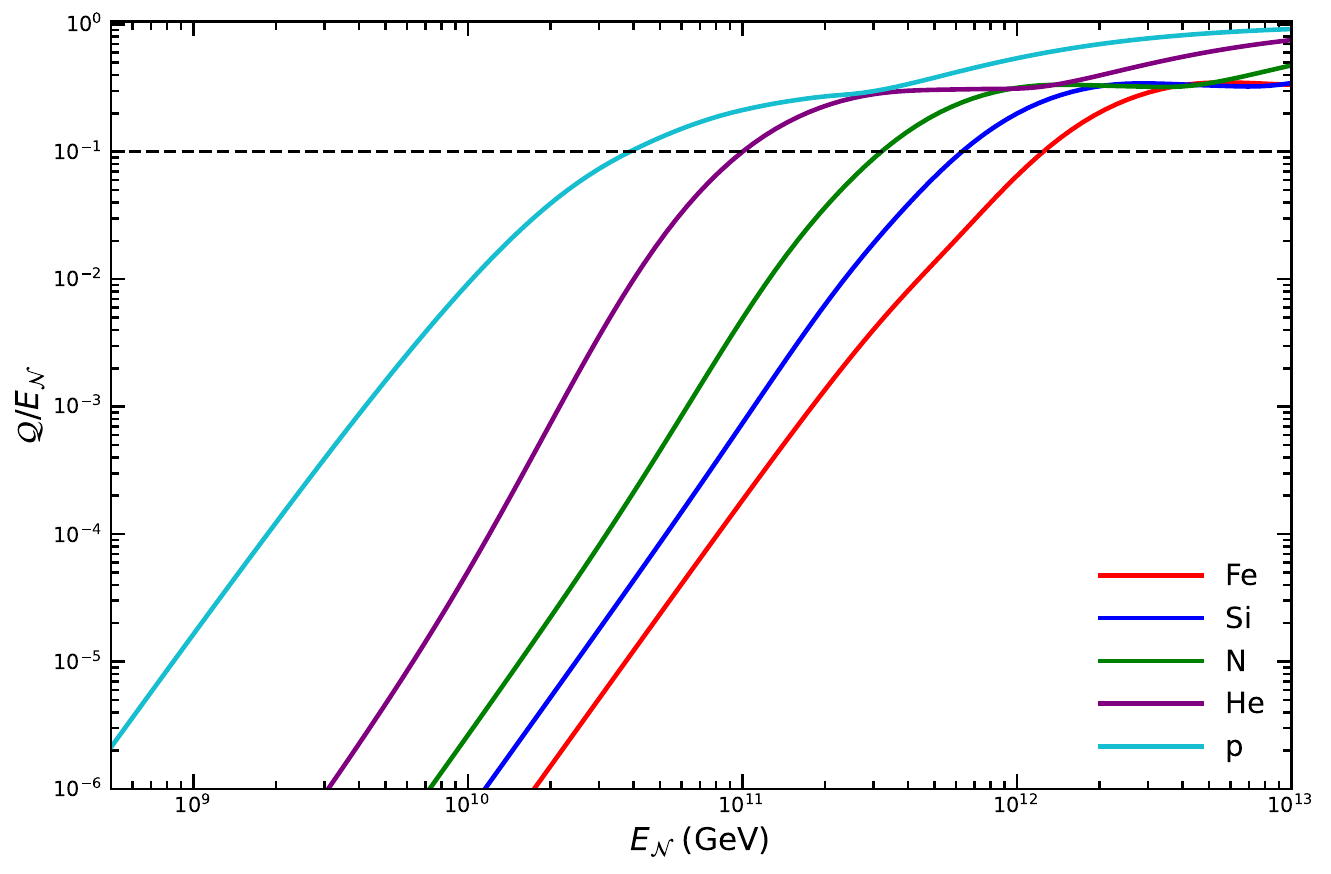}
    \end{minipage}

    \caption{
    Fractional UHECR energy loss induced by C$\nu$B scattering during propagation from $z=6$ to $z=0$ in NO.
    The left panel corresponds to $m_1=0.1~{\rm eV}$ and $\eta=10^8$, while the right panel corresponds to $m_1=0.01~{\rm eV}$ and $\eta=10^9$.
    The different curves denote the representative UHECR components $p$, He, N, Si, and Fe. The horizontal dashed line indicates a fractional energy loss of $10\%$.
    }
    \label{fig:cr_energy_loss}
\end{figure*}

Fig.~\ref{fig:cr_energy_loss} shows the fractional UHECR energy loss induced by C$\nu$B scattering for two representative benchmark cases,
$m_1=0.1~{\rm eV}$ with $\eta=10^8$ and
$m_1=0.01~{\rm eV}$ with $\eta=10^9$.
These overdensities are chosen close to the IceCube exclusion region obtained with the PriNCe UHECR model, which gives the weakest constraints among the UHECR descriptions considered in this work.
In this estimate, all UHECRs are assumed to propagate from $z=6$ to $z=0$, whereas the source distributions considered in this work are concentrated mainly at $z\sim1$--$3$.
We also neglect the additional UHECR energy losses caused by interactions with the CMB and EBL, which would further reduce the UHECR energy and suppress subsequent UHECR--C$\nu$B scattering.
Therefore, the accumulated C$\nu$B-induced energy loss estimated here should be regarded as a conservative upper estimate.
As seen by comparing Fig.~\ref{fig:cr_energy_loss} with Fig.~\ref{fig:cr_flux_sfr_redshift}, fractional energy losses above $10\%$ occur mainly at energies where the corresponding UHECR fluxes are already rapidly decreasing.
The UHECR spectra that provide the dominant contribution to the boosted C$\nu$B flux are therefore only weakly affected by C$\nu$B scattering for the parameter region considered here.

A fully self-consistent treatment would incorporate UHECR--C$\nu$B scattering directly into the UHECR propagation calculation together with the standard propagation processes.
Such a calculation was performed in Ref.~\cite{Azeredo:2026qnc}.
For $m_\nu=0.1~{\rm eV}$, that study found only small modifications of the propagated UHECR spectrum for overdensities below approximately $\eta\sim10^9$, while UHECR--C$\nu$B interactions become comparable to interactions with diffuse photon backgrounds only for $\eta\gtrsim10^{10}$.
These results are consistent with our estimate and support neglecting the modification of the UHECR spectrum in the parameter region relevant to the present analysis.

%\end{appendix}

%%%%%%%%%%%%%%%%%%%%%%%%%%%%%%%%%%%%%%%%%%%%%%%
%%
\newpage
\bibliographystyle{JCAP}
\bibliography{references}

\end{document}